\documentclass{aa}  

\usepackage{graphicx}
\usepackage[varg]{txfonts}
\usepackage{longtable}
\usepackage{lscape}

\newcommand{\Teff}{\ensuremath{T_{\mathrm{eff}}}}
\newcommand{\logg}{\ensuremath{\log g}}

\newcommand{\FeH}{\ensuremath{\left[\mathrm{Fe}/\mathrm{H}\right]}}
\newcommand{\alphaFe}{\ensuremath{\left[\alpha/\mathrm{Fe}\right]}}

\newcommand{\GIRAFFE}{{\tt GIRAFFE}}
\newcommand{\UVES}{{\tt UVES}}

\newcommand{\ATLAS}{{\tt ATLAS9}}

\newcommand{\SYNTHE}{{\tt SYNTHE}}

\newcommand{\WIDTH}{{\tt WIDTH9}}
\newcommand{\MULTI}{{\tt MULTI}}
\newcommand{\IRAF}{{\tt IRAF}}

\begin{document} 

\title{Abundance of zirconium in the globular cluster 47~Tuc: \\a possible Zr--Na correlation?\thanks{Based on observations obtained at the European Southern Observatory (ESO) Very Large Telescope (VLT) at Paranal Observatory, Chile. Table~\ref{app-tab:abund-indiv-stars} is only available in electronic form at the CDS via anonymous ftp to cdsarc.u-strasbg.fr (130.79.128.5) or via http://cdsweb.u-strasbg.fr/cgi-bin/qcat?J/A+A/.}}

\author
	{
	E. Kolomiecas\inst{1}
	\and
    V. Dobrovolskas\inst{1}
    \and
    A. Ku\v{c}inskas\inst{1}
    \and
    P. Bonifacio\inst{2}
    \and
    S. Korotin\inst{3}
    }

\institute
	{
	Institute of Theoretical Physics and Astronomy, Vilnius University, Saul\.{e}tekio al. 3, Vilnius, LT-10257, Lithuania\\
	\email{edgaras.kolomiecas@ff.vu.lt}
	\and
	GEPI, Observatoire de Paris, Universit\'{e} PSL, CNRS, 5 Place Jules Janssen, 92190 Meudon, France
	\and
	Crimean Astrophysical Observatory, Nauchny 298409, Crimea
	}

\date{Received: date; accepted: date}


\abstract
{}
{We determined abundances of Na and Zr in the atmospheres of 237 RGB stars in Galactic globular cluster (GGC) 47~Tuc (NGC 104), with a primary objective of investigating possible differences between the abundances of Zr in the first generation (1P) and second generation (2P) stars.}
{For the abundance analysis, we used archival \UVES/\GIRAFFE\ spectra obtained during three different observing programmes. Abundances were determined from two \ion{Na}{i} and 
three \ion{Zr}{i} lines, using 1D hydrostatic \ATLAS\ model atmospheres. The target stars for the abundance analysis were limited to those with $4200\leq\Teff\leq4800$\,K. This is the largest sample of GGC stars in which Na and Zr abundances have been studied so far.}
{While our mean [Na/Fe] and [Zr/Fe] ratios agree well with those determined in the earlier studies, we find a weak but statistically significant correlation in the ${\rm [Zr/Fe]} - {\rm [Na/Fe]}$ plane. A comparison of the mean [Zr/Fe] abundance ratios in the 1P and 2P stars suggests a small but statistically significant Zr over-abundance in the 2P stars, $\Delta {\rm [Zr/Fe]}_{\rm 2P-1P}\approx+0.06$\,dex.
Also, our analysis shows that stars enriched in both Zr and Na are more centrally concentrated. However, we find no correlation between their distance from the cluster centre and their full spatial velocity, as indicated by the velocity dispersions at different mean values of [Zr/Fe] and [Na/Fe].
While there may be some influence of CN line blends on the determined Zr abundances, it seems very unlikely that the detected Zr--Na correlation, for the slightly higher Zr abundances in the 2P stars, would be caused by the CN blending alone.
}
{The obtained results indicate that, in 47~Tuc, some amount of Zr should have been synthesised by the same polluters that enriched 2P stars with the light elements. While sizeable amounts of Zr may be synthesised by both AGB stars ($M\sim1.5-5\,{\rm M}_{\odot}$) and massive rotating stars ($M\sim12-25\,{\rm M}_{\odot}$, $v_{\rm rot}>150$\,km\,s$^{-1}$), our data alone do not allow us to distinguish which of the two scenarios, or whether or not a combination of both, could have operated in this GGC.
}


\authorrunning{Kolomiecas et al.}
\titlerunning{Abundance of Zr in 47~Tuc}

\maketitle

\section{Introduction}

Nearly all Galactic globular clusters (GGCs) studied until now exhibit spreads and correlations (or anti-correlations) between the abundances of light chemical elements; for example the O--Na, Mg--Al \citep{CBG09a}, and Li--O \citep{PBM05,SBP10} anti-correlations, and the Na--Li \citep{BPM07} correlation.
Evolutionary scenarios proposed to explain these trends typically assume that stars in the GGCs formed during several star formation episodes and that the second-generation (2P) stars were polluted with light chemical elements synthesised in the first-generation (1P) stars. Various candidate polluters have been proposed, including fast-rotating massive stars \citep[FRMS; e.g.][]{DCS07}, asymptotic giant branch (AGB) stars \citep[e.g.][]{DAVD16}, and super-massive stars \citep[$\sim10^4\,{\rm M}_{\odot}$, SMS; e.g.][]{DH14,GCK18}.

In contrast to the light elements, variations in the abundances of neutron-capture elements (s-process and r-process elements) are less commonly reported and have been observed only in a few GGCs which belong to a class of so-called anomalous or Type-II clusters \citep{MMK15,MMR19}. Type-II clusters are amongst the most massive ones ($>10^{6}$\,M$_{\odot}$) and are characterised by significant spreads in the abundance of Fe and Fe-group elements \citep[e.g.][]{MMR19}. Although the remaining GGCs, namely Type-I, are generally homogeneous in the s-process abundances, there have been claims that relations between the abundances of s-process and light elements may exist in some Type~I GGCs as well \citep[see e.g.][]{GLS13,YMG13}. For example, \citet{GLS13} suggested the existence of a Na--Ba correlation in the red horizontal branch (RHB) stars of one of the closest GGCs, 47~Tuc. However, this finding was not corroborated by further investigation of this cluster \citep[e.g.][]{cordero14,TSA14}. Unfortunately, studies of the s-process elements in 47~Tuc (and in the majority of the other GGCs) have mostly been based on relatively small stellar samples. This makes it difficult to make a robust statistical assessment of the abundance spreads and correlations.

To partly fill this gap, we determined abundances of Na and Zr in 237 red giant branch (RGB) stars in 47~Tuc. This is the largest stellar sample used to study s-process elements in any GGC so far. As discussed in \citet{prantzos18}, evolution of Zr in the Galaxy is well reproduced by nucleosynthesis in massive rotating stars. Theoretical models predict that light s-process elements (Sr, Y, Zr) are produced in AGB stars ($1.5-5\,{\rm M}_\odot$) or massive stars ($>9\,{\rm M}_\odot$) during weak s-process nucleosynthesis, while the heavier s-process elements (Ba, La, etc.) are synthesised mostly in lower-mass AGB stars, via the main s-process \citep[e.g.][]{C18}. Thus, the detection of correlations between the abundances of light and s-process elements may allow more stringent constraints to be placed on the nature of possible polluters. 


\section{Observational data and atmospheric parameters\label{sect:obs-data}}

\subsection{Spectroscopic data and determination of atmospheric parameters\label{sect:spec-data-atmosph-pars}}

Abundances of Fe, Na, and Zr were determined using \texttt{VLT}/\GIRAFFE\ spectra that were obtained during three observing programmes (Table~\ref{tab:obs-journal}) and downloaded for the analysis from the ESO Advanced Data Products archive\footnote{\url{http://archive.eso.org/wdb/wdb/adp/phase3_spectral/form}}. The data were taken during three observing programmes: 072.D-0777(A) (PI: P.~Fran\c{c}ois); 073.D-0211(A) (PI: E.~Carretta); 088.D-0026(A) (PI: I.~McDonald). The median sky spectrum was subtracted from each individual stellar spectrum, and the remaining spectra were then continuum normalised using the \texttt{IRAF} \texttt{splot} task \citep{T86}. Each object in the programme 088.D-0026(A) was observed multiple times, therefore the sky-subtracted spectra from the individual exposures were co-added together. The typical signal-to-noise ratios in the spectra were ${\rm S/N}\sim70-150$ at 620\,nm.


\begin{table}[tb]
	\begin{center}
		\caption{Spectroscopic data used in this work.
			\label{tab:obs-journal}}
		\resizebox{\hsize}{!}{%
			\begin{tabular}{cccccccc}
				\hline\hline
				\noalign{\smallskip}
				Programme       & Date of      & Setting & $\lambda_{\mathrm{central}}$, & R  & Exposure, & Number of \\
				& observation &         & nm                            &    & s         & targets     \\
				\hline\noalign{\smallskip}
				072.D-0777(A)   & 2003-10-21  & HR13    & 627.3   & 26400     & 1500         & 112       \\
				& 2003-10-21  & HR13    & 627.3   & 26400     & 3600         & 121       \\
				073.D-0211(A)   & 2004-07-07  & HR13    & 627.3   & 22500     & 1600         & 113       \\
				088.D-0026(A)   & 2011-11-26  & HR13    & 627.3   & 26400     & $3\times700$ & 113       \\
				\hline
			\end{tabular}}
	\end{center}
\end{table}

The effective temperatures of the target RGB stars were determined using photometric observations from \citet{BS09} and the $\Teff-(V-I)$ calibration of \citet{RM05}. Although GAIA EDR3 photometry for the target stars was also available, we decided to use photometric data from \citet{BS09} in order to retain homogeneity and consistency with our previous studies of 47 Tuc (see e.g. \citealt{DKB14, CKK17}). The average difference between the effective temperatures determined using $(V-I)$ colour indices and those obtained from the GAIA EDR3 photometry and colour-effective temperature calibration from \citet{MBM21} is $\sim -59$\,K and does not affect the results obtained in this study. To check the reliability of our \Teff\ estimates, we determined Fe abundances in 111 RGB stars in the 073.D-0211(A) sample using a set of \ion{Fe}{i} lines (see Sect.~\ref{app-sect:Fe-abund}) and effective temperatures from photometry. On average, the residual slopes of the iron abundance versus the line excitation potential were smaller than 0.03\,dex/eV with a typical $1\sigma$ slope uncertainty of 0.04 dex/eV, suggesting good agreement between the photometric and spectroscopic effective temperatures.

Surface gravities of the target stars were determined using the classical relation between stellar mass, luminosity, effective temperature, and surface gravity. We assumed an identical mass of 0.89\,M$_{\odot}$ for all target RGB stars, as determined from the Yonsei-Yale isochrone\footnote{Version 2 from \url{http://www.astro.yale.edu/demarque/yyiso.html}} (12 Gyr, [M/H] = $-0.68$; \citealt{YY2}). We did not use spectroscopic gravities: although there were four \ion{Fe}{ii} lines available in the wavelength range of \GIRAFFE\ spectra, accurate equivalent width ($W$) measurements were possible for only two or three lines per target star spectrum which was insufficient for reliable gravity determination. Nevertheless, the mean iron abundance obtained for the entire sample of stars using \ion{Fe}{i} and \ion{Fe}{ii} lines agrees to within 0.01\,dex. For any particular star, the largest difference between the two Fe abundance estimates never exceeded 0.1\,dex. This suggests good agreement between the surface gravities determined from photometry and those obtained via the ionisation balance condition.

\begin{figure}[tb]
	\resizebox{\hsize}{!}{\includegraphics{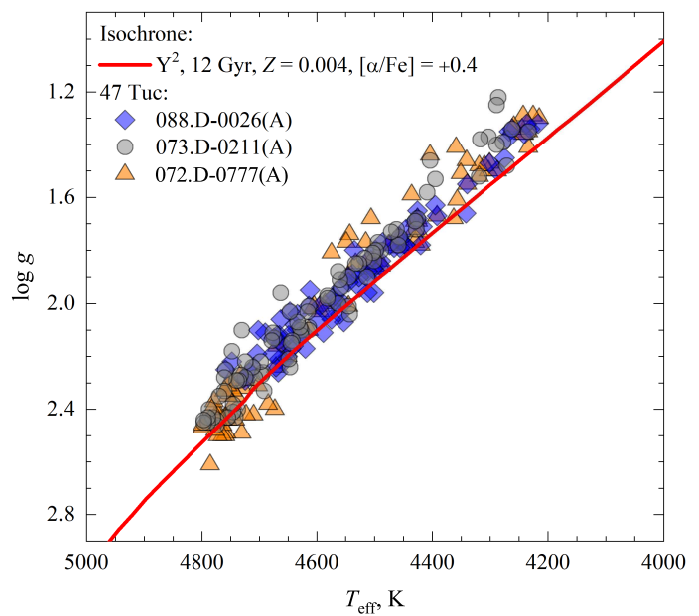}}
	\caption{Atmospheric parameters of the target RGB stars. Stars observed in different programmes are shown in different symbols and colours: orange triangles -- 072.D-0777(A), grey circles -- 073.D-0211(A), blue diamonds -- 088.D-0026(A). The Yonsei-Yale isochrone of 12 Gyr and [M/H] = $-$0.68, and [$\alpha$/Fe] = +0.4 is shown for comparison as a red solid line \citep{YY2}.}
	\label{app-fig:47Tuc-hrd}
\end{figure}

After visual inspection of the target star positions in the $\Teff-\log g$ diagram, we discarded 36 horizontal branch (HB) and AGB stars in the sample of 072.D-0777(A). We also discarded four HB stars from the 088.D-0026(A) sample and one HB star from the sample of 073.D-0211(A).
The determined effective temperatures and gravities of the target stars are shown in Fig.~\ref{app-fig:47Tuc-hrd} and are listed in Table~\ref{app-tab:abund-indiv-stars}.

\begin{figure}[tb]
	\resizebox{\hsize}{!}{\includegraphics{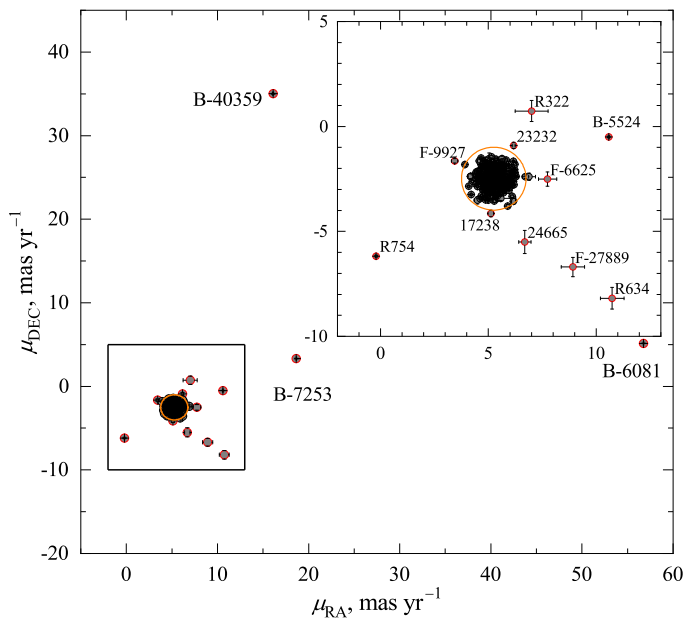}}
	\caption{Proper motions of the target RGB stars from the Gaia EDR3 catalogue \citep{G21}. Identifications of non-member stars are shown next to the symbols. The region that encloses motions falling within 1.5\,mas\,yr$^{-1}$ from the mean cluster proper motion is marked by a orange circle. Stars falling outside this region were rejected from the abundance analysis. Zoomed-in region marked by a square is shown in the top right corner.}
	\label{app-fig:47Tuc-pm}
\end{figure}

\begin{figure*}
	\resizebox{\hsize}{!}{\includegraphics{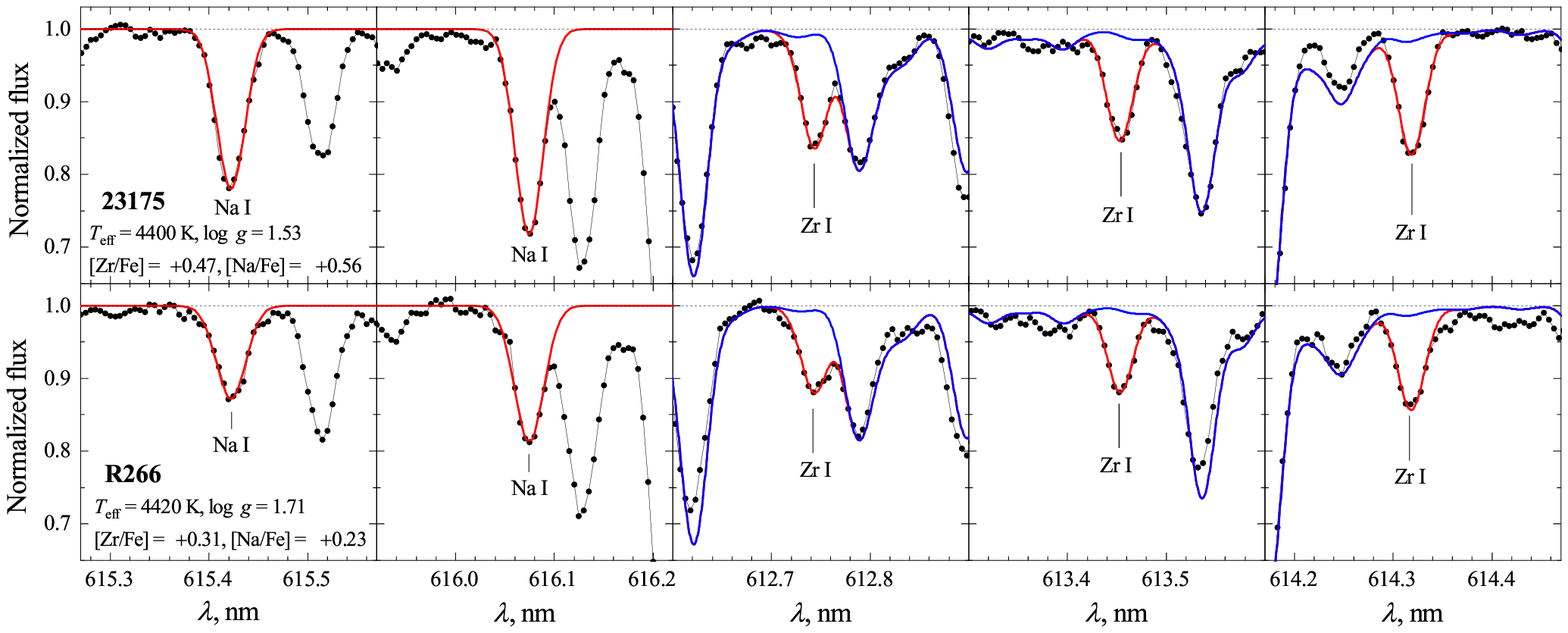}}
	\caption{Observed (dots) and best-fitted synthetic profiles of the \ion{Na}{i} (NLTE) and \ion{Zr}{i} (LTE) lines (red solid lines) in the \GIRAFFE\ spectra of two RGB stars in 47~Tuc: Na-rich (23175, 2P, top row) and Na-poor (R266, 1P, bottom row). Identification numbers of each star, their atmospheric parameters, and the determined Na and Zr abundances are provided in the leftmost panels of each row. The blue lines show synthetic spectrum computed with A(Zr) = 0 to show the possible influence of CN and other blends (further discussed in Sect.~\ref{sect:CN-blends}.).
	}
	\label{fig:spectral-lines}
\end{figure*}

\subsection{Radial velocities, proper motions, and cluster membership\label{sect:kinematics}}

Radial velocities of the target stars were measured from the \texttt{VLT}/\GIRAFFE\ spectra using the cross-correlation method with the \IRAF\ task {\tt fxcor}. For cross-correlation, we used the \ATLAS\ model with $\Teff=4500$\,K, $\logg=1.90$, $\FeH=-0.7$ to compute a synthetic spectrum with ${\rm R} = 22\,000$ in the wavelength range of $612-640$\,nm using \SYNTHE. We obtained a mean value for radial velocity of $-17.4$\,km\,s$^{-1}$ ($\sigma=7.9$\,km\,s$^{-1}$) with a typical measurement error of $\pm$0.1\,km\,s$^{-1}$ for individual stars. Stars with radial velocities falling within $\pm 3\sigma$ of the mean cluster radial velocity were assigned as cluster members.

Full spatial velocities of our target stars were obtained by combining their radial velocities (determined by us) with their proper motions from the Gaia EDR3 catalogue \citep{G21}. We matched the coordinates of the target stars (taken from \citealt{BS09}) with those of stars in the Gaia EDR3 catalogue using a $1^{\prime\prime}$ matching radius. The average proper motion of 47~Tuc, $\mu_{\rm RA} = 5.25$\,mas\,yr$^{-1}$ and $\mu_{\rm DEC} = -2.53$\,mas\,yr$^{-1}$, was taken from \citet{baumgardt19} and the proper motions of individual stars were computed relative to the average proper motion of the cluster. Following \citet{MMM18}, targets were selected as cluster members if their proper motions did not deviate by more than 1.5\,mas\,yr$^{-1}$ from the mean cluster proper motion. Based on this consideration, 13 stars were rejected from further analysis (Fig.~\ref{app-fig:47Tuc-pm}). 

Finally, we converted the proper motions from angular units to tangential velocities in km\,s$^{-1}$ assuming the distance of 4.5\,kpc to 47~Tuc. The obtained radial and tangential velocities were used to compute full spatial velocities.
After rejecting the likely non-members based on kinematical considerations, we further restricted our target sample to stars with $4200\leq\Teff\leq4800$\,K. This was done in order to limit the influence of uncertainties on the abundances determined at the lower and higher ends of \Teff\ range (see Sect.~\ref{sect:abund}).

\section{Determination of abundances\label{sect:abund}}

Elemental abundances were determined using the 1D hydrostatic \ATLAS\ model atmospheres \citep{K93,S05} enhanced in the $\alpha$-element abundances (O, Ne, Mg, Si, S, Ar, Ca, Ti) by $\alphaFe=+0.4$. In cases where target star spectra were available from several observing programmes, abundances of Fe, Na, and Zr were determined from each individual spectrum and average abundances were used for further analysis. The equivalent width method was used for obtaining abundances of Fe and Zr (but also see Sect.~\ref{sect:abund_Zr} below) while the Na abundances were measured by performing spectral line synthesis.

The stars used in the abundance analysis were limited to those with effective temperatures in the range $4200 \leq \Teff \leq 4800$\,K. This choice was made because: (a)  uncertainties in the continuum determination and a possible influence of molecular blends become higher at lower temperatures and lead to larger errors in Zr abundances; (b) uncertainties in the Zr abundances increase significantly due to weaker \ion{Zr}{i} lines at higher temperatures; and (c) at higher $\Teff$, \ion{Zr}{i} lines become increasingly too weak to be detected in stars with the lowest Zr abundance. The final sample thus consists of 237 RGB stars (Sect.~\ref{app-sect:final-sample}). 

Examples of the \ion{Na}{i} and \ion{Zr}{i} line fits are shown in Fig.~\ref{fig:spectral-lines} where we also provide a synthetic spectrum computed with Zr abundance set to zero to indicate a possible influence of CN lines on the determined Zr abundances (this effect is discussed in Sect.~\ref{sect:CN-blends}). Abundances of Fe, Na, and Zr determined in the individual stars are provided in Table~\ref{app-tab:abund-indiv-stars}, and their mean values and star-to-star spreads are listed in Table~\ref{tab:abundance-ranges} (Sect.~\ref{sect:mean-abundances}), with mean abundances in the 1P and 2P stars given in Table~\ref{tab:abundance-ratios-1p-2p} (Sect.~\ref{app-sect:abund-list-1P-2P}). Details of the abundance determination process are provided in Sects.~\ref{sect:abund_Fe}-\ref{sect:abund_Zr} below.

%

\begin{figure}[tb]
	\resizebox{\hsize}{!}{\includegraphics{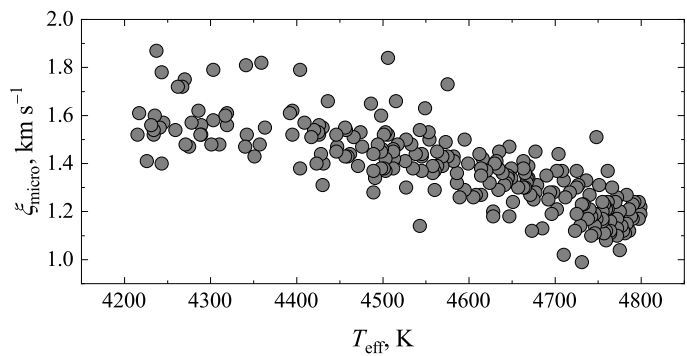}}
	\caption{Microturbulence velocity in the target RGB stars, $\xi_{\rm micro}$, versus the effective temperature of individual stars.}
	\label{app-fig:vmic-teff}
\end{figure}

\subsection{Abundance of Fe\label{sect:abund_Fe}}

Local thermodynamic equilibrium (LTE) was assumed in the analysis of Fe abundances which were determined using the equivalent width method. Abundances were obtained using a set of 17--28 \ion{Fe}{i} lines ($612.79-691.67$\,nm), where the lower level excitation potentials of the lines are in the range of 2.18--4.61\,eV (Table~\ref{app-tab:iron-list}). 

Microturbulence velocities, $\xi_{\rm micro}$, were determined alongside Fe abundances by enforcing a zero trend of \ion{Fe}{i} abundances with the line equivalent width. We discarded strong lines ($W>15$\,pm) because of their lower sensitivity to $\xi_{\rm micro}$. The obtained $\xi_{\rm micro}$ are plotted versus \Teff\ in Fig.~\ref{app-fig:vmic-teff}. The typical accuracy of the determined $\xi_{\rm micro}$ values was $\pm$0.2\,km\,s$^{-1}$ and was constrained by the number of \ion{Fe}{i} lines (17-28) available.



Iron abundances determined in the individual RGB stars are provided in Table~\ref{app-tab:abund-indiv-stars}. The obtained abundances show no strong dependence on either $\xi_{\rm micro}$ or \Teff. In addition, no significant discrepancies are seen in abundances determined using data from different observing programmes (Sect.~\ref{app-sect:Fe-abund}).

\subsection{Abundance of Na\label{sect:abund_Na}}

In the case of Na, abundances were measured assuming non-local thermodynamic equilibrium (NLTE) in the spectral line synthesis computations that were performed using the \MULTI\ code \citep{carlsson,KAL99}. The Na model atom used in the analysis consisted of 20 levels of \ion{Na}{i} and the ground state of \ion{Na}{ii}, with 46 radiative transitions between the different atomic levels. Collision rates with electrons and hydrogen were taken from \citet{IGE08} and \citet{BBD10} respectively \citep[see][for details]{DKB14}.
Two \ion{Na}{i} lines located at 615.4225 and 616.0747\,nm were used in the analysis. Synthetic line profiles were computed with the \MULTI\ code and were broadened with a Voigt profile to account for the instrumental and macroturbulence broadening. We used microturbulence velocities that were determined individually for each target star. The abundance of Na and macroturbulence velocity were then determined iteratively to achieve the best fit of the synthetic line profiles to the observed spectra (Fig.~\ref{fig:spectral-lines}). 

No significant systematic differences are seen in Na abundances obtained from the two \ion{Na}{i} lines or those determined using data from different observing programmes. There is a hint of a weak dependence of Na abundance on \Teff\ but it does not exceed 0.1\,dex (see Sect~\ref{app-sect:Na-abund}). The Na abundances determined in the target RGB stars are provided in Table~\ref{app-tab:abund-indiv-stars}.

%

\subsection{Abundance of Zr\label{sect:abund_Zr}}

\subsubsection{Zr abundance from \ion{Zr}{i} lines\label{sect:abund_Zr-I}}

Abundances of Zr were determined with the \WIDTH\ code \citep{S05,K05_2} using three \ion{Zr}{i} lines located at 612.7475, 613.4585, and 614.3252\,nm (Table~\ref{app-tab:line-list}). Line equivalent widths were measured by fitting Gaussian profiles with the \texttt{splot} task in \IRAF. 
Because the 612.7475\,nm \ion{Zr}{i} line is blended with the \ion{Fe}{i} 612.7906 nm line, the equivalent width of the former was determined using the \texttt{IRAF} task \texttt{deblend}. In addition, all \ion{Zr}{i} lines used in our study were affected by weak CN blends. We did not take this effect into account when measuring line equivalent widths but we did investigate its possible influence on the determined Zr abundances and did not find it to be critical (see Sect.~\ref{sect:CN-blends} for details). The final Zr abundances were obtained by averaging the measurements of all three \ion{Zr}{i} lines.

For the abundance determination, we used an updated \ion{Zr}{i} ionisation energy of 6.634\,eV from \citet{HHM86} instead of the default value of 6.840\,eV implemented in the original versions of the Kurucz \ATLAS, \WIDTH, and \SYNTHE\ packages \citep{S05}. This led to a systematic change in zirconium abundance of $+0.29$\,dex at $\Teff=4200$\,K and $+0.25$ dex at $\Teff=4800$\,K.

There is good agreement between Zr abundances obtained from the three \ion{Zr}{i} lines and those determined using data from different observing programmes. The obtained abundances do not show trends with the microturbulence velocity or \Teff\ (see Sect.~\ref{app-sect:Zr-abund}). Abundances of Zr  obtained for the individual target RGB stars are provided in Table~\ref{app-tab:abund-indiv-stars}. 


It is important to stress that in our analysis, the hyperfine structure splitting (HFS) of \ion{Zr}{i} lines was not taken into account. To assess its influence, we synthesised all three \ion{Zr}{i} lines with and without taking HFS into account. For this, we used the \ATLAS\ model atmosphere with $\Teff=4320$\,K, $\log g =1.52$, $\xi_{\rm micro}=1.56$\,km\,s$^{-1}$, $\FeH=-0.78$, and $\alphaFe=+0.4$. The effective temperature of the model atmosphere is similar to those of the `coolest' stars in our sample ($\Teff\approx4200$\,K). Because at a given Zr abundance \ion{Zr}{i} lines are stronger in stars with lower $\Teff$, one may anticipate larger HFS-related effects at the lower end of the \Teff\ scale. Spectral line synthesis was done with the \SYNTHE\ package. Line oscillator strengths for the individual HFS components were taken from \citet{TSA14} while for the computation of line profiles without HFS these components were added up for each \ion{Zr}{i} line (Table~\ref{app-tab:zr-hfs-test}). We used ${\rm A(Zr)}=2.16$ for both HFS and non-HFS lines. The equivalent widths of synthetic lines were measured by fitting the Gaussian profile, and abundances of Zr were determined from the obtained equivalent widths using the \WIDTH\ code. As seen from Table~\ref{app-tab:zr-hfs-test}, the largest abundance difference of +0.02\,dex is seen in the case of the 612.7475\,nm line, while the difference for lines at 613.4585 and 614.3252\,nm is smaller than $0.01$\,dex. At higher $\Teff$ values, the differences are even smaller because the \ion{Zr}{i} lines are weaker. We therefore conclude that HFS of the \ion{Zr}{i} lines does not have a significant impact on Zr abundances determined in our study.

\begin{table}
	\begin{center}
		\caption{Impact of HFS on the determination of Zr abundance from the \ion{Zr}{i} lines used in this study.
			\label{app-tab:zr-hfs-test}}
		\resizebox{8cm}{!}{%
			\begin{tabular}{cccccc}
				\hline\hline
				\noalign{\smallskip}
				& \multicolumn{2}{c}{HFS}    & \multicolumn{2}{c}{no-HFS}   & $\Delta {\rm A(Zr)}$ \\
				$\lambda$, nm   & \textit{W}, pm  & A(Zr)           & \textit{W}, pm   & A(Zr)            & (HFS--noHFS)  \\
				\hline\noalign{\smallskip}
				\multicolumn{6}{c}{$\Teff=4320$\,K} \\
				612.7475        & 5.57     & 2.177           & 5.46      & 2.159            & +0.018   \\
				613.4585        & 5.61     & 2.184           & 5.59      & 2.181            & +0.003   \\
				614.3252        & 6.07     & 2.174           & 6.02      & 2.166            & +0.008   \\
				\hline
			\end{tabular}
		}
	\end{center}
\end{table}

The 088.D-0026(A) sample contains one star ---`R265' or `Lee 4710'--- which is strongly enriched in s-process elements \citep{CHJ15}. For this star, we determined a zirconium abundance of [Zr/Fe] = $0.79\pm0.15$ (error is standard deviation due to line-to-line abundance scatter), which is in very good agreement with the value of $0.83\pm0.17$ obtained by \citet{CHJ15}. The atmospheric parameters of the star determined in both studies also agree very well: we obtained $\Teff=4452$ K, $\log g=1.82$, $\xi_\mathrm{micro} = 1.55$\,km\,s$^{-1}$, and $\FeH=-0.79$ while \citet{CHJ15} derived $\Teff=4475$ K, $\log g=1.60$, $\xi_\mathrm{micro} = 1.65$\,km\,s$^{-1}$, and $\FeH=-0.78$. Due to its anomalous abundance patterns, Lee~4710 was excluded from further analysis.

\subsubsection{Zr abundance from \ion{Zr}{ii} lines\label{sect:abund_Zr-II}}

To some extent, the robustness and reliability of Zr abundances obtained from \ion{Zr}{i} lines may be assessed by comparing them with those determined using \ion{Zr}{ii} lines. Unfortunately, no existing \texttt{VLT/GIRAFFE} observations of 47~Tuc extend over the wavelengths of \ion{Zr}{ii} lines. We therefore used \texttt{VLT/UVES} spectra of 13 RGB stars in 47~Tuc obtained by \citet{TSA14} whose wavelength range covers several \ion{Zr}{ii} lines. Although the authors focused on the coolest RGB stars in 47~Tuc, there are three stars with $4200\leq\Teff\leq4800$\,K (our \Teff\ scale) that have been observed both in their \UVES\ and our \GIRAFFE\ samples.
To the best of our knowledge, these are the only archival \UVES\ spectra of our \GIRAFFE\ targets which cover the wavelength range of \ion{Zr}{ii} lines. 

The results of our analysis of the three stars suggest good agreement between Zr abundances obtained using \ion{Zr}{i} and \ion{Zr}{ii} lines (see Sect.~\ref{app-sect:Zrii-abund} for details). Even in the extreme cases, the differences do not exceed $0.10$\,dex. For each star, the corresponding averages obtained from \ion{Zr}{i} and \ion{Zr}{ii} lines show even better agreement, with the differences being smaller than 0.03\,dex. We therefore conclude that, despite the small number of stars used in this test, the consistency of Zr abundances obtained from individual \ion{Zr}{i} and \ion{Zr}{ii} lines nevertheless suggests that the influence of various effects (NLTE, line blends, line data, systematic errors in the analysis procedures, etc.) on the determined Zr abundances is likely minor (the case of CN blends is discussed separately in Sect.~\ref{sect:CN-blends}).

\subsection{Abundance determination errors\label{sect:abund_errors}}

To estimate the errors in the determined abundances of Fe, Na, and Zr, we followed the prescription provided in \citet{CKK17}. Briefly, we account for the uncertainties in stellar parameters, continuum determination, and line profile fitting, with the final abundance errors determined by adding error components in quadrature. Detailed descriptions of the methodology, error determination procedure, and the determined abundance errors are provided in Sect.~\ref{app-sect:abund-err}, with typical values being in the range of $0.09-0.16$\,dex (Tables~\ref{app-tab:Fe-abund-errors} - \ref{app-tab:Zr-abund-errors}).

\section{Results and Discussion\label{sect:results-and-disc}}

\subsection{Mean abundances of Fe, Na, and Zr in the target RGB stars in 47~Tuc\label{sect:mean-abundances}}

For the sample of 237 stars ($4200\leq\Teff\leq4800$\,K, Sect.~\ref{sect:abund}), we obtained a mean $\langle{\rm [Fe/H]}\rangle=-0.75\pm 0.05$ (Table~\ref{tab:abundance-ranges}; the error denotes standard deviation due to the star-to-star abundance scatter). This agrees well with the values determined in earlier studies (e.g. $\langle{\rm [Fe/H]}\rangle=-0.74\pm0.05$ by \citealt[][for 114 RGB stars]{CBG09a}; $\langle{\rm [Fe/H]}\rangle=-0.77\pm0.08$ by \citealt[][for 44 RGB stars]{WPC17}), as well as with the mean iron abundance measured by us for the same target stars using \ion{Fe}{ii} lines ($\langle{\rm [Fe/H]_{\rm Fe~II}}\rangle=-0.73\pm 0.09$\footnote{As was already mentioned in Sect.~\ref{sect:spec-data-atmosph-pars}, typically, only 2-3 \ion{Fe}{ii} lines could be measured reliably in the target star spectra. For this reason, we did not determine stellar gravities using the assumption of ionization equilibrium.}). For reference, we used the solar iron abundance of ${\rm A(Fe)_{\odot}}=7.55 \pm 0.06$ that was determined in Sect.~\ref{app-sect:ref-abnd} utilizing the same set of \ion{Fe}{i} lines as in the analysis of target RGB stars in 47~Tuc. 

As expected, the obtained [Na/Fe] ratios show a considerable range of scatter, $0.05\leq{\rm [Na/Fe]}\leq0.78$, with a mean value of $\langle {\rm [Na/Fe]} \rangle = 0.41 \pm 0.16$ (Table~\ref{tab:abundance-ranges}; we remind that abundances of Na were determined assuming NLTE and those of Fe using LTE). The scatter range, as well as the mean [Na/Fe] value, agree well with those obtained in other studies; for example $\Delta {\rm [Na/Fe]=0.76}$ and $\langle {\rm [Na/Fe]} \rangle = 0.47\pm0.15$ (\citealt{CBG09a}, 147 RGB stars), $\Delta {\rm [Na/Fe]=0.56}$ and $\langle {\rm [Na/Fe]} \rangle = 0.36\pm0.18$ (\citealt{WPC17}, 27 RGB stars).

For Zr, the range of star-to-star scatter is similar to that obtained for Na, ${0.11\leq\rm [Zr/Fe]}\leq0.68$, with a mean value of $\langle {\rm [Zr/Fe]}\rangle = 0.35\pm0.09$ (Table~\ref{tab:abundance-ranges}). Again, the  mean value and abundance dispersion due star-to-star scatter agree well with those obtained in earlier studies; for example $\langle {\rm [Zr/Fe]} \rangle = 0.52\pm0.03$ (\citealt{WCM10}, 5 AGB stars), and $\langle {\rm [Zr/Fe]} \rangle = 0.41\pm0.19$ (\citealt{TSA14}, 13 RGB stars). A larger difference is seen when comparing with the results of \citet[][average in seven RGB/AGB stars]{WCS06} who obtain $\langle {\rm [Zr/Fe]} \rangle = 0.69\pm0.15$.


\begin{table}[tb]
	\begin{center}
		\caption{Minimum, maximum, and mean Fe, Na, and Zr abundance ratios determined in the sample of 237 RGB stars in 47~Tuc.
			\label{tab:abundance-ranges}}
		\resizebox{\hsize}{!}{%
			\begin{tabular}{lcccccccc}
				\hline\hline
				\noalign{\smallskip}
				&  ${\rm [X/H]}_{\rm min}$  &  ${\rm [X/H]}_{\rm max}$  &  $\langle{\rm [X/H]}\rangle$  &  ${\rm [X/Fe]}_{\rm min}$  &  ${\rm [X/Fe]}_{\rm max}$  &  $\langle{\rm [X/Fe]}\rangle$ \\
				\hline\noalign{\smallskip}
				Fe~I (LTE)    &  $-0.87$  &  $-0.54$    &  $-0.75\pm 0.05$    &     --     &     --    &           --          \\				
				Na~I (NLTE)   &  $-0.70$  &  $0.03$     &  $-0.34\pm 0.16$    &  $0.05$  &  $0.78$  &  $0.41\pm 0.16$  \\				
				Zr~I  (LTE)   &  $-0.64$  &  $-0.07$    &  $-0.40\pm 0.09$    &  $0.11$  &  $0.68$  &  $0.35\pm 0.09$  \\
				\hline
			\end{tabular}
		}
	\end{center}
	Note: Errors denote the standard deviation due to star-to-star abundance variation.	
\end{table}

\subsection{A possible Zr--Na correlation?\label{sect:zr-na-correl}}

\begin{figure}[tb]
	\centering
	\includegraphics[width=8cm]{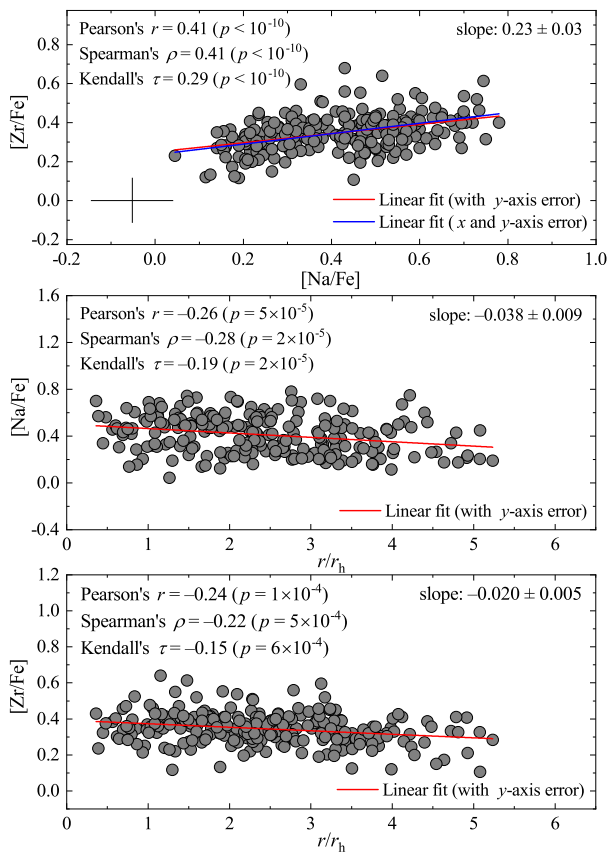}
	\caption{[Zr/Fe] ratios in the sample of 237 RGB stars in 47~Tuc.
			Top panel: [Zr/Fe] versus [Na/Fe] ratios. Middle and bottom panels, respectively: [Zr/Fe] and [Na/Fe] ratios versus projected distance from the cluster centre, $r/r_{\rm h}$. Linear least-square best fits are shown as solid lines, fit parameters, correlation coefficients, and the $p$-values are provided in each individual panel. Typical errors of the individual abundance measurements are indicated in the top panel.}
	\label{fig:zr-correlations}
\end{figure}


The obtained Zr and Na abundances suggest the existence of a weak correlation in the ${\rm [Zr/Fe]}-{\rm [Na/Fe]}$ plane (Fig.~\ref{fig:zr-correlations}, top). To estimate its statistical significance, we assumed a null hypothesis that the Pearson's and the non-parametric Spearman's and Kendall's correlation coefficients ($r$, $\rho$, and $\tau$, respectively) are zero in this plane, that is, that the two abundance ratios are uncorrelated. Although the determined correlation coefficients suggest only a weak relation between the Zr and Na abundances, the two-tailed probability, $p$, for obtaining such Student's $t$-values in the given data set by chance is indeed extremely small, as seen from the determined $p$-values in Table~\ref{tab:p-values}.

\begin{figure}[tb]
	\centering
	\includegraphics[width=9.1cm]{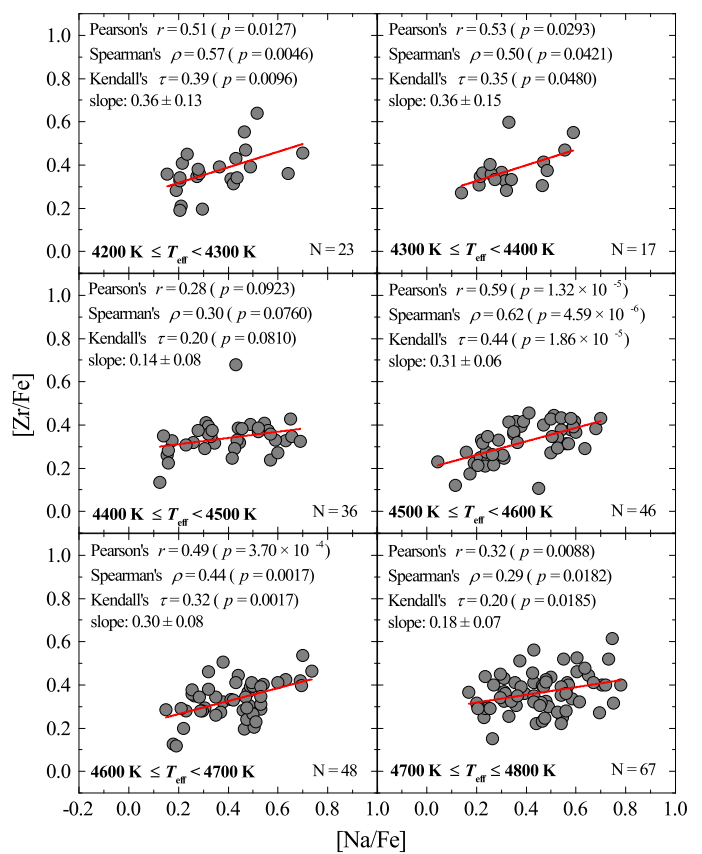}
	\caption{ ${\rm [Zr/Fe]}-{\rm [Na/Fe]}$ plots for stars divided into bins of $\Delta\Teff=100$\,K ($N$ is the number of stars in each bin). Values of different correlation coefficients and the corresponding $p$-values are provided for each temperature bin.}
	\label{fig:zr-na-correlations-Teff-bins}
\end{figure}

A qualitatively similar picture is seen also in the ${\rm [Zr/Fe]}-{\rm [Na/Fe]}$ plots constructed using stars divided into bins of $\Delta\Teff=100$\,K. Here, we obtained $p<0.05$ in all but one of the  temperature bins, namely $4400\leq\Teff\leq4500$\,K; Fig.~\ref{fig:zr-na-correlations-Teff-bins}.

%

\begin{table}[tb]
	\begin{center}
		\caption{ Pearson, Spearman and Kendall correlation coefficients and the corresponding $p$-values in the [Zr/Fe]--[Na/Fe] and abundance--distance planes.}
		\label{tab:p-values}
		\resizebox{\hsize}{!}{%
			\begin{tabular}{lccccccc}
				\hline\hline
				\noalign{\smallskip}
				&   \multicolumn{2}{c}{Pearson}     &  \multicolumn{2}{c}{Spearman}  &   \multicolumn{2}{c}{Kendall}   \\
				&           $r$          &         $p$        &            $\rho$          &        $p$        &           $\tau$        &        $p$         \\
				\hline\noalign{\smallskip}
				${\rm [Zr/Fe]} - {\rm [Na/Fe]}$                &  $0.41$  & $<10^{-10}$ &  $0.41$  & $<10^{-10}$ & $0.29$ &  $<10^{-10}$ \\      
				${\rm [Na/Fe]} - r/r_{\rm h}$                  & $-0.26$ & $5\times10^{-5}$ & $-0.28$ & $2\times10^{-5}$ & $-0.19$ &  $2\times10^{-5}$ \\                  
				${\rm [Zr/Fe]} - r/r_{\rm h}$                  & $-0.24$ & $1\times10^{-4}$ & $-0.22$ & $5\times10^{-4}$ & $-0.15$ &  $6\times10^{-4}$ \\          
				
				\hline
			\end{tabular}
		}
	\end{center}
\end{table}

Our data also indicate that stars with the highest [Na/Fe] and [Zr/Fe] ratios tend to concentrate towards the cluster centre, as suggested by trends seen in the ${\rm [Zr/Fe]}-r/r_{\rm h}$ and ${\rm [Na/Fe]}-r/r_{\rm h}$ planes (Fig.~\ref{fig:zr-correlations}), where $r$ is the projected distance from the cluster centre of the given target star and $r_{\rm h}$ is the half-light radius of 47~Tuc ($r_{\rm h}=174^{\prime\prime}$, \citealt{T93}). This is in line with the findings of previous studies of this and other GGCs (see e.g. \citealt{BL18}). The $p$-values computed for  all correlation coefficients are extremely small (Table~\ref{tab:p-values}).

\begin{figure}[tb]
	\centering
	\includegraphics[width=8cm]{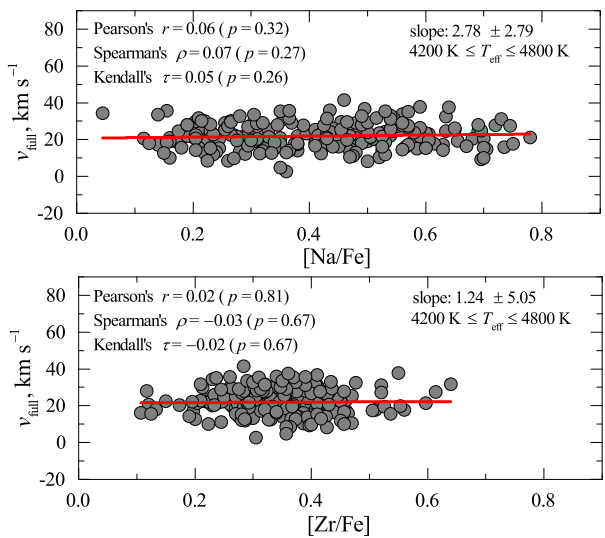}
	\caption{Full spatial velocities of 237 RGB stars in 47~Tuc versus their [Na/Fe] and [Zr/Fe] ratios (top and bottom, respectively).}
	\label{fig:zr-na-rrh}
\end{figure}

\begin{figure}[tb]
	\centering
	\includegraphics[width=7.5cm]{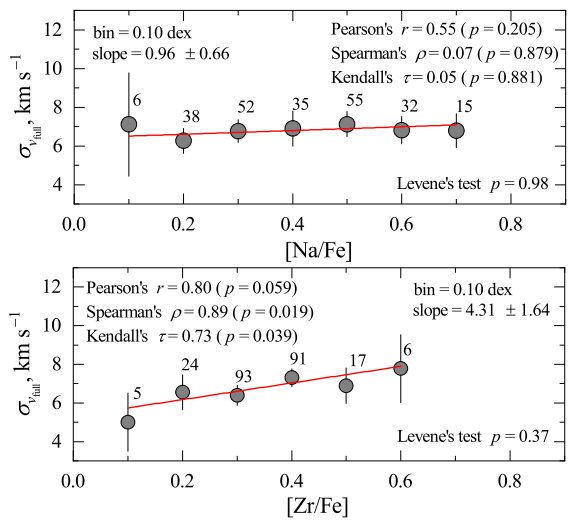}
	\caption{Dispersions of full spatial velocities of the target RGB stars computed in non-overlapping [Zr/Fe] (top panel) and [Na/Fe] (bottom panel) abundance bins of 0.1\,dex. Solid lines are linear least-square fits and the number of stars in each abundance bin is indicated next to each data point. Indicated are the $p$-values that were obtained using different methods (including the Levene's test, see Sect.~\ref{app-sect:Levene-test}), and error bars were computed using the bootstrapping technique. }
	\label{fig:vel-disp}
\end{figure}

We find no correlation between the [Zr/Fe] and [Na/Fe] ratios and the full spatial velocities of the target stars (Fig.~\ref{fig:zr-na-rrh}). Similarly, there is no statistically significant correlation between the two abundance ratios and velocity dispersions computed in the non-overlapping 0.1\,dex-wide abundance bins, as indicated by the results of the $t$-test and the Levene's test (Fig.~\ref{fig:vel-disp}; see also Sect.~\ref{app-sect:Levene-test}; we note that this result is not affected by the size of the binning step and/or whether the bins are overlapping or not). This is in contrast to the findings of \citet{KDB14} who, based on the analysis of 101 TO stars in 47~Tuc, detected a weak but statistically significant correlation between the radial velocity dispersion and [Na/O] and [Li/Na] abundance ratios. One possible explanation for this discrepancy is that [Na/O] shows a significantly larger 1P--2P variation range ($>1$\,dex) than the [Na/Fe] or [Zr/Fe] ratios which may make the detection of a possible correlation more difficult in the latter two cases. Nevertheless, while our stellar sample is more than two times larger, we find no differences in the kinematical properties of 1P and 2P stars as traced by the [Na/Fe] and/or [Zr/Fe] abundance ratios.



\subsection{Influence of CN blends on the determined Zr abundances\label{sect:CN-blends}}

\begin{figure}[tb]
	\centering
	\includegraphics[width=\hsize]{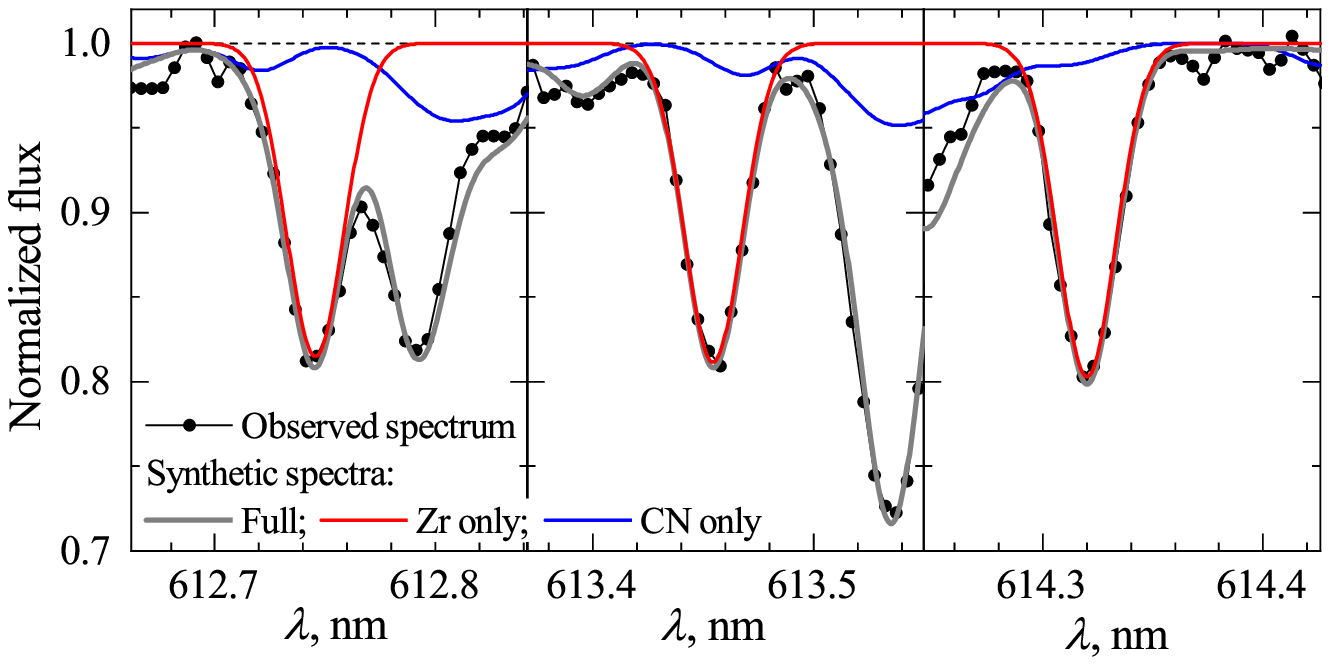}
	\caption{\ion{Zr}{i} lines in \GIRAFFE\ spectrum of the target RGB star 23821 (dots) overlaid with the synthetic spectrum (grey solid line) computed using
		$\Teff=4404$\,K, $\log g=1.46$, $\FeH=-0.76$, $\xi_{\rm micro}=1.38$\,km\,s$^{-1}$, and $\xi_{\rm macro}=4.80$\,km\,s$^{-1}$. Different panels show the environment of \ion{Zr}{i} lines located at 612.7475\,nm (left), 613.4585\,nm (middle), and 614.3252\,nm (right). Red and blue solid lines are synthetic profiles of the individual \ion{Zr}{i} and CN lines, respectively. 
	}
	\label{fig:ZrI-blends-RGB}
\end{figure}

One issue of concern is that all three \ion{Zr}{i} lines used in this study are blended with CN lines (see Fig.~\ref{fig:spectral-lines}, \ref{fig:ZrI-blends-RGB}). Because abundances of C and N differ significantly in the 1P and 2P stars \citep[e.g.][]{CGL05,MMGH20}, the strengths of CN blends will differ as well. This may lead to systematically stronger blended lines (i.e. Zr+CN) and higher Zr abundances in the 2P stars when measured from the line equivalent widths without taking the influence of CN lines into account (i.e. as was done in our study because abundances of C and N are not known for all target stars; see Sect~\ref{sect:CN-blends-test-knownCN}). Because the 2P stars have higher Na abundances, the CN blends may therefore lead to systematically overestimated Zr abundances in these stars, which in turn may produce a spurious correlation between Zr and Na abundances.
We carried out two tests to assess the possible influence of CN blends, the results of which are described below.

\subsubsection{Influence of CN blends estimated using extreme C and N abundances\label{sect:CN-blends-test-extremeCO}}

For the first test, we computed synthetic profiles of the three \ion{Zr}{i} lines at 612.7475, 613.4585, and 614.3252\,nm, with the CN line profiles calculated using CNO abundances determined in the 1P and 2P stars of this GGC from the {\tt APOGEE-2} data by \citet[][82 RGB stars in total]{MMGH20}. For this, we selected three sets of CNO abundances corresponding to the CN-weak, CN-intermediate, and CN-strong lines:

\begin{itemize}
	\item[$\bullet$]
	${\rm [C/Fe]}=+0.20$, ${\rm [N/Fe]}=+0.20$, ${\rm [O/Fe]}=+0.80$ (case-A, "CN-weak", 1P);
	\item[$\bullet$]
	${\rm [C/Fe]}=-0.20$, ${\rm [N/Fe]}=+1.10$, ${\rm [O/Fe]}=+0.60$ (case-B, "CN-intermediate");
	\item[$\bullet$]
	${\rm [C/Fe]}=-0.80$, ${\rm [N/Fe]}=+1.70$, ${\rm [O/Fe]}=+0.10$ (case-C, "CN-strong", 2P).
\end{itemize}

In all three cases, the same Zr abundance was used, namely ${\rm A(Zr)}=2.16$. The \ion{Zr}{i} line profiles were computed using two \ATLAS\ models with $\Teff=4320$\,K, $\log g = 1.52$, $\FeH=-0.78$, and 4675\,K, $\log g=2.13$, and $\FeH=-0.67$ (both with $\alphaFe=+0.4$), representing stars at the `cooler' and `hotter' ends of the effective temperature range in our target sample. We used CN line data from \citet{brooke14} which were taken from the Kurucz's website\footnote{\url{http://kurucz.harvard.edu/molecules.html}}. The equivalent widths of synthetic \ion{Zr}{i} lines were determined by fitting the Gaussian profiles, that is, in the same way as in the measurements of Zr abundances in the target stars (Sect.~\ref{sect:abund_Zr-I}). The measured equivalent widths were then used with the Kurucz \WIDTH\ code to derive Zr abundances.


\begin{table*}
	\begin{center}
		\caption{Impact of CN line blends on the determination of Zr abundances from \ion{Zr}{i} lines at different CNO abundances (see Sect.~\ref{sect:CN-blends-test-extremeCO} for details).
			\label{tab:zri-cn-test}}
				\resizebox{12cm}{!}{%
		\begin{tabular}{cccccccccc}	
			\hline\hline
			\noalign{\smallskip}
			& \multicolumn{2}{c}{case-A} & \multicolumn{2}{c}{case-B} & \multicolumn{2}{c}{case-C}  & $\Delta$A(Zr) &  $\Delta$A(Zr)  & $\Delta$A(Zr) \\
			$\lambda$, nm  & $W$, pm  & A(Zr)           & $W$, pm  & A(Zr)           & $W$, pm  & A(Zr)            & (C--A)     &  (B--A)      &  (C--B) \\
			\hline\noalign{\smallskip}
			\multicolumn{10}{c}{$\Teff=4320$\,K, $\log g = 1.52$, $\FeH=-0.78$} \\
			612.7475      & 6.09     &  2.26           &   6.20   &   2.28          &   6.41   &  2.31            & +0.05     &  +0.02      &  +0.03 \\
			613.4585      & 5.82     &  2.22           &   6.02   &   2.25          &   6.19   &  2.27            & +0.05     &  +0.03      &  +0.02 \\
			614.3252      & 6.54     &  2.25           &   6.68   &   2.27          &   6.77   &  2.28            & +0.03     &  +0.02      &  +0.01 \\	               
			\noalign{\smallskip}
			\multicolumn{10}{c}{$\Teff=4675$\,K, $\log g=2.13$, and $\FeH=-0.67$} \\
			612.7475      & 2.15     &  2.26           &   2.32   &   2.31          &   2.43   &  2.33            & +0.07     &  +0.05      &  +0.02 \\
			613.4585      & 2.00     &  2.25           &   2.18   &   2.30          &   2.30   &  2.33            & +0.08     &  +0.05      &  +0.03 \\
			614.3252      & 2.53     &  2.29           &   2.65   &   2.32          &   2.72   &  2.34            & +0.05     &  +0.03      &  +0.02 \\
			\hline
		\end{tabular}
				}		
	\end{center}
\end{table*}

\begin{figure}[tb]
	\centering
	\includegraphics[width=\hsize]{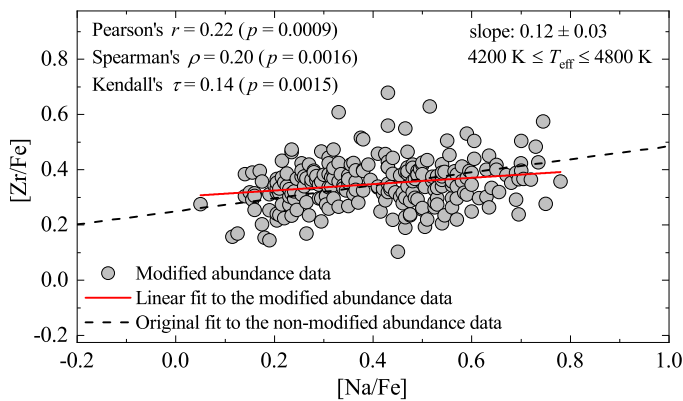}
	\caption{Influence of CN lines on the Zr--Na correlation. To account for the influence of CN blends, the original Zr abundances obtained from \ion{Zr}{i} lines in Sect.~\ref{sect:abund_Zr-I} were modified by adding a correction which changed linearly from $+0.04$\,dex at ${\rm [Na/Fe]}=0.10$ to $-0.04$\,dex at ${\rm [Na/Fe]}=0.75$. A linear fit to the modified data (filled grey circles) is shown as a red solid line. The black dashed line is linear fit to the non-modified abundances in the ${\rm [Zr/Fe]} - {\rm [Na/Fe]}$ plane shown in Fig.~\ref{fig:zr-correlations} (top panel) shifted to match the mean [Zr/Fe] value in the modified data at ${\rm [Na/Fe]}=0.42$.
	}
	\label{fig:zr-abund-CN-adjusted}
\end{figure}

\begin{table}
	\begin{center}
		\caption{Impact of CN line blends on the determination of Zr abundances from \ion{Zr}{ii} lines at different CNO abundances (see Sect.~\ref{sect:CN-blends-test-extremeCO} for details).
			\label{tab:zrii-cn-test}}
			\resizebox{7.5cm}{!}{%
		\begin{tabular}{cccccc}
			\hline\hline
			\noalign{\smallskip}
			& \multicolumn{2}{c}{case-A} & \multicolumn{2}{c}{case-C}  & $\Delta {\rm A(Zr)}$ \\
			$\lambda$, nm & $W$, pm  & A(Zr)           & $W$, pm  & A(Zr)            & (C--A)     \\
			\hline\noalign{\smallskip}
			\multicolumn{6}{l}{$\Teff=4320$\,K} \\
			511.2270         & 3.86     & 2.23            & 4.85     & 2.42             & $+0.19$   \\
			535.0089         & 1.56     & 2.24            & 1.44     & 2.20             & $-0.04$   \\
			535.0350         & 2.50     & 2.17            & 2.46     & 2.16             & $-0.01$   \\
			\hline
		\end{tabular}}
	\end{center}
\end{table}

The obtained results show that while the influence of CN blends is non-negligible, even in the extreme cases it does not exceed 0.08\,dex for the individual \ion{Zr}{i} lines (Table~\ref{tab:zri-cn-test}). The average C--A (2P--1P) corrections obtained for the three \ion{Zr}{i} lines are 0.04\,dex and 0.07\,dex at $\Teff=4320$\,K and 4675\,K, respectively. We stress that CNO abundances selected for cases A and C represent the extreme values observed in this GGC. However, the CNO abundance variation will be smaller for the majority of stars, and therefore the actual mean corrections will also be somewhat smaller. For further tests, we assumed a conservative change in the slope of the [Zr/Fe]--[Na/Fe] plane due to CN blends of $\approx+0.11.$.


To eliminate the influence of CN line blends, we corrected Zr abundances determined from the three \ion{Zr}{i} lines in Sect.~\ref{sect:abund_Zr-I} by removing a contribution to the slope in the ${\rm [Zr/Fe]} - {\rm [Na/Fe]}$ plane that may be caused by CN blends alone. To this end, we added a correction that changed linearly from $+0.04$\,dex at ${\rm [Na/Fe]}=0.10$ to $-0.04$\,dex at ${\rm [Na/Fe]}=0.75$ which corresponds to the change in slope of $\approx-0.11$.
This left us with a residual slope in the ${\rm [Zr/Fe]} - {\rm [Na/Fe]}$ equal to $\approx+0.12$ which still gives a change of $+0.07$\,dex in Zr abundance over the range of $0.10\leq{\rm [Na/Fe]}\leq0.75$.

Using this modified data set, we computed Pearson, Spearman, and Kendall-$\tau$ correlation coefficients in the ${\rm [Zr/Fe]} - {\rm [Na/Fe]}$ plane. The Student's $t$-test shows that the probability of obtaining such $t$-values by chance is lower than $p=0.002$ (Fig.~\ref{fig:zr-abund-CN-adjusted}). The results of this test therefore support the notion that the Zr--Na correlation seen in our data is in fact real, despite a small possible influence of CN blends.

We did not search for a possible Zr--Na correlation using Zr abundances determined from \ion{Zr}{ii} lines because these abundances were determined for only three stars in our sample (Sect~\ref{sect:abund_Zr-II}).
Nevertheless, we estimated the influence of CN blends on these lines using the same methodology as described above. The obtained results (Table~\ref{tab:zrii-cn-test}) show that the influence of CN blends on the 511.2270\,nm line is significant and may lead to overestimation of Zr abundances by up to $\approx 0.19$\,dex. The remaining two lines, 535.0089\,nm and 535.0350\,nm, are weakly affected and therefore may be used as good Zr abundance indicators regardless of CNO abundance variations (we note that the 535.0350\,nm is also influenced by weak \ion{V}{ii} and \ion{Ti}{i} blends but the effect is minor; see Sect.~\ref{app-sect:Zrii-abund}).

\subsubsection{Influence of CN blends in the subsample of RGB targets with known C and N abundances\label{sect:CN-blends-test-knownCN}}

\begin{figure}[tb]
	\centering
	\includegraphics[width=\hsize]{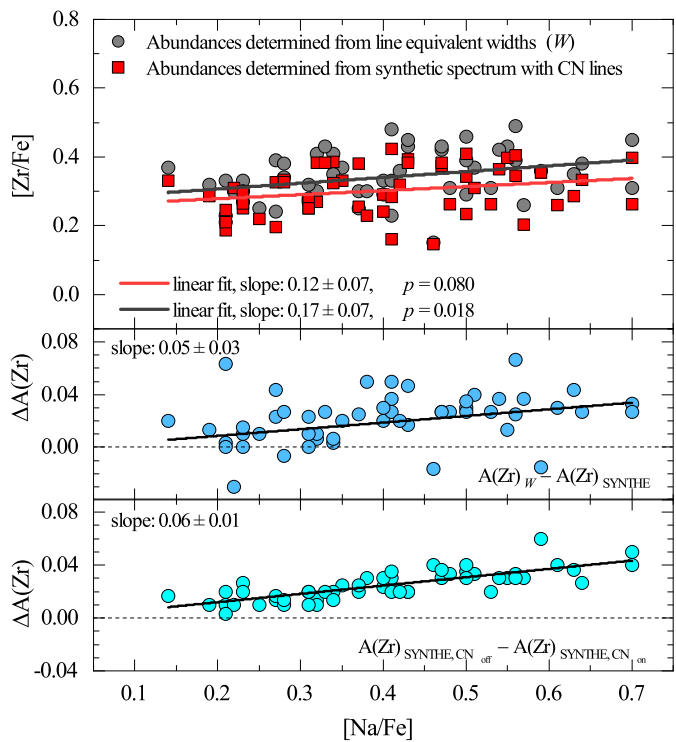}
	\caption{Comparison of Zr abundances obtained in the subsample of 54 RGB stars with and without taking the influence of CN lines into account. Top: [Zr/Fe] abundance ratios obtained using the equivalent width method (Sect.~\ref{sect:abund_Zr-I}, grey circles) and spectral line synthesis (Sect.~\ref{sect:CN-blends-test-knownCN}, red squares). Middle: Difference between Zr abundances obtained using equivalent width and spectral line synthesis methodologies. Bottom: Difference between Zr abundances obtained with the line synthesis methodology when switching the CN lines off and on. In all panels, solid lines are linear best fits to the data.
	}
	\label{fig:zr-abund-CN-comparison}
\end{figure}

For the second test, we selected a subsample of our target RGB stars for which C and N abundances were determined by \citet{MMGH20} using {\tt APOGEE-2} spectra. This resulted in a subset of 54 stars. We then determined Zr abundances from the same three \ion{Zr}{i} lines as was done in Sect.~\ref{sect:abund_Zr-I} but this time using the spectral synthesis methodology to eliminate the influence of CN blends on the determined abundances. For this, we used the \SYNTHE\ package together with the model atmospheres used in the analysis of these stars in Sect.~\ref{sect:abund}, and the CN line list taken from \citet{brooke14}. 
The obtained Zr abundances are shown in Fig.~\ref{fig:zr-abund-CN-comparison}.

Indeed, the slope in the ${\rm [Zr/Fe]} - {\rm [Na/Fe]}$ plane becomes smaller when using Zr abundances obtained by taking the influence of CN lines into account and is equal to $\approx+0.12$ (Fig.~\ref{fig:zr-abund-CN-comparison}). This value is about two times smaller than that obtained using Zr abundances determined with the equivalent width method, of namely $\approx0.23$ (Sect.~\ref{sect:zr-na-correl}). Part of the difference stems from the fact that for the subsample of 54 RGB stars, the slope in the ${\rm [Zr/Fe]} - {\rm [Na/Fe]}$ plane is lower than that obtained for 237 stars, $\approx+0.17$ vs. $\approx+0.23$. The rest of the change in the slope, $\Delta=+0.05$, is caused by CN lines alone, as is seen in Fig.~\ref{fig:zr-abund-CN-comparison}.

Thus, the Zr--Na correlation remains prominent when Zr abundances are determined with the line synthesis methodology, that is, by eliminating the influence of CN lines. We obtain  $p=0.08$ when using this approach on our subsample of 54 RGB stars. If adjusted for the smaller subsample size, this value would become smaller, $p=0.04$, and would suggest that the Zr--Na correlation may in fact be real.

\subsection{Mean values of Na and Zr abundances in the 1P and 2P stars in 47~Tuc\label{app-sect:abund-list-1P-2P}}

As the detected Zr--Na correlation is relatively weak, we can also compare mean abundances in the 1P and 2P stars in order to check whether they are indeed statistically different. For this, following \citet{CBG09a} we classified stars with ${\rm [Na/Fe]} = [{\rm [Na/Fe]}_{\rm min}, {\rm [Na/Fe]}_{\rm min}+0.3]$ as 1P stars, and those with the higher [Na/Fe] abundance ratios as 2P. In each population, we then computed the mean abundance ratios and the errors of the mean. The obtained results are provided in Table~\ref{tab:abundance-ratios-1p-2p}.

Using the mean 1P and 2P abundances, we ran the two-sample $t$-test to verify whether the possible 2P--1P differences are indeed statistically significant. We also did the same test using the modified [Zr/Fe] ratios obtained in Sect.~\ref{sect:CN-blends-test-extremeCO}. We did not perform such a test for the subsample of 54 target RGB stars with Zr abundances determined using the line synthesis approach (Sect.~\ref{sect:CN-blends-test-knownCN}) because in this case the sample size was too small for a meaningful comparison of the 2P--1P difference in mean Zr abundance.

When comparing the mean [Na/Fe] abundance ratios in the 1P ($N=93$) and 2P ($N=144$) stars, the probability of obtaining the $t$-values by chance is $p<10^{-15}$, which supports the notion that the difference that we see between the mean Na abundances in the 1P and 2P stars is real. A similar situation is also seen for the differences in the mean [Zr/Fe] abundance ratios where the $p$-values obtained using original Zr abundances (Sect.~\ref{sect:abund_Zr-I}) and abundances adjusted to account for CN blends (Sect.~\ref{sect:CN-blends-test-extremeCO}) do not exceed $p=0.07$.


\begin{table}[tb]
	\begin{center}
		\caption{Mean [Na/Fe] and [Zr/Fe] abundance ratios in the 1P and 2P stars in 47~Tuc and the corresponding $p$-values.
			\label{tab:abundance-ratios-1p-2p}}
		\resizebox{8cm}{!}{%
			\begin{tabular}{lccc}
				\hline\hline
				\noalign{\smallskip}
				& $\langle{\rm [X/Fe]}\rangle_{\rm 1P}$ & $\langle{\rm [X/Fe]}\rangle_{\rm 2P}$ & $p$-value \\
				\hline\noalign{\smallskip}
				\ion{Na}{i} (NLTE)   &  $0.247\pm 0.006$  &  $0.517\pm 0.009$  & $<10^{-15}$  \\                               
				\ion{Zr}{i} (LTE)     &  $0.310\pm 0.008$  &  $0.372\pm 0.007$  & $<10^{-7}$    \\
				\ion{Zr}{i} (LTE),    &  $0.331\pm 0.008$  &  $0.361\pm 0.007$ & 0.007     \\
				CN-adjusted   &   &   &   \\
				\hline
		\end{tabular}}
	\end{center}
	Note: Uncertainties are standard errors of the mean. The last line provides the values obtained with Zr abundances in 237 RGB stars adjusted to account for the influence of CN blends (see Sect.~\ref{sect:CN-blends-test-extremeCO}.
\end{table}

\subsection{Implications for nucleosynthesis in the GGCs}

Taken together, the evidence we find suggests that there may indeed be a statistically significant correlation between ${\rm [Zr/Fe]}$ and ${\rm [Na/Fe]}$. No earlier studies hinted towards the existence of a Zr--Na correlation, either in 47~Tuc or in other GGCs. On the one hand, this is unsurprising given the small number of stars per GGC studied in these papers ($N<14$). On the other hand, an analysis of the data obtained by \citet[][]{TSA14} suggests a weak correlation in the ${\rm [Zr/Fe] - {\rm [Na/Fe]}}$ plane although the authors did not find it to be statistically significant (see their Fig.~A.2). Importantly, abundances from \citet{TSA14} follow the same trend as seen in our data in the ${\rm [Zr/Fe] - {\rm [Na/Fe]}}$ plane, even if the mean [Na/Fe] and [Zr/Fe] ratios obtained in the two studies differ slightly (Sect.~\ref{app-sect:Zrii-abund}).

Theoretical modelling of s-process element yields shows that Zr may be synthesised both in AGB stars \citep[$M\sim1.5-5\,{\rm M_\odot}$;][]{CSP15} and massive rotating stars \citep[$M\sim12-25\,{\rm M_\odot}$, $v_{\rm rot}>150$\,km\,s$^{-1}$;][]{LC18}. It is therefore impossible to provide reliable constraints on the possible polluters based on the abundances of Zr alone.

Interestingly, analysis of 106 red horizontal branch (RHB) stars in 47~Tuc by \citet{GLS13} revealed a weak but statistically significant correlation in the ${\rm [Ba/Fe] - {\rm [Na/O]}}$ plane. In contrast, our recent study of Ba abundance in the same sample of RGB stars as that used in our Zr analysis does not corroborate this finding, as we detect no statistically significant correlation in the ${\rm [Ba/Fe] - {\rm [Na/Fe]}}$ plane \citep{DKK21}, despite the fact that our target sample was twice as large as that used in \citet{GLS13}. 
To the best of our knowledge, there are no other s-process elements with similar correlations detected in 47~Tuc (although there may be hints to such correlations in other Type~I GGCs, e.g. Sr--Na and Y--Na relations in M~4, \citealt{SSG16,VG11}; but see also \citealt{DOCL13}).



This evidence is therefore not sufficient to identify the possible enrichment scenario that has operated in this GGC. It is important to stress that existing abundance determinations of other s-process elements are not very helpful here because they show no evidence for correlations with the abundances of light elements, as seen in the case of Zr. Thus, they may reflect abundances of the primordial gas cloud that was enriched prior to the formation of 2P stars.
Future investigations of differences between the heavy s-process elements in the 1P and 2P stars of this and other GGCs would therefore be very desirable, especially for large samples of GGC stars as this is critical for a reliable detection of weak correlations between the elemental abundances. In particular, it would be very interesting to check whether such differences also exist for Sr, Y, and other elements of the so-called `first s-process peak' which are synthesised under similar conditions to Zr. In addition, detection of 1P--2P differences for heavier s-process elements in the Ba and Pb peaks may allow more stringent constraints to be placed on the possible polluters, because s-process elements in the different peaks are synthesised under different conditions and in different polluters.

\section{Conclusions and outlook}

Our analysis of Na and Zr abundances in 237 RGB stars in 47~Tuc suggests the existence of a weak Na--Zr correlation.
The data also reveal a small but statistically significant difference between the mean Zr abundances in the 1P and 2P stars, of namely $\Delta {\rm [Zr/Fe]}_{\rm 2P-1P}\approx 0.06$\,dex.
Correlation is seen in the ${\rm [Zr/Fe]}-r/r_{\rm h}$ plane as well, where $r$ is the projected distance from the cluster centre. Although there may be some influence on Zr abundance from blending with CN lines ($\leq0.08$\,dex), the residual correlations remain statistically significant when the estimated trends due to CN blends are removed. 

If the detected Zr--Na correlation is indeed real, it would indicate that at least some fraction of Zr in 47~Tuc should have been synthesised by the same polluters that enriched 2P stars with light elements. Amongst the potential candidate polluters are AGB stars \citep[$M\sim1.5-5\,{\rm M_\odot}$][]{CSP15} and/or massive rotating stars \citep[$M\sim12-25\,{\rm M_\odot}$, $v_{\rm rot}>150$\,km\,s$^{-1}$][]{LC18}, both of which may synthesise Zr in sizeable amounts.
Unfortunately, our Zr data alone do not allow us to distinguish between the pollution scenarios, or to confirm that a combination of them has operated in this GGC. It would therefore be very desirable to investigate whether such 1P--2P differences exist for other s-process elements in this or other GGCs in order to put tighter constraints on the possible polluters and evolutionary scenarios of the GGCs.


\begin{acknowledgements}
%
{We are very grateful to Tamara Mishenina and Sergey Andrievsky for useful comments during the preparation of the manuscript. Our study has benefited from the activities of the "ChETEC" COST Action (CA16117), supported by COST (European Cooperation in Science and Technology) and from the European Union’s Horizon 2020 research and innovation programme under grant agreement No 101008324 (ChETEC-INFRA). This work has made use of the VALD database, operated at Uppsala University, the Institute of Astronomy RAS in Moscow, and the University of Vienna.}
\end{acknowledgements}



\begin{appendix}



\section{Target RGB stars common to different observing programmes\label{app-sect:final-sample}}

The list of targets common to the three observing programmes is provided in Table~\ref{app-tab:common-stars}. Amongst them, there are five stars that were observed in all three datasets. This sample of overlapping stars was used to investigate possible systematic shifts in the abundances determined using different datasets (Sect.~\ref{app-sect:Fe-abund}-\ref{app-sect:Zr-abund}). Abundances obtained using different datasets were averaged.
The final sample for the abundance analysis consisted of 237 individual RGB stars.

\begin{table}[tbh]
	\begin{center}
		\caption{List of stars common to different observing programmes.
			\label{app-tab:common-stars}}
		\resizebox{7cm}{!}{
		\begin{tabular}{lll}
			\hline\hline
			\noalign{\smallskip}
			072.D-0777(A) & 073.D-0211(A) & 088.D-0026(A) \\
			\hline\noalign{\smallskip}
			B-1256   &  --             &   R287  \\  
			F-1389   &   1389          &   --    \\  
			B-3449   &  --             &   R563  \\  
			F-3476   &  --             &   R317  \\  
			F-4373   &   4373          &   --    \\  
			--       &   5172          &   R259  \\  
			B-5362   &  --             &   R253  \\  
			F-7711   &   7711          &   --    \\  
			--       &   7904          &   R443  \\  
			B-7993   &  --             &   R277  \\  
			B-9163   &  --             &   R800  \\  
			--       &   9518          &   R237  \\  
			--       &   9717          &   R682  \\  
			B-9997   &  --             &   R248  \\  
			F-10198  &  --             &   R756  \\  
			F-10527  &  --             &   R782  \\  
			F-12408  &  --             &   R784  \\  
			F-13668  &  13668          &   --    \\  
			B-13795  &  13795          &   R752  \\  
			B-13853  &  --             &   R246  \\  
			B-14583  &  14583          &   --    \\  
			F-15451  &  15451          &   --    \\   
			--       &  15552          &   R381  \\			
			B-16667  &  --             &   R790  \\  
			B-17819  &  --             &   R245  \\  
			--       &  21369          &   R249  \\  
			F-24463  &  24463          &   --    \\  
			B-29146  &  --             &   R256  \\  
			--       &  29490          &   R231  \\  
			B-30463  &  30463          &   --    \\  
			B-30949  &  --             &   R766  \\  
			B-32730  &  32730          &   --    \\  
			B-35878  &  35878          &   --    \\  
			B-38976  &  --             &   R762  \\  
			B-41429  &  --             &   R392  \\  
			B-42866  &  42866          &   R656  \\  
			B-42887  &  --             &   R760  \\  
			F-43632  &  43632          &   R512  \\  
			B-43852  &  43852          &   R704  \\  
			B-43889  &  43889          &   R450  \\
			\hline
		\end{tabular}
	}
	\end{center}
\end{table}

\section{Abundance analysis\label{app-sect:abnd-analysis}}

Abundances of Fe and Zr in the sample of 237 RGB stars were determined using the equivalent width method, assuming LTE in spectral synthesis calculations. In the case of Na, we fitted synthetic line profiles that were computed under the assumption of NLTE. In what follows, we describe some technical aspects of the abundance determination of all three elements, including the line lists, diagnostics of obtained abundances, and a comparison of abundances determined in stars common to several observing programmes.

\subsection{Reference abundances in the Sun and Arcturus\label{app-sect:ref-abnd}}

\begin{table}
	\begin{center}
		\caption{The list of \ion{Fe}{i} and \ion{Fe}{ii} lines used in the abundance analysis.
			\label{app-tab:iron-list}}
		\resizebox{6cm}{!}{
		\begin{tabular}{cccc}
			\hline\hline
			\noalign{\smallskip}
			$\lambda$, nm & $\chi$, eV & log \textit{gf} & Ion. stage\\
			\hline\noalign{\smallskip}
			612.79070 &  4.1400 & --1.398 & \ion{Fe}{i} \\
			615.16180 &  2.1800 & --3.300 & \ion{Fe}{i} \\
			616.53600 &  4.1400 & --1.460 & \ion{Fe}{i} \\
			617.33360 &  2.2200 & --2.810 & \ion{Fe}{i} \\
			618.02040 &  2.7300 & --2.650 & \ion{Fe}{i} \\
			618.79900 &  3.9400 & --1.580 & \ion{Fe}{i} \\
			620.03130 &  2.6100 & --2.310 & \ion{Fe}{i} \\
			621.34300 &  2.2200 & --2.550 & \ion{Fe}{i} \\
			621.92810 &  2.2000 & --2.410 & \ion{Fe}{i} \\
			622.92283 &  2.8450 & --2.805 & \ion{Fe}{i} \\
			623.26410 &  3.6500 & --1.130 & \ion{Fe}{i} \\
			624.63188 &  3.6020 & --0.733 & \ion{Fe}{i} \\
			625.25554 &  2.4040 & --1.687 & \ion{Fe}{i} \\
			626.51340 &  2.1800 & --2.510 & \ion{Fe}{i} \\
			627.02250 &  2.8600 & --2.570 & \ion{Fe}{i} \\
			627.12788 &  3.3320 & --2.703 & \ion{Fe}{i} \\
			630.15012 &  3.6540 & --0.718 & \ion{Fe}{i} \\
			631.58115 &  4.0760 & --1.710 & \ion{Fe}{i} \\
			632.26860 &  2.5900 & --2.280 & \ion{Fe}{i} \\
			633.53308 &  2.1980 & --2.177 & \ion{Fe}{i} \\
			633.68243 &  3.6860 & --0.856 & \ion{Fe}{i} \\
			634.41490 &  2.4300 & --2.890 & \ion{Fe}{i} \\
			638.07430 &  4.1900 & --1.270 & \ion{Fe}{i} \\
			660.91100 &  2.5600 & --2.610 & \ion{Fe}{i} \\
			670.35670 &  2.7600 & --3.010 & \ion{Fe}{i} \\
			672.66660 &  4.6100 & --1.010 & \ion{Fe}{i} \\
			675.01530 &  2.4200 & --2.580 & \ion{Fe}{i} \\
			680.68450 &  2.7300 & --3.090 & \ion{Fe}{i} \\
			681.02630 &  4.6100 & --0.940 & \ion{Fe}{i} \\
			684.36560 &  4.5500 & --0.780 & \ion{Fe}{i} \\
			685.51620 &  4.5600 & --0.570 & \ion{Fe}{i} \\
			685.81500 &  4.6100 & --0.910 & \ion{Fe}{i} \\
			691.66820 &  4.1500 & --1.260 & \ion{Fe}{i} \\
			614.92580 &  3.8900 & --2.720 & \ion{Fe}{ii} \\
			623.83920 &  3.8700 & --2.520 & \ion{Fe}{ii} \\
			624.75570 &  3.8900 & --2.320 & \ion{Fe}{ii} \\
			636.94620 &  2.8910 & --4.160 & \ion{Fe}{ii} \\
			\hline
		\end{tabular}
	}
	\end{center}
\end{table}

\begin{table}
	\begin{center}
		\caption{Parameters of \ion{Na}{i} and \ion{Zr}{i} lines used in the abundance analysis.
			\label{app-tab:line-list}}
		\resizebox{\hsize}{!}{%
			\begin{tabular}{llcrllc}
				\hline\hline
				\noalign{\smallskip}
				Element & $\lambda$, nm  & $\chi$, eV & log \textit{gf} & log $\gamma_{rad}$ & log $\frac{\gamma_4}{N_e}$ & log $\frac{\gamma_6}{N_H}$ \\
				\hline\noalign{\smallskip}
				\ion{Na}{i}  & 615.4225 & 2.102 & $-1.547$ & 7.85                & $-4.39$     & $-7.28$               \\
				\ion{Na}{i}  & 616.0747 & 2.104 & $-1.246$ & 7.85                & $-4.39$     & $-7.28$               \\
				\ion{Zr}{i}  & 612.7475 & 0.154 & $-1.060$ & 7.77                & $-5.69$     & $-7.79$              \\
				\ion{Zr}{i}  & 613.4585 & 0.000 & $-1.277$ & 7.77                & $-5.70$     & $-7.79$              \\
				\ion{Zr}{i}  & 614.3252 & 0.071 & $-1.097$ & 7.77                & $-5.69$     & $-7.79$              \\
				\hline
			\end{tabular}}
		\end{center}
		\small{Oscillator strength values of \ion{Zr}{I} lines were taken from \citet{BGH81} and of \ion{Na}{I} lines from NIST database (\url{https://www.nist.gov/pml/atomic-spectra-database}).}
	\end{table}

To obtain the element-to-iron abundance ratios, [X/Fe], in the sample of RGB stars in 47~Tuc, we determined the reference abundances of Fe, Na, and Zr in the Sun. For this, we used the Kitt Peak Solar Flux atlas \citep{KFB84} and spectral lines of \ion{Fe}{i}, \ion{Na}{i}, and \ion{Zr}{i} that are identical to those that were used in the analysed RGB targets. Abundances were determined with the 1D hydrostatic \ATLAS\ solar model atmosphere that was computed by F.~Castelli\footnote{\url{https://wwwuser.oats.inaf.it/castelli/sun.html}} using the solar abundance table from \citet{AGS05}.

Solar iron abundances were obtained using the equivalent width method.
The \ion{Fe}{i} line list used for the abundance determination is provided in Table~\ref{app-tab:iron-list}. Atomic line data were taken from the VALD3 database \citep{RPK15}. We determined the microturbulence velocity iteratively to obtain zero trend of Fe abundance with the line equivalent width. The obtained solar value, $\xi_\mathrm{micro} = 0.93$\,km\,s$^{-1}$, is in good agreement with the typical values of 0.9--1.0\,km\,s$^{-1}$ found in other studies \citep[e.g.][]{Doyle14}. The average solar Fe abundance, ${\rm A(Fe)}_{\odot}=7.55 \pm 0.01$ ($\sigma = 0.06$; here $\pm0.01$ is the error of the mean, $\sigma$ denotes the standard deviation due to line-to-line abundance variation), which was obtained from 29 \ion{Fe}{i} lines with $W<10.5$\,pm, agrees well with the value of ${\rm A(Fe)}_{\odot}=7.52\,(\sigma = 0.06)$\, from \citet{CLS11}.

To obtain the reference solar Na abundance, we used the same set of \ion{Na}{i} lines and analysis methodology as we used in the analysis of target RGB stars (Sect.~\ref{app-sect:Na-abund}). Atomic parameters of the \ion{Na}{i} lines are listed in Table~\ref{app-tab:line-list}. We obtained an identical abundance from both \ion{Na}{i} lines, ${\rm A(Na)_{\odot}}=6.17$, which agrees well with ${\rm A(Na)_{\odot}}=6.21\pm0.04$ determined by \citet{SAG15_Na}.

The \ion{Zr}{i} lines that were used in the analysis of target RGB stars in 47~Tuc were too weak in the solar spectrum for reliable reference abundance determination. We therefore used solar Zr abundance of A(Zr)$_{\odot} = 2.62$ from \citet{CFL11}.

We also determined abundances of Fe, Na, and Zr in the atmosphere of Arcturus in order to check whether the abundances obtained from the \ion{Fe}{i}, \ion{Na}{i}, and \ion{Zr}{i} lines used in our study agree with those determined in the earlier studies of Arcturus. As in the case of the Sun, we used the same line lists as those used in the analysis of target RGB stars in 47~Tuc\footnote{Note that the \ion{Fe}{i} line sets used to determine Fe abundances in the Sun, Arcturus, and RGB stars in 47~Tuc were not identical but individual lines were always selected from the list in Table~\ref{app-tab:iron-list}.}. The abundances were obtained using the `Visible and Near Infrared Atlas of the Arcturus Spectrum 3727-9300 \AA{}' by \citet{HWV00}. We calculated the model atmosphere of Arcturus using the \ATLAS\ model atmosphere package and $\Teff=4286$\,K and $\log g=1.66$ from \citet{RAP11}, with the solar-scaled chemical composition from \citet{AGS05} and $\alpha$-element enhancement of $+0.4$\,dex.

In case of Arcturus, we determined the microturbulence velocity of $\xi_\mathrm{micro} = 1.78$\,km\,s$^{-1}$ and iron abundance of ${\rm A(Fe)}=7.03 \pm 0.02$ ($\sigma = 0.09$) or $\FeH=-0.52$, with both the microturbulence velocity and abundance obtained using \ion{Fe}{i} lines with $W<16$\,pm. As in the case of the Sun, the mean Fe abundance and microturbulence velocity agrees well with the results obtained in the earlier studies \citep[cf.][]{RAP11,Sheminova15}.

Abundance of Na in Arcturus was determined in NLTE using the same line list and methodology as utilised in the analysis of RGB stars in 47~Tuc (Sect.~\ref{app-sect:Na-abund}). We obtained identical Na abundance from both \ion{Na}{i} lines, ${\rm A(Na)}=5.71$ or ${\rm [Na/Fe]} = 0.06$. This agrees reasonably well with the NLTE estimate of ${\rm A(Na)}=5.80$ derived using a set of 14 \ion{Na}{i} lines by \citet{OAPH20}. In LTE, we obtained ${\rm A(Na)}=5.85$ which agrees very well with ${\rm A(Na)}=5.83\pm0.03$ determined by \citet{RAP11} using the  two \ion{Na}{i} lines utilised in our study (we note that the LTE value obtained by \citealt{OAPH20}, ${\rm A(Na)}=5.91$, is noticeably higher than that determined either by us or \citealt{RAP11}).

A Gaussian fit to the observed profiles of \ion{Zr}{i} lines in the Arcturus spectrum resulted in the equivalent widths of 5.42\,pm, 4.97\,pm, and 5.39\,pm for the lines located at 612.7475\,nm, 613.4585\,nm, and 614.3252\,nm, respectively, with the corresponding 
${\rm A(Zr)}=2.05$, 1.98, and 1.97. The obtained mean Zr abundance is ${\rm [Zr/Fe]}= -0.10$ ($\sigma=0.04$). This agrees reasonably well with ${\rm [Zr/Fe]}= 0.01$ ($\sigma = 0.07$) obtained by \citet{WCF09} from seven \ion{Zr}{i}  lines.

It is important to note that  in the original Kurucz software package (\ATLAS, \WIDTH, \SYNTHE) the value of \ion{Zr}{i} ionisation potential is set to 6.840\,eV \citep{S05}. Throughout this study, we used a revised value of 6.634\,eV from \citet{HHM86} which is set as default in the F.~Castelli version of the Kurucz's package\footnote{\url{https://wwwuser.oats.inaf.it/castelli/sources.html}}. For Arcturus, this lead to an increase in Zr abundance of +0.29\,dex. In case of target RGB stars of 47~Tuc, the corresponding upward change was $+0.25$ to $+0.29$\,dex at $\Teff=4800$\,K and 4200\,K respectively.

\subsection{Determination of Fe, Na, and Zr abundance in RGB stars of 47~Tuc: additional tests}

\subsubsection{Abundance of Fe\label{app-sect:Fe-abund}}

As it was described in Sect.~\ref{sect:abund}, for each target RGB star in 47~Tuc, iron abundances were determined using a set of $17-28$ \ion{Fe}{i} lines located at $612.79-691.67$\,nm, with the line lower level excitation potentials spanning the range of $2.18-4.61$\,eV. Atomic data of \ion{Fe}{i} and \ion{Fe}{ii} lines used in the analysis are provided in Table~\ref{app-tab:iron-list}. 

\begin{figure}[tb]
	\resizebox{\hsize}{!}{\includegraphics{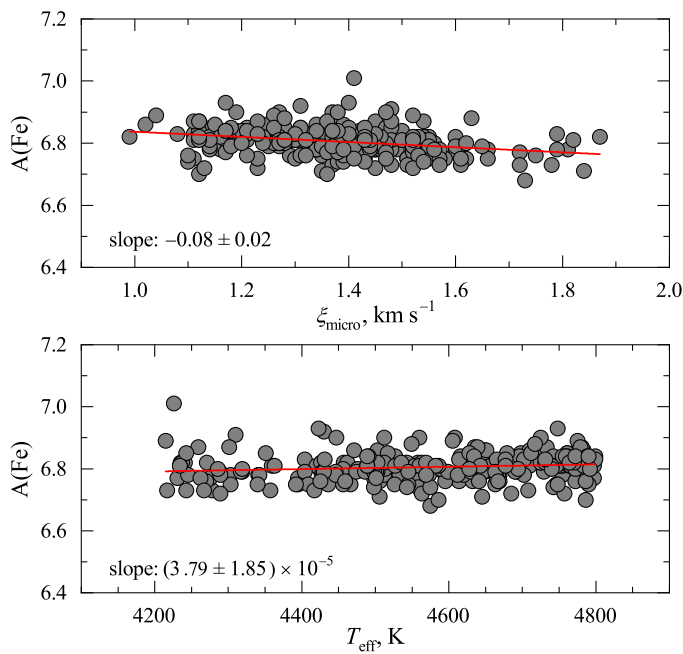}}
	\caption{Abundance of Fe in the target RGB stars, plotted against the microturbulence velocity (top) and effective temperature (bottom).}
	\label{app-fig:fe-abnd}
\end{figure}

\begin{figure}[tb]
	\resizebox{\hsize}{!}{\includegraphics{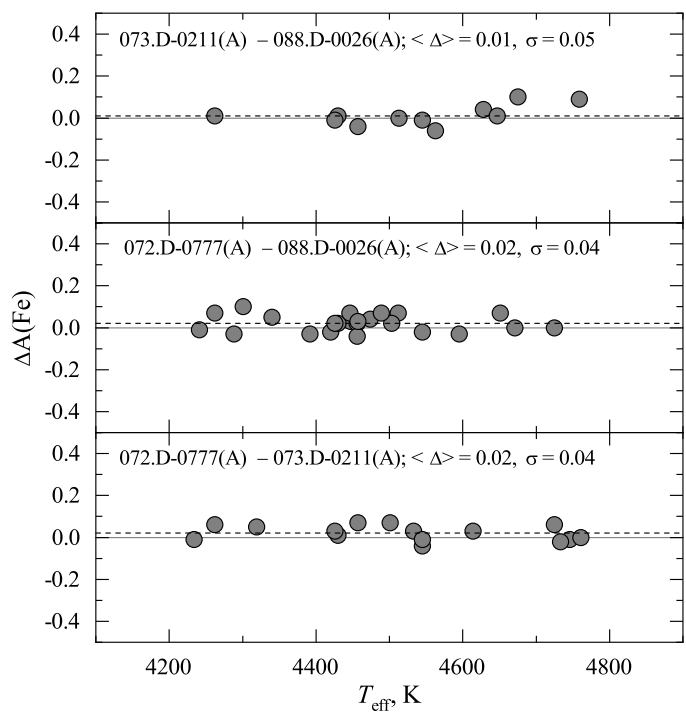}}
	\caption{Differences in the Fe abundances obtained in the target RGB stars that are common to different observing programmes used in this study. In each panel, $\langle\Delta\rangle$ symbol denotes the mean difference between the A(Fe) values obtained using spectra from the different programmes, $\sigma$ is standard deviation from $\langle\Delta\rangle$. 
	}
	\label{app-fig:Fe-abnd-common}
\end{figure}

There is a very weak dependence of the obtained abundances on $\xi_{\rm micro}$ and \Teff\ (Fig.~\ref{app-fig:fe-abnd}). However, even between the extreme ends of the effective temperature scale the average difference in Fe abundance is less than 0.05\,dex. There are no statistically significant differences between the Fe abundances determined using spectra obtained in different observing programmes (Fig.~\ref{app-fig:Fe-abnd-common}).


\subsubsection{Abundance of Na\label{app-sect:Na-abund}}

As indicated in Sect.~\ref{sect:abund}, an NLTE approach was used to determine Na abundance in 47~Tuc. For this purpose, we used spectral synthesis code \MULTI\ \citep{carlsson,KAL99} and \ATLAS\ model atmospheres, together with the model atom of Na that was exploited in our earlier studies \citep[e.g.][]{DKB14}. Two \ion{Na}{i} lines located at 615.4225 and 616.0747\,nm were used in the abundance analysis (Table~\ref{app-tab:line-list}). 



\begin{figure}[tb]
	\resizebox{\hsize}{!}{\includegraphics{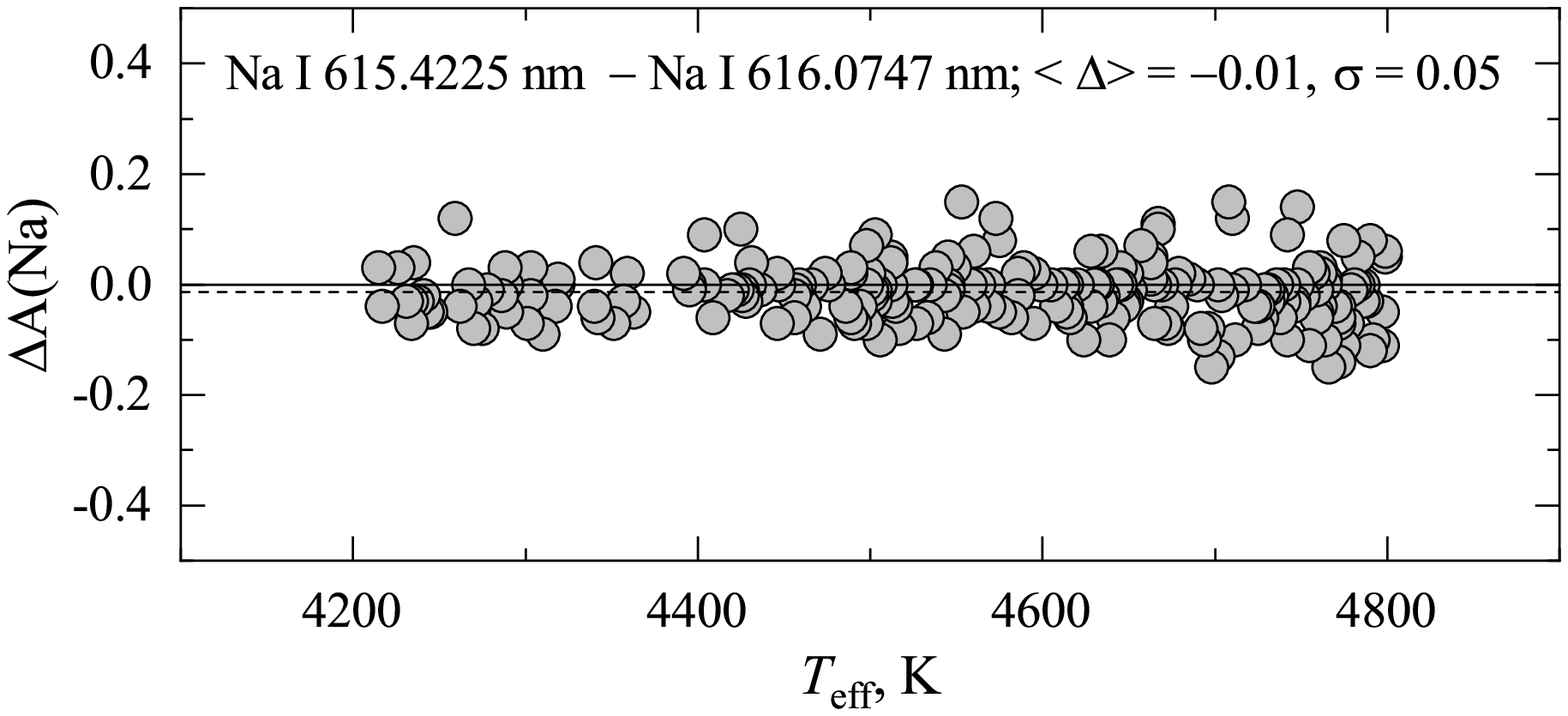}}
	\caption{Differences in the Na abundances obtained from the two \ion{Na}{i} lines used in this study. The $\langle\Delta\rangle$ symbol denotes the mean difference between the A(Na) values obtained using different \ion{Na}{i} lines, and $\sigma$ is the standard deviation from $\langle\Delta\rangle$.
	}
	\label{app-fig:Na-diff-lines}
\end{figure}

\begin{figure}[tb]
	\resizebox{\hsize}{!}{\includegraphics{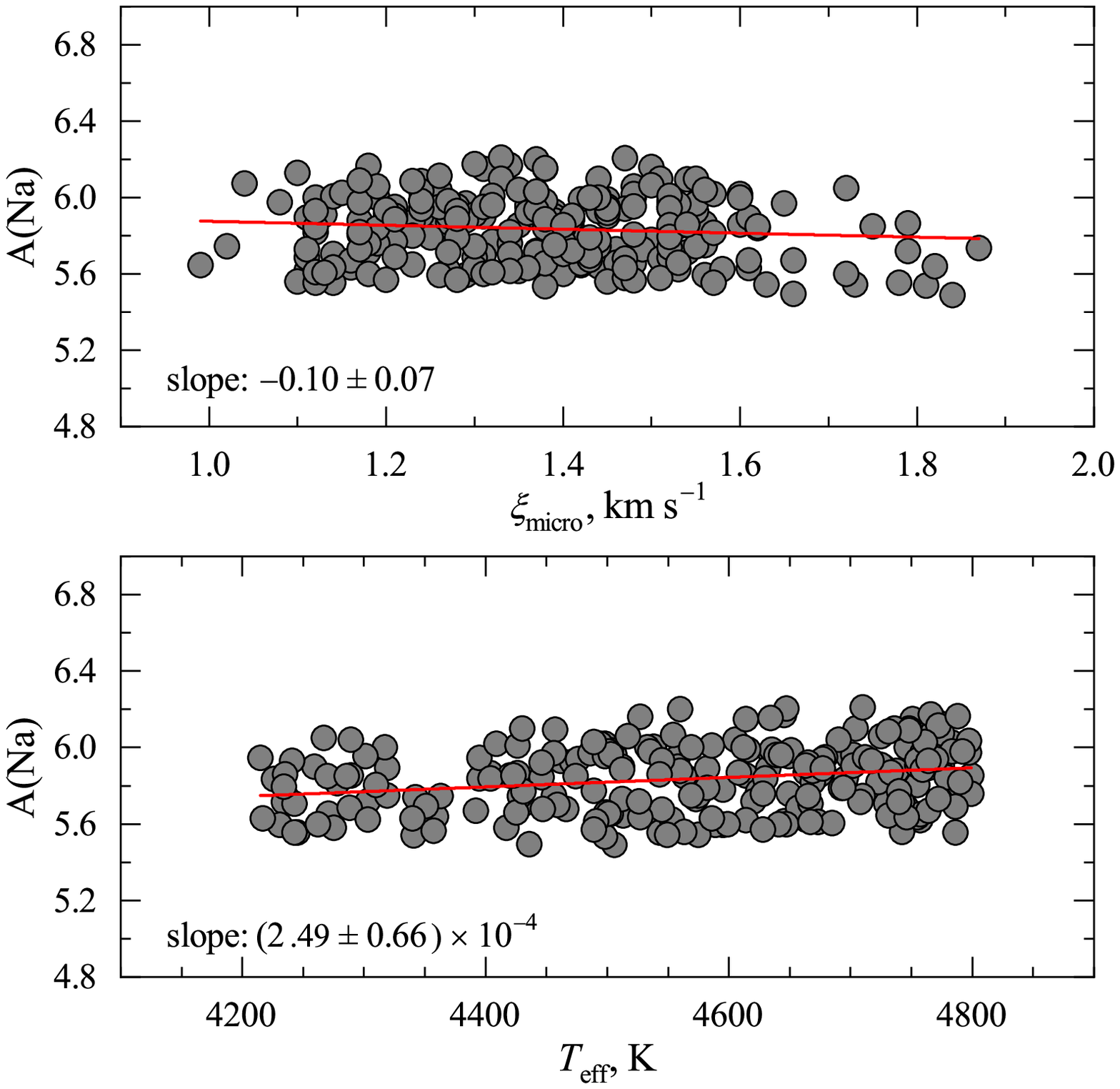}}
	\caption{Abundance of Na in the target RGB stars versus the microturbulence velocity (top) and effective temperature (bottom).}
	\label{app-fig:na-abnd}
\end{figure}

\begin{figure}[tb]
	\resizebox{\hsize}{!}{\includegraphics{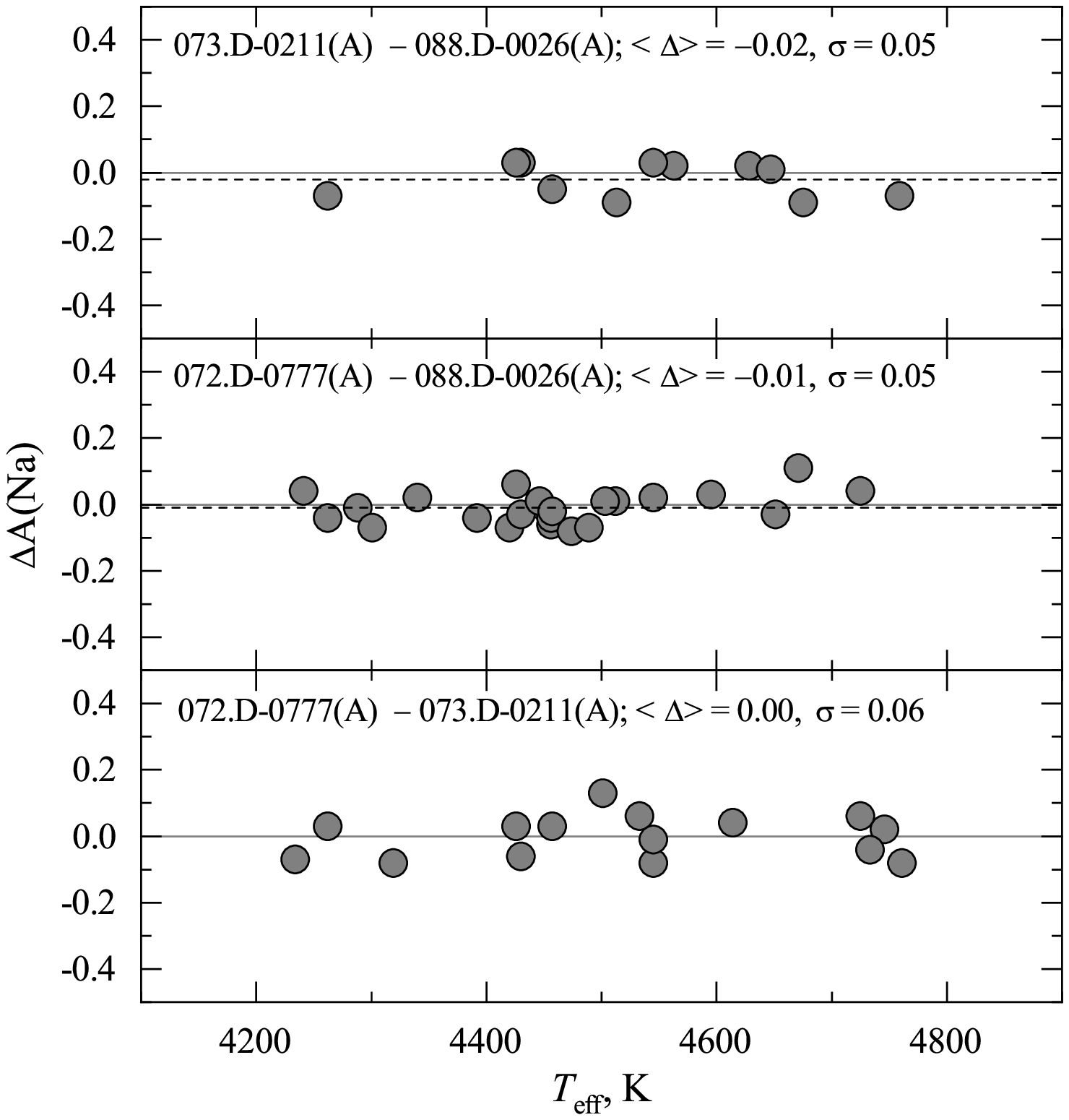}}
	\caption{Differences in the Na abundances obtained in the target RGB stars that are common to different observing programmes. In each panel, the $\langle\Delta\rangle$ symbol denotes the mean difference between the A(Na) values obtained using spectra from the different programmes, and $\sigma$ is the standard deviation from $\langle\Delta\rangle$.
	}
	\label{app-fig:Na-abnd-common}
\end{figure}

There is reasonably good agreement between abundances obtained from the two \ion{Na}{i} lines, with the differences typically below 0.05\,dex (Fig.~\ref{app-fig:Na-diff-lines}). While there may be a very weak dependence of the Na abundance on \Teff, the differences in Na abundance at the extreme ends of \Teff\ scale do not exceed 0.10\,dex (Fig.~\ref{app-fig:na-abnd}). Furthermore, we find no statistically significant differences between Na abundances determined using spectra obtained in the different observing programmes (Fig.~\ref{app-fig:Na-abnd-common}). 


\begin{figure}[tb]
	\resizebox{\hsize}{!}{\includegraphics{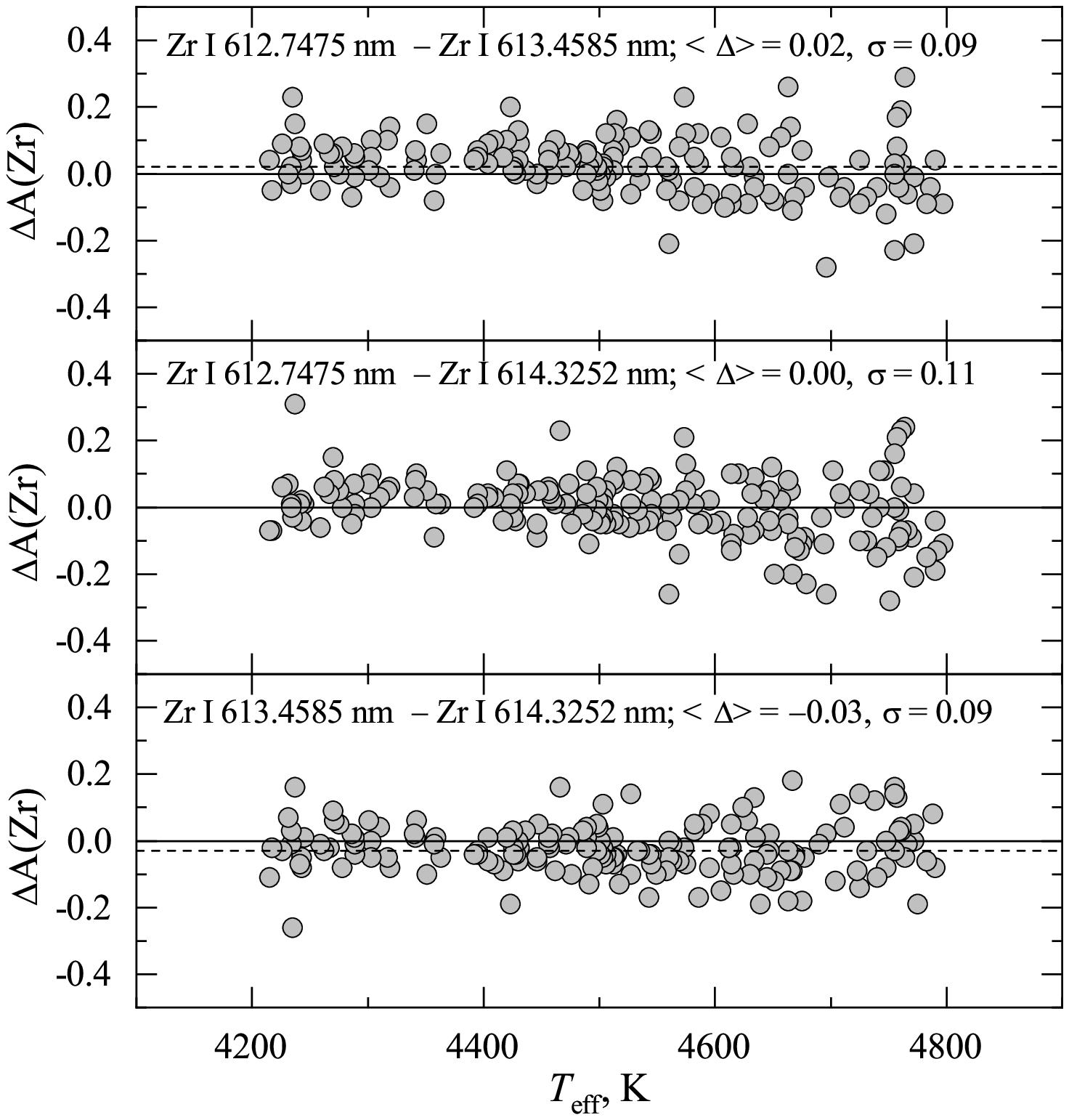}}
	\caption{Differences in the Zr abundances obtained from different \ion{Zr}{i} lines used in this study. The $\langle\Delta\rangle$ symbol denotes the mean difference between the A(Zr) values obtained using different \ion{Zr}{i} lines, and $\sigma$ is the standard deviation from $\langle\Delta\rangle$.
	}
	\label{app-fig:ZrI_abn_diff_lines}
\end{figure}

\subsubsection{Abundance of Zr\label{app-sect:Zr-abund}}


\begin{figure}[tb]
	\resizebox{\hsize}{!}{\includegraphics{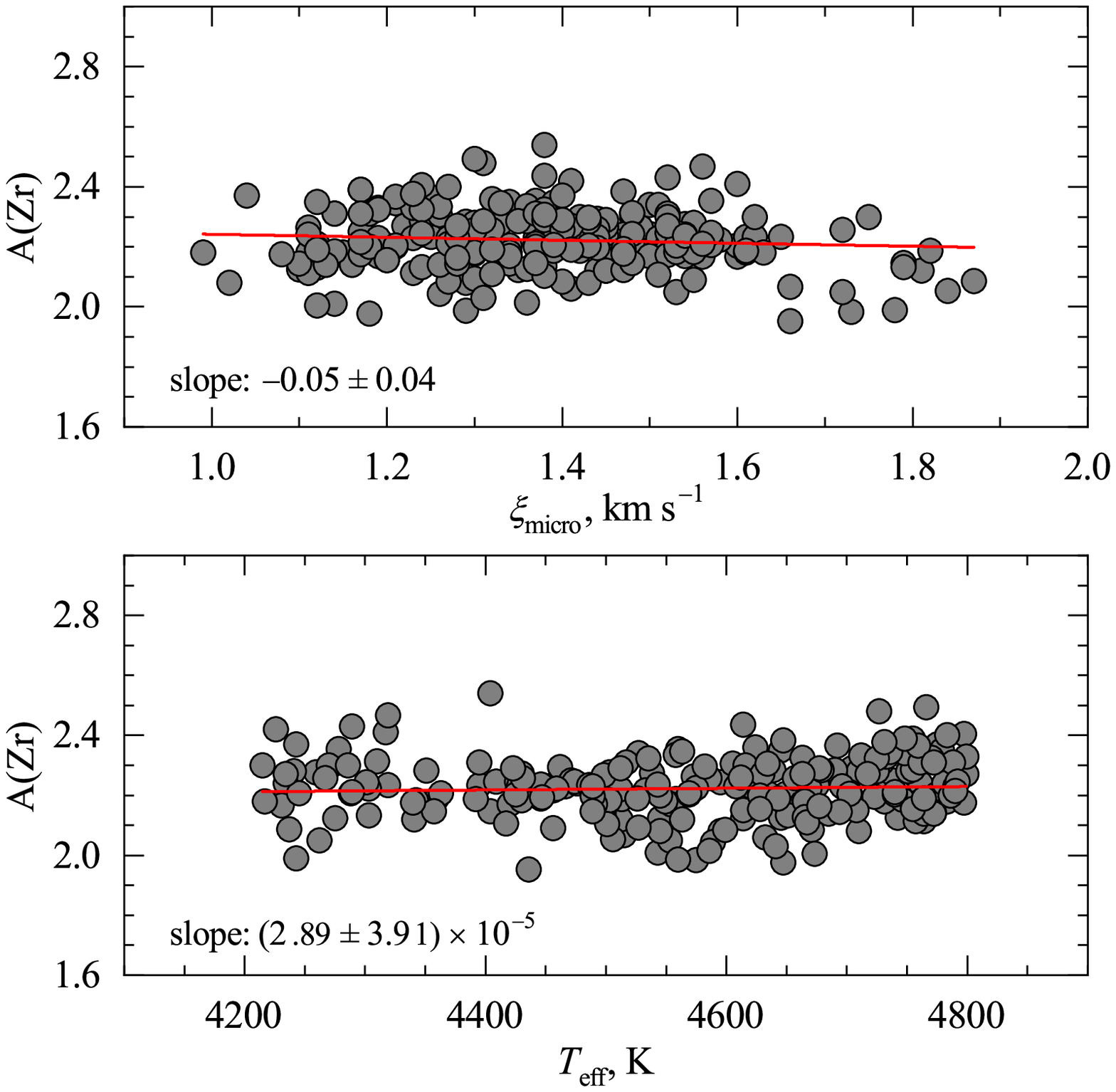}}
	\caption{Abundance of Zr in the target RGB stars versus the microturbulence velocity (top) and effective temperature (bottom).}
	\label{app-fig:zr-abnd}
\end{figure}

\begin{figure}[tb]
	\resizebox{\hsize}{!}{\includegraphics{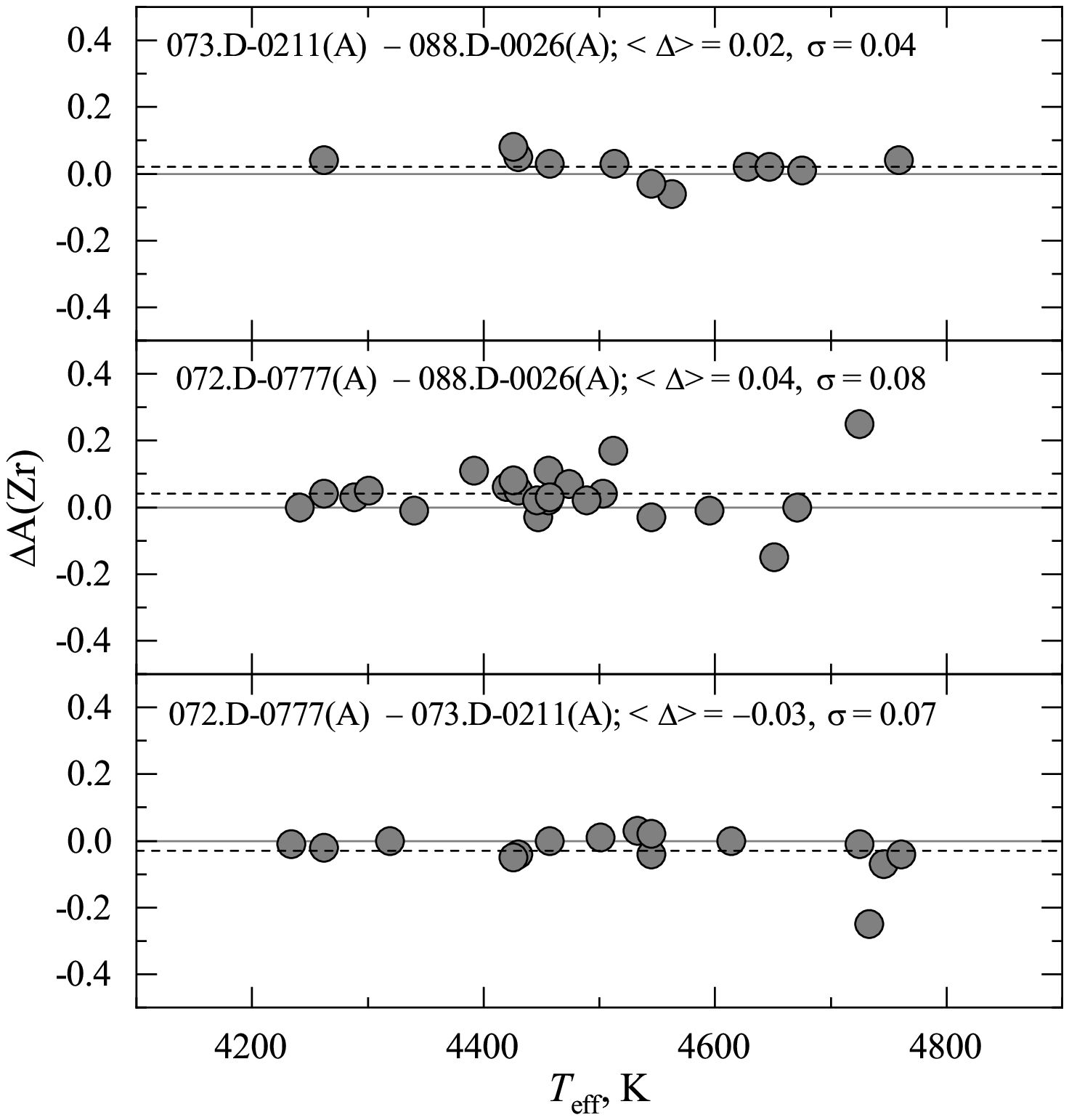}}
	\caption{Differences in the Zr abundances obtained in the target RGB stars that are common to different observing programmes. In each panel, the $\langle\Delta\rangle$ symbol denotes the mean difference between the A(Zr) values obtained using spectra from the different programmes, $\sigma$ is standard deviation from $\langle\Delta\rangle$. 
	}
	\label{app-fig:zr-abnd-common}
\end{figure}

As described in Sect.~\ref{sect:abund_Zr-I} above, Zr abundance was determined for the target RGB stars  using three \ion{Zr}{i} lines located at 612.7475, 613.4585, and 614.3252\,nm (Table~\ref{app-tab:line-list}) under the assumption of LTE. Examples of the observed and best-fitted \ion{Zr}{i} line profiles are shown in Fig.~\ref{fig:spectral-lines}.

Abundances obtained from the three \ion{Zr}{i} lines agree well, with mean differences in A(Zr) obtained from the individual lines of $\leq0.03$\,dex (Fig.~\ref{app-fig:ZrI_abn_diff_lines}). Also, there are no significant trends of Zr abundance with microturbulence velocity or effective temperature (Fig.~\ref{app-fig:zr-abnd}). Finally, there are no significant differences between Zr abundances determined using spectra obtained in the different observing programmes (Fig.~\ref{app-fig:zr-abnd-common}).

\subsubsection{Determination of zirconium abundance using \ion{Zr}{ii} lines\label{app-sect:Zrii-abund}}

As discussed in Sect.~\ref{sect:abund_Zr-I}, Zr abundances in our target RGB stars were determined using three \ion{Zr}{i} lines. Unfortunately, all three lines are contaminated with weak CN blends (see Sect.~\ref{sect:CN-blends} which were not taken into account when determining Zr abundances in Sect.~\ref{sect:abund_Zr-I}.

As discussed in Sect.~\ref{sect:abund_Zr-II}, several \ion{Zr}{ii} lines are available for abundance analysis and some of them are free from CN blends (Sect.~\ref{sect:CN-blends}). Unfortunately, none of the spectral ranges covering \ion{Zr}{ii} lines have been observed with \GIRAFFE\ for our target RGB stars. We therefore used the \UVES\ spectra of three stars in our sample obtained by \citet{TSA14} to compare abundances determined from the \ion{Zr}{i} and \ion{Zr}{ii} lines.

\begin{figure}[tb]
	\centering
	\includegraphics[width=\hsize]{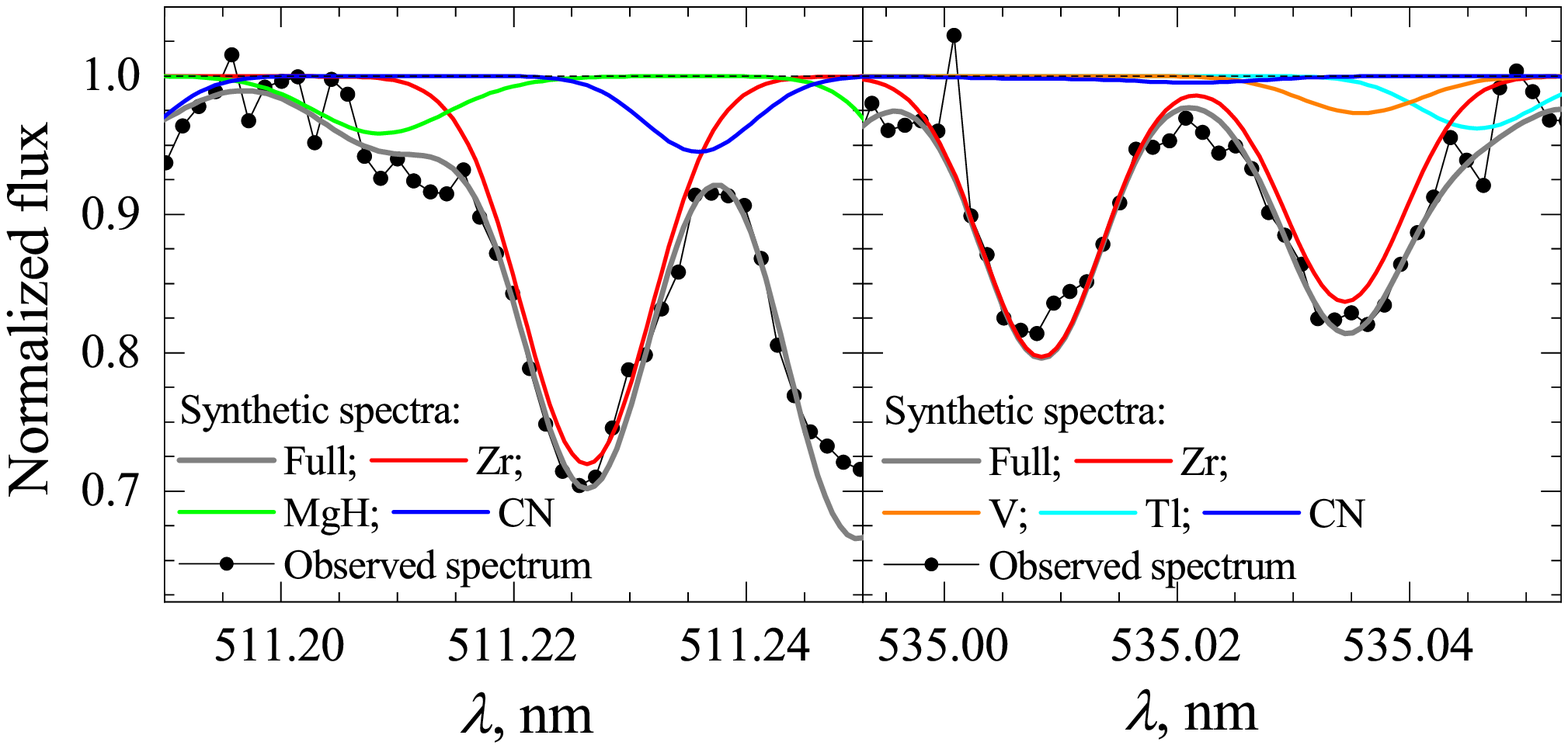}
	\caption{\ion{Zr}{ii} lines in the \UVES\ spectrum of the target star 20885 (dots) overlaid with the synthetic spectrum (grey solid line) computed using the \ATLAS\ model with $\Teff=4359$\,K, $\log g=1.41$, $\FeH=-0.74$, [Zr/Fe] = 0.30 dex, $\xi_{\rm micro}=1.82$\,km\,s$^{-1}$, $\xi_{\rm macro}=3.00$\,km\,s$^{-1}$ and solar-scaled elemental abundances \citep{GS98}.
		The left panel shows the vicinity of \ion{Zr}{ii} 511.2270\,nm line, the right panel that of 535.0089\,nm and of 535.0350\,nm \ion{Zr}{ii} lines. Other solid lines are synthetic profiles of individual lines, left panel: \ion{Zr}{ii} 511.2270\,nm (red line), MgH 511.2080\,nm (green line),  CN 511.2360\,nm (blue line); right panel: \ion{Zr}{ii} 535.0089\,nm and 535.0350\,nm (red line), \ion{V}{ii} 535.0358\,nm (orange line), \ion{Tl}{i} 535.0456\,nm (cyan line), CN 535.0215\,nm (blue line).
	}
	\label{app-fig:47Tuc_ZrII_blends}
\end{figure}

The \UVES\ spectra from \citet{TSA14} were obtained in the 580\,nm setting, with a resolving power of $R = 110\,000$. The spectra of the three common stars were taken from the ESO Advanced Data Products archive\footnote{\url{http://archive.eso.org/wdb/wdb/adp/phase3/spectral/form}}. We determined Zr abundances using the same \ion{Zr}{i} lines as in the analysis of \GIRAFFE\ spectra (Sect.~\ref{sect:abund_Zr-I}), as well as \ion{Zr}{i} 614.0460\,nm line (which becomes too weak to be measured reliably in the \GIRAFFE\ spectra for stars with $\Teff>4450$\,K) and three \ion{Zr}{ii} lines located at 511.2270, 535.0089, and 535.0350\,nm. Abundances from all lines were obtained by fitting synthetic spectra to the observed line profiles. The main reason for this was that all \ion{Zr}{i} and \ion{Zr}{ii} lines are blended with CN lines (Sect.~\ref{sect:CN-blends}). In addition, the 535.0350\,nm \ion{Zr}{ii} line is influenced by the weak \ion{V}{ii} 535.0358\,nm and \ion{Tl}{i} 535.0456\,nm lines, while the 511.2270\,nm \ion{Zr}{ii} line is affected by the nearby \ion{MgH} 511.2080 nm and \ion{Cr}{i} 511.2484 nm lines (Fig.~\ref{app-fig:47Tuc_ZrII_blends}); and the \ion{Zr}{i} 612.7475\,nm line is blended with the \ion{Fe}{i} 612.7906 nm line (Sect.~\ref{sect:abund_Zr-I}). Furthermore, Mg abundance influences the opacity of stellar model atmosphere which leads to a slight change in the electron densities and thus in determined Zr abundances. To account for the latter, we used Mg abundances obtained in the three stars by \citet{TSA14}. Solar-scaled CN abundances, ${\rm A(C)}=7.82$ and ${\rm A(N)}=7.22$, were used to account for the influence of CN blends, with the CN line data taken from \citet[][]{brooke14}; see also Sect.~\ref{sect:CN-blends} for more details about the influence of CN blends on the determined Zr abundances.

\begin{table*}
	\begin{center}
		\caption{Abundances of Zr in the three stars common to the \GIRAFFE\ (used in this work) and \UVES\ samples \citep{TSA14}, determined from \ion{Zr}{i} and \ion{Zr}{ii} lines in the UVES spectra.}
		\label{app-tab:zr-abund-indiv-lines}
		\resizebox{\hsize}{!}{%
			\begin{tabular}{rcccccccccc}
				\hline\hline
				\noalign{\smallskip}
				ID	&	$\Teff$	&	$\log g$   &   $\xi_{\rm micro}$	&  ${\rm A(Zr)}_{\rm I}$  &  ${\rm A(Zr)}_{\rm I}$  &  ${\rm A(Zr)}_{\rm I}$  &  ${\rm A(Zr)}_{\rm I}$ &	${\rm A(Zr)}_{\rm II}$	&${\rm A(Zr)}_{\rm II}$ &  ${\rm A(Zr)}_{\rm II}$   \\
				& K & & km\,s$^{-1}$ & 612.7475 nm & 613.4585 nm & 614.0535 nm & 614.3252 nm & 511.2270 nm & 535.0089 nm & 535.0350 nm   \\
				\hline
				\noalign{\smallskip}
				13396	&	4245	&	1.34   &   1.57   & 2.24            & 2.18         & 2.17*   & 2.16	         &   2.26   &	2.14	&	2.23       \\		
				20885	&	4359	&	1.41   &   1.82   & 2.16	        & 2.13*       & 2.12*   &           --   &   2.18   &	2.16	&	2.20           \\		
				29861	&	4217	&	1.32   &   1.61   & 2.12*	       & 2.18         & --        & 2.18          &   2.20	 &	2.15	&	2.22        \\			
				\hline
				\noalign{\smallskip}
			\end{tabular}
		}
	\end{center}
	Note: Atmospheric parameters used in the abundance analysis were those determined in our study (Sect~\ref{sect:spec-data-atmosph-pars}).
	Abundances marked with the asterisks were determined from the weak and/or noisy lines.
\end{table*}

\begin{table*}
	\begin{center}
		\caption{Mean abundances of Fe, Na, and Zr in the three common stars obtained by us from the \GIRAFFE\ and \UVES\ spectra, and those determined by \citet{TSA14} from the same \UVES\ spectra.}
		\label{app-tab:zri-zrii-abund-common-stars}
		\resizebox{\hsize}{!}{%
			\begin{tabular}{r|cccccccc|cccccc}
				\hline\hline
				&  \multicolumn{8}{c}{This study}  &  \multicolumn{6}{c}{\citet{TSA14}}  \\
				ID	&	$\Teff$	&	$\log g$   &   $\xi_{\rm micro}$	&  [Fe/H]  &  [Na/Fe] &	[Zr$_{\rm I}$/Fe] & [Zr$_{\rm I}$/Fe]  &  [Zr$_{\rm II}$/Fe]  & $\Teff$ &	$\log g$   & $\xi_{\rm micro}$ &  [Fe/H]  &  [Na/Fe] &	[Zr$_{\rm I}$/Fe]   \\
				& K & & km\,s$^{-1}$ &  \GIRAFFE\  &  \GIRAFFE\  &  \GIRAFFE\  &  \UVES\  &  \UVES\  & & & & \UVES\  &  \UVES\  &  \UVES\  \\
				\hline
				13396	&	4245   &   1.34   &   1.57   &  $-$0.77  & 0.16 &   0.36   &  0.34  & 0.36 & 4190 &	 1.45  & 1.60  &  $-$0.83  &  0.07  &  0.25    \\			
				20885	&	4359   &   1.41   &   1.82   &  $-$0.74  & 0.21 &   0.31   &  0.27  & 0.30 & 4260 &	 1.35  & 1.90  &  $-$0.84  &  0.11  &  0.18   \\
				29861	&	4217   &   1.32   &   1.61   &  $-$0.82  & 0.28 &   0.38   &  0.36  & 0.39 & 4160 &	 1.20  & 1.50  &  $-$0.84  &  0.10  &  0.23  \\
				\hline
			\end{tabular}
		}
	\end{center}
	Note: Abundances obtained by us from the \ion{Zr}{i} 614.0460\,nm line were not used in the computation of the mean Zr abundances.
\end{table*}

\begin{figure}[tb]
	\centering
	\includegraphics[width=\hsize]{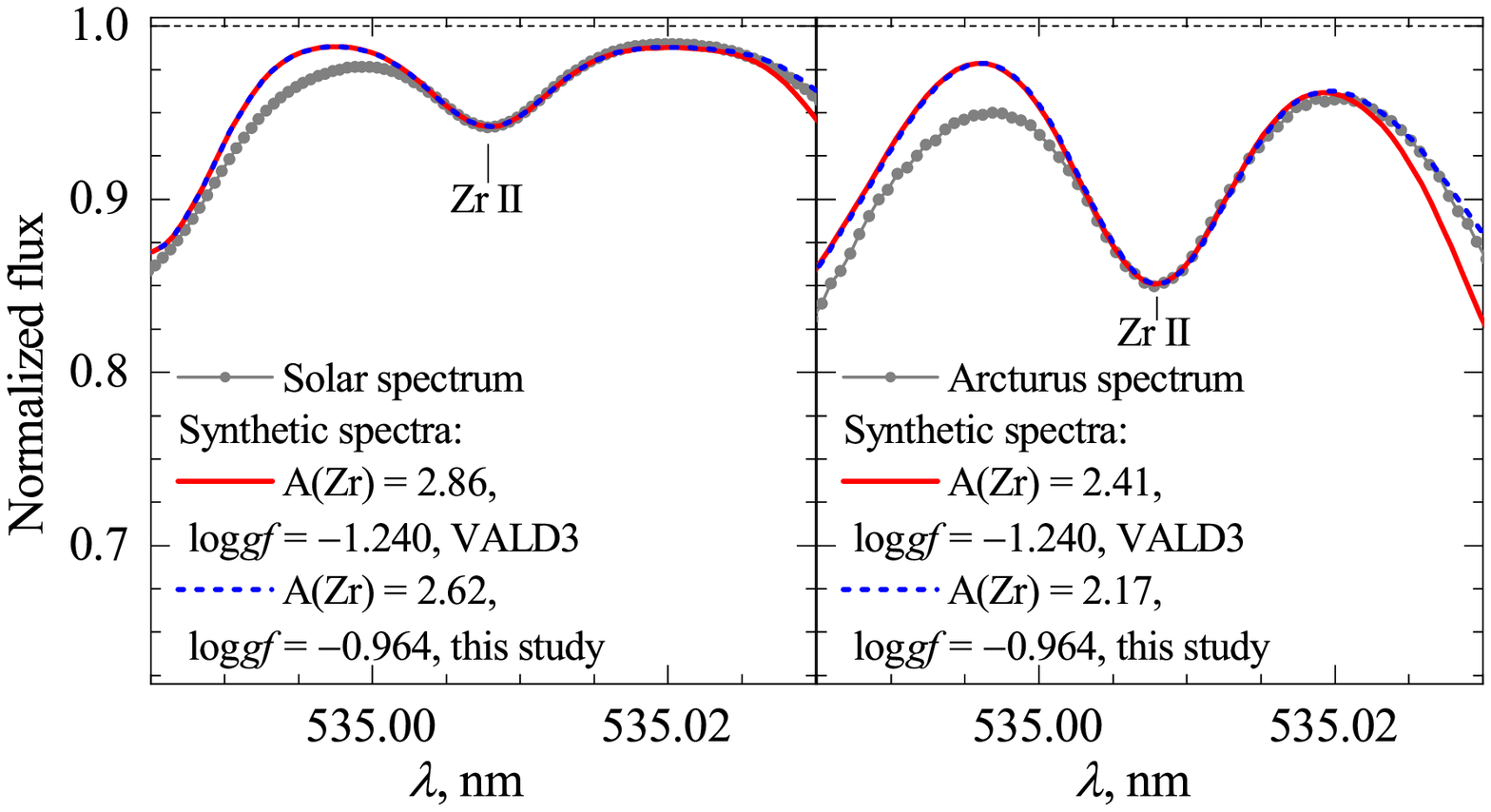}
	\caption{Left: Fit of the synthetic \ion{Zr}{ii} 535.0089\,nm line to that observed in the solar spectrum from \citet{K05}. The red solid line shows the best fit obtained using the VALD3 oscillator strength, and the blue dashed line is the fit obtained using ${\rm A(Zr)}=2.62$ from \citet{CFL11} and a modified oscillator strength (oscillator strengths and Zr abundances used/determined are indicated in the figure). For the line synthesis we used $\xi_{\rm micro}=1.00$\,km\,s$^{-1}$ and $\xi_{\rm macro}=3.80$\,km\,s$^{-1}$ from \citet{Doyle14}. Right: Best fit of the synthetic \ion{Zr}{ii} 535.0089\,nm line to that observed in the spectrum of Arcturus from \citet{HWV00}. Spectral line synthesis was done using $\xi_{\rm micro}=1.70$\,km\,s$^{-1}$ and $\xi_{\rm macro}=5.20$\,km\,s$^{-1}$ from \citet{Sheminova15}.
	}
	\label{app-fig:ZrII-Arcturus-Sun}
\end{figure}

\begin{figure}[tb]
	\resizebox{\hsize}{!}{\includegraphics{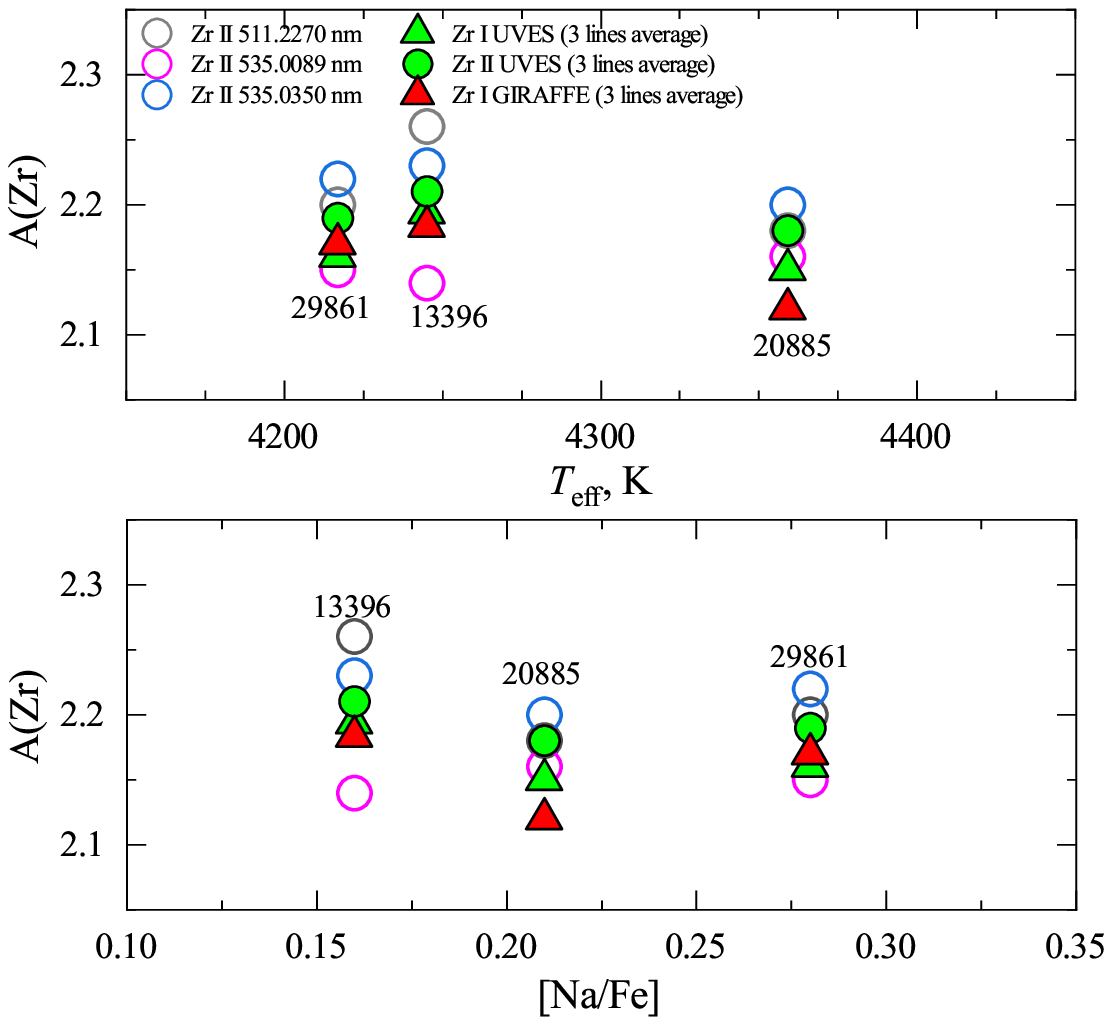}}
	\caption{Abundances of Zr in the three RGB stars common to the \GIRAFFE\ (this study) and \UVES\ samples \citep{TSA14} determined by us using \ion{Zr}{i} and \ion{Zr}{ii} lines in the \GIRAFFE\ and \UVES\ spectra (Table~\ref{app-tab:zr-abund-indiv-lines} and \ref{app-tab:zri-zrii-abund-common-stars}, Sect.~\ref{app-sect:Zrii-abund}) and plotted against the effective temperature (top) and [Na/Fe] ratio (bottom). The mean abundances obtained using \ion{Zr}{i} lines in the \GIRAFFE\ and \UVES\ spectra are shown by solid red and green triangles, respectively, while the mean abundances obtained from \ion{Zr}{ii} lines are plotted as green circles; and abundances determined from individual \ion{Zr}{ii} lines in the \UVES\ spectra are shown as hollow circles.}
	\label{app-fig:zr-zrii-abund}
\end{figure}

\begin{figure}[tb]
	\centering
	\includegraphics[width=\hsize]{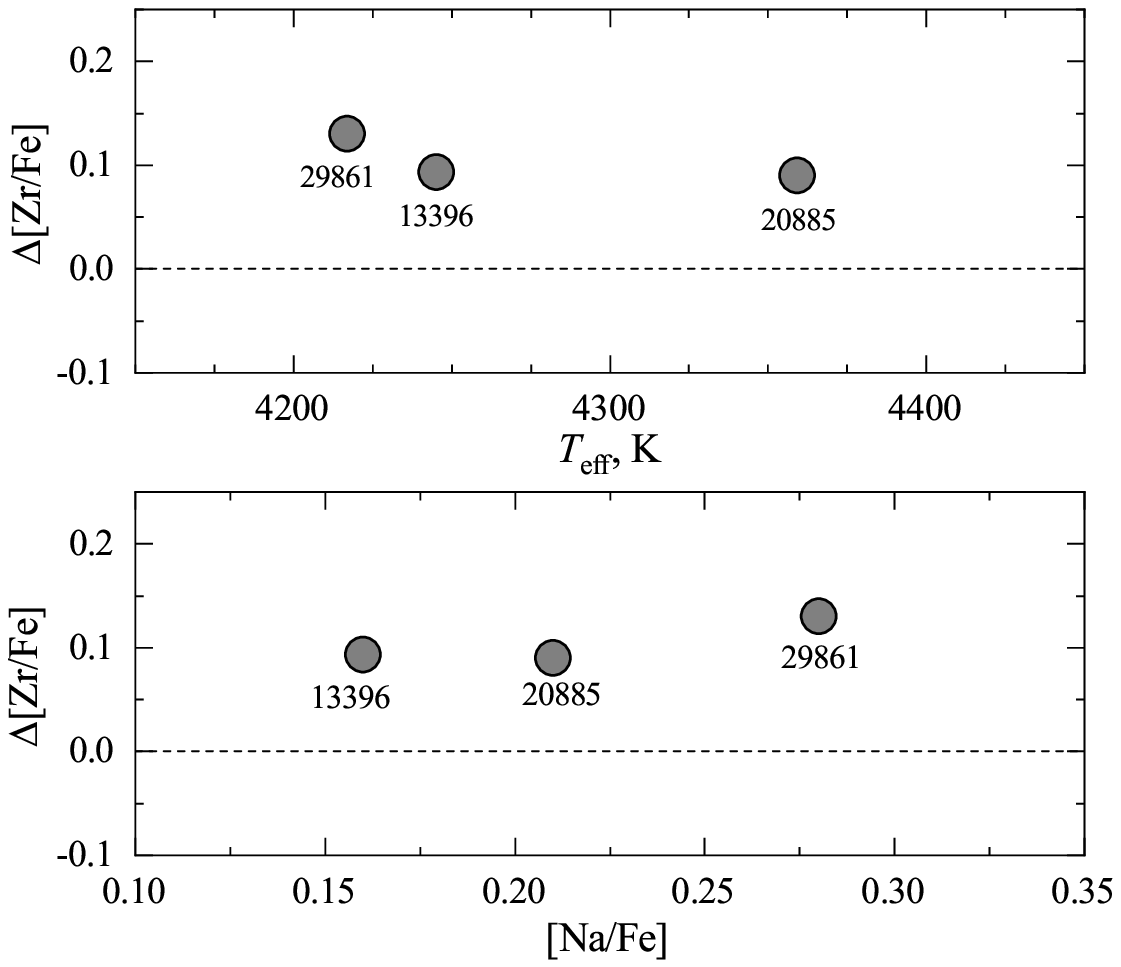}
	\caption{Differences between Zr abundances determined in this study and those obtained by \citet{TSA14} from \ion{Zr}{i} lines in the the \UVES\ spectra, plotted against the effective temperature (top) and [Na/Fe] values (bottom). 
	}
	\label{app-fig:zr-abund-our-thygesen-teff}
\end{figure}

For abundance analysis with the \UVES\ spectra we had to modify oscillator strength of the 535.0089\,nm line because the VALD3 oscillator strength led to solar Zr abundance that was $\sim 0.28$\,dex higher than the reference value of ${\rm A(Zr)} = 2.62$ from \citet[][cf. Fig.~\ref{app-fig:ZrII-Arcturus-Sun}, left panel]{CFL11}; the solar model atmosphere used in the analysis was the same as that used in Sect.~\ref{app-sect:ref-abnd}. We therefore modified the oscillator strength of this line so that we would obtain ${\rm A(Zr)} = 2.62$ from this line in the Kurucz solar spectrum (Fig.~\ref{app-fig:ZrII-Arcturus-Sun}, left panel). With the new value of $\log gf = -0.964$, the \ion{Zr}{ii} 535.0089 nm line gave ${\rm A(Zr)}=2.17$ in Arcturus (Fig.~\ref{app-fig:ZrII-Arcturus-Sun}, right panel; the model atmosphere of Arcturus was identical to that used in Sect.~\ref{app-sect:ref-abnd}). Using ${\rm A(Fe)}=-0.52$ determined in Sect.~\ref{app-sect:ref-abnd}, this gives ${\rm [Zr/Fe]}=+0.07$. The latter value agrees well with ${\rm [Zr/Fe]}=+0.01 \pm 0.07$ determined by \citet[][]{WCF09} but it is slightly higher than the average Zr abundance in Arcturus that we determined using \ion{Zr}{i} lines, ${\rm [Zr/Fe]}=-0.10$ (Sect.~\ref{app-sect:ref-abnd}).

For the three target RGB stars in 47~Tuc that have been observed both in the \GIRAFFE\ and \UVES\ samples, abundances determined from individual \ion{Zr}{i} and \ion{Zr}{ii} lines in the \UVES\ spectra agree to $\leq0.06$\,dex (Table~\ref{app-tab:zr-abund-indiv-lines}; Fig.~\ref{app-fig:zr-zrii-abund}). The mean abundances obtained from \ion{Zr}{i} lines in the \GIRAFFE\ and \UVES\ spectra agree even better, with the differences being less than $0.03$\,dex (Table~\ref{app-tab:zri-zrii-abund-common-stars}, Fig.~\ref{app-fig:zr-zrii-abund}).

On average, abundances that we obtained using \ion{Zr}{i} lines in the \UVES\ spectra are $\sim 0.10$\,dex higher than those determined by \citet[][cf. Table~\ref{app-tab:zri-zrii-abund-common-stars}, Fig.~\ref{app-fig:zr-abund-our-thygesen-teff}]{TSA14}. Although there are small differences in the effective temperatures and gravities used in the two studies, they alone cannot account for more than $\sim0.02$\,dex of this discrepancy. On average, the microturbulence velocities obtained by \citet{TSA14} are $\sim 0.04$\,km\,s$^{-1}$ lower than those used in our study (Table~\ref{app-tab:zri-zrii-abund-common-stars}). This would lead to Zr abundances that are by $\sim0.1$\,dex higher than those that would be obtained with our $\xi_{\rm micro}$ values. However, the main difference between the two analyses is that we used an updated ionisation potential for \ion{Zr}{i}, 6.634\,eV from \citet{HHM86} instead of the default value of 6.840\,eV that is implemented with the older version of Kurucz package (Sect.~\ref{sect:abund}). This alone would lead to our Zr abundances being $\sim0.29$\,dex higher. Thus, the latter two effects (working in different directions) would be fully able to account for the difference in Zr abundances obtained by us and \citet{TSA14}.

\subsubsection{Errors in the determined Fe, Na, and Zr abundances\label{app-sect:abund-err}}

For estimating typical uncertainties in the determined abundances of Fe, Na, and Zr, we followed the procedure described in \citet{CKK17}. With this, we accounted for the following error sources:

\begin{itemize}
	\item
	The error in abundance due to uncertainty in the determined effective temperature was estimated by computing \Teff\ values that were increased and decreased by the amount which corresponded to the uncertainty in determined photometric magnitudes. A conservative estimate of $\sigma(V) = \sigma(I) = 0.03$ was used, which translated to an uncertainty of $\pm 60$\,K in the effective temperature. This value was used to estimate abundance uncertainty caused by errors in the determination of \Teff.
	\item
	An abundance error due to uncertainty in the determination of surface gravity, $\sigma(\log g) = \pm 0.2$, was obtained by taking into account individual uncertainties in the estimates of \Teff, luminosity, and stellar mass.
	\item
	We used the RMS variation of microturbulence velocity, $\pm 0.2$\,km\,s$^{-1}$, as a representative uncertainty in $\xi_{\rm micro}$. This value was used to estimate the abundance error due to variation in $\xi_{\rm micro}$.
	\item
	An abundance error due to uncertainties in the continuum determination, $\sigma({\rm cont})$, was obtained by measuring the continuum flux dispersion, $\sigma({\rm flux})$, in the wavelength intervals that are located close to the investigated spectral lines and that are line-free:
	\begin{equation}
	\sigma({\rm cont}) = \dfrac{\sigma({\rm flux})}{\sqrt{N}},
	\end{equation}
	\noindent where \textit{N} is the number of wavelength points. The resulting error was used to increase and decrease the continuum level, after which the line fitting was done again and the resulting difference in the abundance was taken as the abundance error due to uncertainty in the continuum determination.
	\item
	To estimate errors in the line profile fitting, we computed RMS variation between the observed and best-fitted line profiles. The line equivalent width was then increased and decreased by this value to obtain the difference in the corresponding abundances of the given element. The latter were used as abundance uncertainties due to errors in the line profile fitting.
\end{itemize}

The final abundance errors were obtained by adding individual error components in quadratures.
To obtain abundance errors for our target stars, we selected a subsample of 8 stars for Zr, Na and Fe abundance error estimation, with their effective temperatures distributed evenly over the entire \Teff\ range of 237 RGB targets. Using the prescription above, we determined typical abundance errors of Zr, Na, and Fe for these stars. In this procedure, we  assumed median equivalent widths of these lines at the given \Teff, i.e. no variation of the line strength was taken into account. 
The determined abundance errors are provided in Tables~\ref{app-tab:Fe-abund-errors}-\ref{app-tab:Zr-abund-errors}.

\begin{table}
	\begin{center}
		\caption{Typical Fe abundance measurement errors.
			\label{app-tab:Fe-abund-errors}}
		\resizebox{\hsize}{!}{%
			\begin{tabular}{cccccccc}
				\hline\hline
				\noalign{\smallskip}
				\textit{T}$_{\mathrm{eff}}$ & Star ID & $\sigma$(\textit{T}$_{\mathrm{eff}}$)  & $\sigma$(log\textit{g})  & $\sigma$($\xi_\mathrm{t}$)  & $\sigma$(cont)  & $\sigma$(fit)  &$\sigma$$_\mathrm{i}$(total)  \\
				&  &    dex     & dex                            &  dex  & dex & dex  & dex            \\
				\hline\noalign{\smallskip}
				4195	&	19042	&	0.03	&	0.02	&	0.12	&	0.10	&	0.02	&	0.16	\\
				4317	&	23335	&	0.03	&	0.02	&	0.12	&	0.10	&	0.02	&	0.16	\\
				4395	&	23175	&	0.03	&	0.02	&	0.12	&	0.10	&	0.02	&	0.16	\\
				4459	&	13047	&	0.03	&	0.02	&	0.12	&	0.10	&	0.03	&	0.16	\\
				4560	&	27381	&	0.03	&	0.02	&	0.11	&	0.10	&	0.03	&	0.16	\\
				4645	&	30297	&	0.03	&	0.02	&	0.11	&	0.10	&	0.04	&	0.16	\\
				4712	&	10496	&	0.04	&	0.02	&	0.11	&	0.10	&	0.05	&	0.16	\\
				4797	&	26403	&	0.04	&	0.02	&	0.11	&	0.10	&	0.05	&	0.16	\\			
				\hline
			\end{tabular}
		}
	\end{center}
\end{table}

\begin{table}
	\begin{center}
		\caption{Typical Na abundance measurement errors.
		}
		\label{app-tab:Na-abund-errors}
		\resizebox{\hsize}{!}{%
			\begin{tabular}{cccccccc}
				\hline\hline
				\noalign{\smallskip}
				\textit{T}$_{\mathrm{eff}}$ & Star ID & $\sigma$(\textit{T}$_{\mathrm{eff}}$)  & $\sigma$(log\textit{g})  & $\sigma$($\xi_\mathrm{t}$)  & $\sigma$(cont)  & $\sigma$(fit)  &$\sigma$$_\mathrm{i}$(total)  \\
				&  &    dex     & dex                            &  dex  & dex & dex  & dex            \\
				\hline\noalign{\smallskip}
				4195	&	19042	&	0.06	&	0.01	&	0.05	&	0.03	&	0.03	&	0.09	\\
				4317	&	23335	&	0.06	&	0.01	&	0.05	&	0.03	&	0.04	&	0.09	\\
				4395	&	23175	&	0.06	&	0.01	&	0.04	&	0.03	&	0.04	&	0.09	\\
				4459	&	13047	&	0.06	&	0.01	&	0.04	&	0.03	&	0.04	&	0.09	\\
				4560	&	27381	&	0.06	&	0.01	&	0.04	&	0.03	&	0.05	&	0.09	\\
				4645	&	30297	&	0.06	&	0.01	&	0.03	&	0.03	&	0.05	&	0.09	\\
				4712	&	10496	&	0.06	&	0.01	&	0.03	&	0.03	&	0.05	&	0.09	\\
				4797	&	26403	&	0.06	&	0.01	&	0.03	&	0.03	&	0.05	&	0.09	\\
				
				\hline
			\end{tabular}
		}
	\end{center}
\end{table}

\begin{table}
	\begin{center}
		\caption{Typical Zr abundance measurement errors.
		}
		\label{app-tab:Zr-abund-errors}
		\resizebox{\hsize}{!}{%
			\begin{tabular}{cccccccc}
				\hline\hline
				\noalign{\smallskip}
				\textit{T}$_{\mathrm{eff}}$ & Star ID & $\sigma$(\textit{T}$_{\mathrm{eff}}$)  & $\sigma$(log\textit{g})  & $\sigma$($\xi_\mathrm{t}$)  & $\sigma$(cont)  & $\sigma$(fit)  &$\sigma$$_\mathrm{i}$(total)  \\
				&  &    dex     & dex                            &  dex  & dex & dex  & dex            \\
				\hline\noalign{\smallskip}
				4195	&	19042	&	0.11	&	0.02	&	0.09	&	0.03	&	0.04	&	0.15	\\
				4317	&	23335	&	0.11	&	0.01	&	0.07	&	0.03	&	0.04	&	0.14	\\
				4395	&	23175	&	0.11	&	0.01	&	0.04	&	0.03	&	0.05	&	0.13	\\
				4459	&	13047	&	0.11	&	0.01	&	0.03	&	0.03	&	0.05	&	0.13	\\
				4560	&	27381	&	0.11	&	0.00	&	0.03	&	0.03	&	0.05	&	0.13	\\
				4645	&	30297	&	0.11	&	0.00	&	0.02	&	0.03	&	0.05	&	0.13	\\
				4712	&	10496	&	0.11	&	0.00	&	0.01	&	0.03	&	0.05	&	0.12	\\
				4797	&	26403	&	0.11	&	0.00	&	0.01	&	0.03	&	0.05	&	0.12	\\			
				\hline
			\end{tabular}
		}
	\end{center}
\end{table}

\section{Statistical significance of possible correlations between the full spatial velocity dispersions and [Na/Fe] and [Zr/Fe] abundance ratios\label{app-sect:Levene-test}}

As in the case of testing the statistical significance of the possible correlations in the abundance--abundance and abundance--radial distance planes, we computed the $p$-values using Pearson's, Spearman's, and Kendall's correlation coefficients in the $\sigma_{v_{\rm full}} - {\rm [Na/Fe]}$ and $\sigma_{v_{\rm full}} - {\rm [Zr/Fe]}$ planes. In the former case, the we obtain $p>0.2$ which indicates that there is no correlation between the full spatial velocity dispersion and [Na/Fe] ratio. The values obtained for the Spearman's and Kendall's coefficients are sufficiently small ($p<0.04$) to claim the possible existence of a weak correlation in the $\sigma_{v_{\rm full}} - {\rm [Zr/Fe]}$ plane, though the $p$-value obtained for the Pearson's correlation coefficient is somewhat higher, at $p=0.059$.

To further verify the possible existence of the correlation in the $\sigma_{v_{\rm full}} - {\rm [Zr/Fe]}$ plane, we tested the statistical significance of differences between the variances of full spatial velocities of individual stars in the 0.1\,dex wide [Zr/Fe] and [Na/Fe] abundance bins (Sect.~\ref{sect:results-and-disc} and Fig.~\ref{fig:vel-disp}). This was done using the Levene's test \citep{L1960, BF1974} which was applied to verify the assumption of the null hypothesis that the full spatial velocities in the different [Zr/Fe] and [Na/Fe] abundance bins have equal variances.The obtained $p$-values were $\geq0.37$ both in the $\sigma(v_{\rm full}) - {\rm [Na/Fe]}$ and $\sigma(v_{\rm full}) - {\rm [Zr/Fe]}$ planes thereby indicating that in both cases the null hypothesis could not be rejected.
Based on these results we therefore conclude that there are no statistically significant correlations between the dispersions of full spatial velocities and [Na/Fe] and [Zr/Fe] abundance ratios in the target RGB stars of 47~Tuc.

%
%
%

\section{Abundances of Fe, Na, and Zr determined in the target RGB stars in 47~Tuc\label{app-sect:abund-list-all}}

Abundances of Fe, Na, and Zr determined in the individual target RGB stars in 47~Tuc (Sect.~\ref{sect:abund_Fe}, \ref{sect:abund_Na}, and \ref{sect:abund_Zr-I}, respectively) are listed in Table~\ref{app-tab:abund-indiv-stars}, together with the atmospheric parameters of individual targets determined in Sect.~\ref{sect:spec-data-atmosph-pars}.

\clearpage
\onecolumn
\begin{center}
	\setlength{\tabcolsep}{5pt}
	\tiny
	\begin{longtable}{cccccccccccc}
		\caption{Abundances of Fe, Na, and Zr determined in the sample of 237 stars in 47~Tuc. 
			\label{app-tab:abund-indiv-stars}}\\
		\hline\hline
		\noalign{\smallskip}
		GAIA Source ID  & $\Teff$ & $\log g$  & $\xi_{\rm micro}$ &  A(Fe)  &  [Fe/H] &  A(Na)  &  [Na/Fe]  &  A(Zr)  &  [Zr/Fe] & ID & Obs. programme  \\
		&  K     &           &      \,km\,s$^{-1}$         &         &         &         &           &         &       &    \\
		\hline\noalign{\smallskip}
		\endfirsthead
		\multicolumn{12}{c}%
		{\tablename\ \thetable\ -- \textit{Continued from previous page}} \\
		\hline\hline
		\noalign{\smallskip}
		GAIA Source ID  & $\Teff$ & $\log g$  & $\xi_{\rm micro}$ &  A(Fe)  &  [Fe/H] &  A(Na)  &  [Na/Fe]  &  A(Zr)  &  [Zr/Fe] & ID & Obs. programme  \\
		&  K     &           &      \,km\,s$^{-1}$         &         &         &         &           &         &       &    \\
		\hline\noalign{\smallskip}
		\endhead
		\hline \multicolumn{12}{r}{\textit{Continued on next page}} \\
		\endfoot
		\hline
		\endlastfoot
		4689624698313319040	&	4746	&	2.38	&	1.14	&	6.81	&	--0.74	&	5.63	&	0.20	&	2.19	&	0.31	&	4373	&	073.D-0211(A)	\\
		&		&		&	1.14	&	6.80	&	--0.75	&	5.65	&	0.23	&	2.12	&	0.25	&	F-4373	&	072.D-0777(A)	\\
		4689625144980306432	&	4675	&	2.13	&	1.25	&	6.88	&	--0.67	&	5.81	&	0.31	&	2.26	&	0.31	&	5172	&	073.D-0211(A)	\\
		&		&		&	1.43	&	6.78	&	--0.77	&	5.90	&	0.50	&	2.25	&	0.40	&	R259	&	088.D-0026(A)	\\
		4689613840634674048	&	4271	&	1.48	&	1.48	&	6.82	&	--0.73	&	5.66	&	0.22	&	2.30	&	0.41	&	5277	&	073.D-0211(A)	\\
		4689614837066802944	&	4630	&	2.09	&	1.40	&	6.79	&	--0.76	&	5.84	&	0.43	&	2.27	&	0.41	&	6808	&	073.D-0211(A)	\\
		4689624801392182784	&	4759	&	2.25	&	1.21	&	6.88	&	--0.67	&	5.85	&	0.35	&	2.38	&	0.43	&	7904	&	073.D-0211(A)	\\
		&		&		&	1.30	&	6.79	&	--0.76	&	5.92	&	0.51	&	2.34	&	0.48	&	R443	&	088.D-0026(A)	\\
		4689629405598401664	&	4738	&	2.40	&	1.17	&	6.83	&	--0.72	&	5.88	&	0.43	&	2.25	&	0.35	&	9268	&	073.D-0211(A)	\\
		4689637205260940416	&	4513	&	1.90	&	1.42	&	6.80	&	--0.75	&	5.60	&	0.18	&	2.20	&	0.33	&	9518	&	073.D-0211(A)	\\
		&		&		&	1.42	&	6.80	&	--0.75	&	5.69	&	0.27	&	2.17	&	0.30	&	R237	&	088.D-0026(A)	\\
		4689626210141359616	&	4628	&	2.12	&	1.18	&	6.87	&	--0.68	&	5.77	&	0.28	&	2.21	&	0.27	&	9717	&	073.D-0211(A)	\\
		&		&		&	1.29	&	6.83	&	--0.72	&	5.75	&	0.30	&	2.19	&	0.29	&	R682	&	088.D-0026(A)	\\
		4689623594517841024	&	4696	&	2.27	&	1.19	&	6.84	&	--0.71	&	5.84	&	0.38	&	2.23	&	0.32	&	10397	&	073.D-0211(A)	\\
		4689626313210555904	&	4712	&	2.24	&	1.26	&	6.85	&	--0.70	&	5.95	&	0.48	&	2.33	&	0.41	&	10496	&	073.D-0211(A)	\\
		4689629779253913984	&	4543	&	2.04	&	1.14	&	6.82	&	--0.73	&	5.56	&	0.12	&	2.01	&	0.12	&	10994	&	073.D-0211(A)	\\
		4689632669773515392	&	4649	&	2.21	&	1.34	&	6.79	&	--0.76	&	5.76	&	0.35	&	2.21	&	0.35	&	11008	&	073.D-0211(A)	\\
		4689626381940026752	&	4708	&	2.28	&	1.37	&	6.73	&	--0.82	&	5.72	&	0.37	&	2.15	&	0.35	&	11169	&	073.D-0211(A)	\\
		4689626416300118400	&	4790	&	2.45	&	1.17	&	6.84	&	--0.71	&	5.82	&	0.36	&	2.22	&	0.31	&	11675	&	073.D-0211(A)	\\
		4689637578914081024	&	4616	&	2.03	&	1.42	&	6.84	&	--0.71	&	5.99	&	0.53	&	2.30	&	0.39	&	11709	&	073.D-0211(A)	\\
		4689625862273037696	&	4781	&	2.43	&	1.11	&	6.87	&	--0.68	&	5.73	&	0.24	&	2.24	&	0.30	&	11756	&	073.D-0211(A)	\\
		4689627099193944064	&	4517	&	1.83	&	1.48	&	6.85	&	--0.70	&	6.06	&	0.59	&	2.31	&	0.39	&	11927	&	073.D-0211(A)	\\
		4689637819432237952	&	4459	&	1.78	&	1.43	&	6.76	&	--0.79	&	5.70	&	0.32	&	2.22	&	0.39	&	13047	&	073.D-0211(A)	\\
		4689638373492020736	&	4614	&	2.10	&	1.27	&	6.79	&	--0.76	&	5.60	&	0.19	&	2.16	&	0.30	&	13668	&	073.D-0211(A)	\\
		&		&		&	1.23	&	6.82	&	--0.73	&	5.64	&	0.20	&	2.14	&	0.25	&	F-13668	&	072.D-0777(A)	\\
		4689627167942570240	&	4742	&	2.43	&	1.18	&	6.84	&	--0.71	&	5.72	&	0.26	&	2.20	&	0.29	&	13718	&	073.D-0211(A)	\\
		4689625896603074816	&	4736	&	2.29	&	1.33	&	6.83	&	--0.72	&	6.10	&	0.65	&	2.35	&	0.45	&	13939	&	073.D-0211(A)	\\
		4689627236649879936	&	4755	&	2.43	&	1.17	&	6.81	&	--0.74	&	5.81	&	0.38	&	2.39	&	0.51	&	14487	&	073.D-0211(A)	\\
		4689625965322548992	&	4614	&	2.01	&	1.38	&	6.87	&	--0.68	&	6.13	&	0.64	&	2.43	&	0.49	&	15552	&	073.D-0211(A)	\\
		&		&		&	1.53	&	6.78	&	--0.77	&	6.16	&	0.76	&	2.47	&	0.62	&	R381	&	088.D-0026(A)	\\
		4689627240923227392	&	4698	&	2.22	&	1.28	&	6.85	&	--0.70	&	5.96	&	0.49	&	2.23	&	0.31	&	16462	&	073.D-0211(A)	\\
		4689646070074296960	&	4694	&	2.26	&	1.28	&	6.80	&	--0.75	&	5.79	&	0.37	&	2.15	&	0.28	&	16597	&	073.D-0211(A)	\\
		4689625965322548096	&	4679	&	2.14	&	1.31	&	6.85	&	--0.70	&	5.95	&	0.48	&	2.29	&	0.37	&	16777	&	073.D-0211(A)	\\
		4689637926807831040	&	4725	&	2.22	&	1.19	&	6.90	&	--0.65	&	6.06	&	0.54	&	2.32	&	0.35	&	16957	&	073.D-0211(A)	\\
		4689625965322552832	&	4278	&	1.39	&	1.57	&	6.73	&	--0.82	&	5.82	&	0.47	&	2.35	&	0.55	&	17193	&	073.D-0211(A)	\\
		4689637716353064960	&	4471	&	1.73	&	1.39	&	6.86	&	--0.69	&	5.93	&	0.45	&	2.25	&	0.32	&	17926	&	073.D-0211(A)	\\
		4689627447091792640	&	4761	&	2.28	&	1.21	&	6.85	&	--0.70	&	5.95	&	0.48	&	2.35	&	0.43	&	18374	&	073.D-0211(A)	\\
		4689637789368802816	&	4628	&	2.10	&	1.20	&	6.80	&	--0.75	&	5.57	&	0.15	&	2.16	&	0.29	&	19992	&	073.D-0211(A)	\\
		4689638785808484224	&	4582	&	1.97	&	1.43	&	6.81	&	--0.74	&	5.79	&	0.36	&	2.30	&	0.42	&	20019	&	073.D-0211(A)	\\
		4689638781505129216	&	4582	&	1.98	&	1.41	&	6.80	&	--0.75	&	5.89	&	0.47	&	2.27	&	0.40	&	20387	&	073.D-0211(A)	\\
		4689627481441306624	&	4748	&	2.18	&	1.17	&	6.93	&	--0.62	&	6.09	&	0.54	&	2.39	&	0.39	&	20701	&	073.D-0211(A)	\\
		4689638781505112576	&	4498	&	1.80	&	1.45	&	6.79	&	--0.76	&	5.96	&	0.55	&	2.27	&	0.41	&	20983	&	073.D-0211(A)	\\
		4689627584520466944	&	4494	&	1.77	&	1.48	&	6.81	&	--0.74	&	6.02	&	0.59	&	2.21	&	0.33	&	22726	&	073.D-0211(A)	\\
		4689637995535079168	&	4395	&	1.53	&	1.52	&	6.77	&	--0.78	&	5.95	&	0.56	&	2.31	&	0.47	&	23175	&	073.D-0211(A)	\\
		4689627584530658048	&	4634	&	2.07	&	1.38	&	6.84	&	--0.71	&	6.16	&	0.70	&	2.31	&	0.40	&	23211	&	073.D-0211(A)	\\
		4689639266835463552	&	4317	&	1.38	&	1.60	&	6.79	&	--0.76	&	6.00	&	0.59	&	2.41	&	0.55	&	23335	&	073.D-0211(A)	\\
		4689639473003821440	&	4404	&	1.46	&	1.38	&	6.79	&	--0.76	&	5.84	&	0.43	&	2.54	&	0.68	&	23821	&	073.D-0211(A)	\\
		4689642496660354048	&	4303	&	1.37	&	1.79	&	6.78	&	--0.77	&	5.72	&	0.32	&	2.13	&	0.28	&	23879	&	073.D-0211(A)	\\
		4689641946904328576	&	4770	&	2.35	&	1.24	&	6.82	&	--0.73	&	5.94	&	0.50	&	2.33	&	0.44	&	24533	&	073.D-0211(A)	\\
		4689641775105775488	&	4788	&	2.40	&	1.18	&	6.84	&	--0.71	&	6.17	&	0.71	&	2.31	&	0.40	&	24553	&	073.D-0211(A)	\\
		4689639473003144704	&	4289	&	1.25	&	1.52	&	6.72	&	--0.83	&	5.86	&	0.52	&	2.43	&	0.64	&	25226	&	073.D-0211(A)	\\
		4689638059950649728	&	4409	&	1.58	&	1.57	&	6.75	&	--0.80	&	6.02	&	0.65	&	2.25	&	0.43	&	25961	&	073.D-0211(A)	\\
		4689639473007322496	&	4286	&	1.22	&	1.62	&	6.80	&	--0.75	&	5.85	&	0.43	&	2.30	&	0.43	&	26058	&	073.D-0211(A)	\\
		4689627790688724096	&	4643	&	2.14	&	1.32	&	6.81	&	--0.74	&	5.96	&	0.53	&	2.19	&	0.31	&	26281	&	073.D-0211(A)	\\
		4689639125125572864	&	4797	&	2.45	&	1.24	&	6.81	&	--0.74	&	6.04	&	0.61	&	2.41	&	0.53	&	26403	&	073.D-0211(A)	\\
		4689641878184846464	&	4790	&	2.43	&	1.23	&	6.75	&	--0.80	&	5.80	&	0.43	&	2.23	&	0.41	&	26713	&	073.D-0211(A)	\\
		4689639507362821760	&	4731	&	2.10	&	1.23	&	6.83	&	--0.72	&	6.09	&	0.64	&	2.38	&	0.48	&	26870	&	073.D-0211(A)	\\
		4689639399980545280	&	4560	&	1.91	&	1.37	&	6.90	&	--0.65	&	6.20	&	0.68	&	2.35	&	0.38	&	27381	&	073.D-0211(A)	\\
		4689639125102514176	&	4289	&	1.40	&	1.56	&	6.78	&	--0.77	&	6.04	&	0.64	&	2.21	&	0.36	&	27385	&	073.D-0211(A)	\\
		4689642629818274176	&	4797	&	2.44	&	1.17	&	6.77	&	--0.78	&	5.98	&	0.59	&	2.18	&	0.34	&	27499	&	073.D-0211(A)	\\
		4689627859408196864	&	4692	&	2.33	&	1.25	&	6.79	&	--0.76	&	5.79	&	0.38	&	2.37	&	0.51	&	27755	&	073.D-0211(A)	\\
		4689639610432530304	&	4645	&	2.15	&	1.35	&	6.71	&	--0.84	&	5.62	&	0.29	&	2.13	&	0.35	&	30297	&	073.D-0211(A)	\\
		4689628026902152448	&	4501	&	1.84	&	1.52	&	6.75	&	--0.80	&	5.92	&	0.55	&	2.17	&	0.35	&	30463	&	073.D-0211(A)	\\
		&		&		&	1.42	&	6.82	&	--0.73	&	6.05	&	0.61	&	2.18	&	0.29	&	B-30463	&	072.D-0777(A)	\\
		4689639747880182784	&	4746	&	2.41	&	1.11	&	6.75	&	--0.80	&	5.69	&	0.32	&	2.18	&	0.36	&	30952	&	073.D-0211(A)	\\
		4689628061260674304	&	4558	&	1.94	&	1.39	&	6.74	&	--0.81	&	5.90	&	0.54	&	2.19	&	0.38	&	31190	&	073.D-0211(A)	\\
		4689639438642541696	&	4503	&	1.81	&	1.37	&	6.86	&	--0.69	&	5.99	&	0.51	&	2.28	&	0.35	&	31426	&	073.D-0211(A)	\\
		4689640331996429696	&	4462	&	1.72	&	1.44	&	6.82	&	--0.73	&	5.93	&	0.49	&	2.29	&	0.40	&	31525	&	073.D-0211(A)	\\
		4689639644800976896	&	4428	&	1.72	&	1.44	&	6.79	&	--0.76	&	5.75	&	0.34	&	2.23	&	0.37	&	31560	&	073.D-0211(A)	\\
		4689638888887480576	&	4677	&	2.11	&	1.28	&	6.81	&	--0.74	&	5.89	&	0.46	&	2.16	&	0.28	&	31638	&	073.D-0211(A)	\\
		4689643390013939968	&	4775	&	2.44	&	1.20	&	6.81	&	--0.74	&	5.91	&	0.48	&	2.21	&	0.33	&	33009	&	073.D-0211(A)	\\
		4689639674858450944	&	4563	&	1.88	&	1.39	&	6.82	&	--0.73	&	5.85	&	0.41	&	2.35	&	0.46	&	34033	&	073.D-0211(A)	\\
		4689640568211564416	&	4527	&	1.85	&	1.50	&	6.84	&	--0.71	&	6.16	&	0.70	&	2.34	&	0.43	&	35160	&	073.D-0211(A)	\\
		4689639159462279552	&	4740	&	2.29	&	1.21	&	6.81	&	--0.74	&	5.77	&	0.34	&	2.20	&	0.32	&	35454	&	073.D-0211(A)	\\
		4689639846657238912	&	4533	&	1.85	&	1.48	&	6.76	&	--0.79	&	5.97	&	0.59	&	2.20	&	0.37	&	35878	&	073.D-0211(A)	\\
		&		&		&	1.45	&	6.79	&	--0.76	&	6.03	&	0.62	&	2.23	&	0.37	&	B-35878	&	072.D-0777(A)	\\
		4689639885318980224	&	4614	&	2.09	&	1.35	&	6.76	&	--0.79	&	5.86	&	0.48	&	2.12	&	0.29	&	37036	&	073.D-0211(A)	\\
		4689640091468599936	&	4647	&	2.24	&	1.18	&	6.79	&	--0.76	&	5.60	&	0.19	&	1.98	&	0.12	&	38289	&	073.D-0211(A)	\\
		4689640602571357312	&	4489	&	1.80	&	1.37	&	6.84	&	--0.71	&	6.03	&	0.57	&	2.15	&	0.24	&	39101	&	073.D-0211(A)	\\
		4689640980528658048	&	4775	&	2.46	&	1.04	&	6.89	&	--0.66	&	6.08	&	0.57	&	2.37	&	0.41	&	42243	&	073.D-0211(A)	\\
		4689637617577709696	&	4310	&	1.48	&	1.48	&	6.91	&	--0.64	&	5.81	&	0.28	&	2.31	&	0.33	&	B-12128	&	072.D-0777(A)	\\
		4689624354715978112	&	4340	&	1.55	&	1.47	&	6.77	&	--0.78	&	5.64	&	0.25	&	2.17	&	0.33	&	B-1256	&	072.D-0777(A)	\\
		&		&		&	1.53	&	6.72	&	--0.83	&	5.62	&	0.28	&	2.18	&	0.39	&	R287	&	088.D-0026(A)	\\
		4689614970198135296	&	4245	&	1.34	&	1.57	&	6.78	&	--0.77	&	5.56	&	0.16	&	2.21	&	0.36	&	B-13396	&	072.D-0777(A)	\\
		4689626652510887168	&	4262	&	1.34	&	1.72	&	6.81	&	--0.74	&	5.60	&	0.17	&	2.05	&	0.17	&	B-13795	&	072.D-0777(A)	\\
		&		&		&	1.77	&	6.75	&	--0.80	&	5.57	&	0.20	&	2.07	&	0.25	&	13795	&	073.D-0211(A)	\\
		&		&		&	1.78	&	6.74	&	--0.81	&	5.64	&	0.28	&	2.03	&	0.22	&	R752	&	088.D-0026(A)	\\
		4689637510186504064	&	4392	&	1.67	&	1.61	&	6.73	&	--0.82	&	5.65	&	0.30	&	2.24	&	0.44	&	B-13853	&	073.D-0211(A)	\\
		&		&		&	1.57	&	6.76	&	--0.79	&	5.69	&	0.31	&	2.13	&	0.30	&	R246	&	088.D-0026(A)	\\
		4689633799342000640	&	4357	&	1.61	&	1.48	&	6.73	&	--0.82	&	5.57	&	0.22	&	2.15	&	0.35	&	B-14013	&	073.D-0211(A)	\\
		4689638403539699584	&	4319	&	1.52	&	1.56	&	6.82	&	--0.73	&	5.71	&	0.27	&	2.26	&	0.37	&	B-14583	&	073.D-0211(A)	\\
		&		&		&	1.56	&	6.77	&	--0.78	&	5.79	&	0.40	&	2.26	&	0.42	&	14583	&	073.D-0211(A)	\\
		4689639919678548736	&	4259	&	1.33	&	1.54	&	6.87	&	--0.68	&	5.90	&	0.41	&	2.28	&	0.34	&	B-14653	&	073.D-0211(A)	\\
		4689637888151712640	&	4506	&	1.68	&	1.84	&	6.71	&	--0.84	&	5.49	&	0.16	&	2.05	&	0.27	&	B-15797	&	073.D-0211(A)	\\
		4689645928342350976	&	4420	&	1.78	&	1.54	&	6.80	&	--0.75	&	5.82	&	0.40	&	2.18	&	0.31	&	B-16667	&	073.D-0211(A)	\\
		&		&		&	1.40	&	6.82	&	--0.73	&	5.89	&	0.45	&	2.12	&	0.23	&	R790	&	088.D-0026(A)	\\
		4689638476571141248	&	4503	&	1.86	&	1.42	&	6.79	&	--0.76	&	5.66	&	0.25	&	2.21	&	0.35	&	B-16865	&	073.D-0211(A)	\\
		4689638648369724288	&	4425	&	1.69	&	1.53	&	6.76	&	--0.79	&	5.66	&	0.28	&	2.20	&	0.37	&	B-17189	&	073.D-0211(A)	\\
		4689638334822345984	&	4512	&	1.84	&	1.38	&	6.93	&	--0.62	&	5.89	&	0.34	&	2.37	&	0.37	&	B-17819	&	073.D-0211(A)	\\
		&		&		&	1.38	&	6.86	&	--0.69	&	5.88	&	0.40	&	2.20	&	0.27	&	R245	&	088.D-0026(A)	\\
		4689638545290915584	&	4404	&	1.44	&	1.79	&	6.83	&	--0.72	&	5.87	&	0.42	&	2.15	&	0.25	&	B-18661	&	073.D-0211(A)	\\
		4689626893035580032	&	4506	&	1.84	&	1.52	&	6.76	&	--0.79	&	5.96	&	0.58	&	2.26	&	0.43	&	B-19269	&	073.D-0211(A)	\\
		4689638918944031616	&	4359	&	1.41	&	1.82	&	6.81	&	--0.74	&	5.64	&	0.21	&	2.19	&	0.31	&	B-20885	&	073.D-0211(A)	\\
		4689627030474839936	&	4476	&	1.78	&	1.47	&	6.80	&	--0.75	&	5.94	&	0.52	&	2.24	&	0.37	&	B-23032	&	073.D-0211(A)	\\
		4689627756329297024	&	4431	&	1.72	&	1.40	&	6.83	&	--0.72	&	5.77	&	0.32	&	2.26	&	0.36	&	B-25554	&	073.D-0211(A)	\\
		4689641976960762240	&	4549	&	1.77	&	1.63	&	6.88	&	--0.67	&	5.55	&	0.05	&	2.18	&	0.23	&	B-26147	&	073.D-0211(A)	\\
		4689628065566610176	&	4503	&	1.87	&	1.53	&	6.83	&	--0.72	&	5.97	&	0.52	&	2.20	&	0.30	&	B-30949	&	073.D-0211(A)	\\
		&		&		&	1.41	&	6.81	&	--0.74	&	5.96	&	0.53	&	2.16	&	0.28	&	R766	&	088.D-0026(A)	\\
		4689628129980153216	&	4234	&	1.35	&	1.52	&	6.80	&	--0.75	&	5.76	&	0.34	&	2.26	&	0.39	&	B-32730	&	073.D-0211(A)	\\
		&		&		&	1.55	&	6.81	&	--0.74	&	5.83	&	0.40	&	2.27	&	0.39	&	32730	&	073.D-0211(A)	\\
		4689639812297492608	&	4319	&	1.48	&	1.61	&	6.79	&	--0.76	&	5.90	&	0.49	&	2.23	&	0.37	&	B-33689	&	073.D-0211(A)	\\
		4689628129977913728	&	4234	&	1.41	&	1.54	&	6.81	&	--0.74	&	5.72	&	0.29	&	2.24	&	0.36	&	B-33886	&	073.D-0211(A)	\\
		4689639713520390144	&	4491	&	1.84	&	1.34	&	6.84	&	--0.71	&	5.62	&	0.16	&	2.17	&	0.26	&	B-36153	&	073.D-0211(A)	\\
		4689639954038489984	&	4436	&	1.59	&	1.66	&	6.75	&	--0.80	&	5.50	&	0.13	&	1.95	&	0.13	&	B-36270	&	073.D-0211(A)	\\
		4689640435075431040	&	4423	&	1.73	&	1.40	&	6.93	&	--0.62	&	5.86	&	0.31	&	2.29	&	0.29	&	B-37056	&	073.D-0211(A)	\\
		4689645245438479360	&	4515	&	1.77	&	1.66	&	6.78	&	--0.77	&	5.67	&	0.27	&	2.07	&	0.22	&	B-3769	&	073.D-0211(A)	\\
		4689640121528464128	&	4341	&	1.46	&	1.81	&	6.78	&	--0.77	&	5.54	&	0.14	&	2.12	&	0.27	&	B-38890	&	073.D-0211(A)	\\
		4689639988398164736	&	4489	&	1.86	&	1.44	&	6.84	&	--0.71	&	5.74	&	0.28	&	2.20	&	0.29	&	B-38976	&	073.D-0211(A)	\\
		&		&		&	1.39	&	6.77	&	--0.78	&	5.81	&	0.42	&	2.18	&	0.34	&	R762	&	088.D-0026(A)	\\
		4689644592602980608	&	4243	&	1.29	&	1.78	&	6.73	&	--0.82	&	5.56	&	0.21	&	1.99	&	0.19	&	B-4047	&	073.D-0211(A)	\\
		4689643660581944192	&	4301	&	1.47	&	1.48	&	6.92	&	--0.63	&	5.92	&	0.38	&	2.27	&	0.28	&	B-41429	&	073.D-0211(A)	\\
		&		&		&	1.62	&	6.82	&	--0.73	&	5.99	&	0.55	&	2.22	&	0.33	&	R392	&	088.D-0026(A)	\\
		4689640774370214528	&	4235	&	1.32	&	1.60	&	6.82	&	--0.73	&	5.86	&	0.42	&	2.20	&	0.31	&	B-41595	&	073.D-0211(A)	\\
		4689643630522760960	&	4215	&	1.30	&	1.52	&	6.89	&	--0.66	&	5.95	&	0.44	&	2.30	&	0.34	&	B-41654	&	073.D-0211(A)	\\
		4689643733610139648	&	4430	&	1.69	&	1.55	&	6.80	&	--0.75	&	6.07	&	0.65	&	2.19	&	0.32	&	B-42866	&	073.D-0211(A)	\\
		&		&		&	1.47	&	6.79	&	--0.76	&	6.13	&	0.72	&	2.23	&	0.37	&	42866	&	073.D-0211(A)	\\
		&		&		&	1.50	&	6.78	&	--0.77	&	6.10	&	0.70	&	2.14	&	0.29	&	R656	&	088.D-0026(A)	\\
		4689640984830516992	&	4446	&	1.78	&	1.52	&	6.82	&	--0.73	&	5.85	&	0.41	&	2.20	&	0.31	&	B-42887	&	073.D-0211(A)	\\
		&		&		&	1.54	&	6.75	&	--0.80	&	5.84	&	0.47	&	2.18	&	0.36	&	R760	&	088.D-0026(A)	\\
		4689622877246364288	&	4543	&	1.74	&	1.54	&	6.74	&	--0.81	&	5.86	&	0.50	&	2.24	&	0.43	&	B-4374	&	073.D-0211(A)	\\
		4689643767969828352	&	4457	&	1.75	&	1.54	&	6.85	&	--0.70	&	6.10	&	0.63	&	2.25	&	0.33	&	B-43852	&	073.D-0211(A)	\\
		&		&		&	1.49	&	6.78	&	--0.77	&	6.07	&	0.67	&	2.25	&	0.40	&	43852	&	073.D-0211(A)	\\
		&		&		&	1.45	&	6.82	&	--0.73	&	6.12	&	0.68	&	2.22	&	0.33	&	R704	&	088.D-0026(A)	\\
		4689625144989910784	&	4447	&	1.78	&	1.47	&	6.81	&	--0.74	&	5.68	&	0.25	&	2.18	&	0.30	&	B-5362	&	073.D-0211(A)	\\
		&		&		&	1.47	&	6.78	&	--0.77	&	5.68	&	0.28	&	2.21	&	0.36	&	R253	&	088.D-0026(A)	\\
		4689626107062230528	&	4243	&	1.36	&	1.40	&	6.85	&	--0.70	&	5.71	&	0.24	&	2.37	&	0.45	&	B-7330	&	073.D-0211(A)	\\
		4689625793515738368	&	4430	&	1.77	&	1.31	&	6.92	&	--0.63	&	5.70	&	0.16	&	2.27	&	0.28	&	B-7391	&	073.D-0211(A)	\\
		4689625759155999232	&	4226	&	1.29	&	1.41	&	7.01	&	--0.54	&	5.84	&	0.21	&	2.42	&	0.34	&	B-7752	&	073.D-0211(A)	\\
		4689625729105416704	&	4456	&	1.76	&	1.43	&	6.81	&	--0.74	&	5.72	&	0.29	&	2.27	&	0.39	&	B-7993	&	073.D-0211(A)	\\
		&		&		&	1.45	&	6.79	&	--0.76	&	5.78	&	0.37	&	2.16	&	0.30	&	R277	&	088.D-0026(A)	\\
		4689626343279687168	&	4351	&	1.51	&	1.43	&	6.85	&	--0.70	&	5.70	&	0.23	&	2.28	&	0.36	&	B-8742	&	073.D-0211(A)	\\
		4689623530081721216	&	4363	&	1.68	&	1.55	&	6.81	&	--0.74	&	5.74	&	0.31	&	2.21	&	0.33	&	B-8871	&	073.D-0211(A)	\\
		4689623770599380864	&	4575	&	1.81	&	1.73	&	6.68	&	--0.87	&	5.55	&	0.25	&	1.98	&	0.23	&	B-9121	&	073.D-0211(A)	\\
		4689629474317872256	&	4456	&	1.79	&	1.55	&	6.73	&	--0.82	&	5.95	&	0.60	&	2.10	&	0.30	&	B-9163	&	073.D-0211(A)	\\
		&		&		&	1.42	&	6.77	&	--0.78	&	5.99	&	0.60	&	2.08	&	0.24	&	R800	&	088.D-0026(A)	\\
		4689618238680828800	&	4303	&	1.50	&	1.58	&	6.75	&	--0.80	&	5.63	&	0.26	&	2.22	&	0.40	&	B-9254	&	073.D-0211(A)	\\
		4689637239620659968	&	4474	&	1.78	&	1.53	&	6.81	&	--0.74	&	5.81	&	0.38	&	2.28	&	0.40	&	B-9997	&	073.D-0211(A)	\\
		&		&		&	1.57	&	6.77	&	--0.78	&	5.89	&	0.50	&	2.21	&	0.37	&	R248	&	088.D-0026(A)	\\
		4689625827883561216	&	4608	&	2.02	&	1.27	&	6.90	&	--0.65	&	5.87	&	0.35	&	2.23	&	0.26	&	F-10396	&	073.D-0211(A)	\\
		4689627069134206080	&	4742	&	2.44	&	1.10	&	6.74	&	--0.81	&	5.56	&	0.20	&	2.13	&	0.32	&	F-11617	&	073.D-0211(A)	\\
		4689627825048445312	&	4731	&	2.49	&	0.99	&	6.82	&	--0.73	&	5.65	&	0.21	&	2.18	&	0.29	&	F-11867	&	073.D-0211(A)	\\
		4689830753653279104	&	4669	&	2.16	&	1.34	&	6.78	&	--0.77	&	5.82	&	0.42	&	2.18	&	0.33	&	F-121	&	073.D-0211(A)	\\
		4689632974708827776	&	4651	&	2.20	&	1.24	&	6.89	&	--0.66	&	5.97	&	0.46	&	2.09	&	0.13	&	F-12408	&	073.D-0211(A)	\\
		&		&		&	1.49	&	6.82	&	--0.73	&	6.00	&	0.56	&	2.24	&	0.35	&	R784	&	088.D-0026(A)	\\
		4689633700565630848	&	4777	&	2.48	&	1.14	&	6.78	&	--0.77	&	6.01	&	0.61	&	2.31	&	0.46	&	F-12518	&	073.D-0211(A)	\\
		4689633696284068480	&	4786	&	2.61	&	1.12	&	6.70	&	--0.85	&	5.56	&	0.24	&	2.21	&	0.44	&	F-12919	&	073.D-0211(A)	\\
		4689633082090295808	&	4673	&	2.40	&	1.12	&	6.81	&	--0.74	&	5.61	&	0.18	&	2.01	&	0.13	&	F-13235	&	073.D-0211(A)	\\
		4689637819439945856	&	4772	&	2.42	&	1.17	&	6.84	&	--0.71	&	5.73	&	0.27	&	2.31	&	0.40	&	F-13569	&	073.D-0211(A)	\\
		4689626446376048000	&	4799	&	2.46	&	1.19	&	6.83	&	--0.72	&	5.76	&	0.31	&	2.33	&	0.43	&	F-13631	&	073.D-0211(A)	\\
		4689628099926335232	&	4723	&	2.42	&	1.12	&	6.87	&	--0.68	&	5.82	&	0.33	&	2.20	&	0.26	&	F-13663	&	073.D-0211(A)	\\
		4689644730042488192	&	4545	&	2.01	&	1.44	&	6.76	&	--0.79	&	5.60	&	0.22	&	2.11	&	0.28	&	F-1389	&	073.D-0211(A)	\\
		&		&		&	1.34	&	6.80	&	--0.75	&	5.68	&	0.26	&	2.15	&	0.28	&	1389	&	073.D-0211(A)	\\
		4689637823736084736	&	4779	&	2.37	&	1.12	&	6.87	&	--0.68	&	5.85	&	0.36	&	2.35	&	0.41	&	F-13977	&	073.D-0211(A)	\\
		4689634044162968192	&	4685	&	2.38	&	1.13	&	6.72	&	--0.83	&	5.61	&	0.27	&	2.14	&	0.35	&	F-14223	&	073.D-0211(A)	\\
		4689638442211483136	&	4761	&	2.33	&	1.24	&	6.85	&	--0.70	&	6.09	&	0.62	&	2.24	&	0.32	&	F-14454	&	073.D-0211(A)	\\
		4689638442211457920	&	4725	&	2.28	&	1.30	&	6.85	&	--0.70	&	5.94	&	0.47	&	2.30	&	0.38	&	F-15451	&	073.D-0211(A)	\\
		&		&		&	1.30	&	6.79	&	--0.76	&	5.88	&	0.47	&	2.31	&	0.45	&	15451	&	073.D-0211(A)	\\
		4689626519378816768	&	4746	&	2.33	&	1.19	&	6.82	&	--0.73	&	6.00	&	0.56	&	2.17	&	0.28	&	F-16258	&	073.D-0211(A)	\\
		4689638712785607680	&	4663	&	2.15	&	1.30	&	6.75	&	--0.80	&	5.63	&	0.26	&	2.18	&	0.36	&	F-18549	&	073.D-0211(A)	\\
		4689627683309862784	&	4772	&	2.44	&	1.13	&	6.85	&	--0.70	&	5.91	&	0.44	&	2.14	&	0.22	&	F-22407	&	073.D-0211(A)	\\
		4689641946904350720	&	4761	&	2.33	&	1.37	&	6.84	&	--0.71	&	5.99	&	0.53	&	2.26	&	0.35	&	F-24463	&	073.D-0211(A)	\\
		&		&		&	1.21	&	6.84	&	--0.71	&	6.07	&	0.61	&	2.30	&	0.39	&	24463	&	073.D-0211(A)	\\
		4689642840257564032	&	4757	&	2.49	&	1.11	&	6.83	&	--0.72	&	5.62	&	0.17	&	2.27	&	0.37	&	F-27938	&	073.D-0211(A)	\\
		4689831853164838784	&	4783	&	2.40	&	1.27	&	6.81	&	--0.74	&	5.98	&	0.55	&	2.40	&	0.52	&	F-2861	&	073.D-0211(A)	\\
		4689642737178089472	&	4764	&	2.50	&	1.16	&	6.82	&	--0.73	&	5.67	&	0.23	&	2.14	&	0.25	&	F-30311	&	073.D-0211(A)	\\
		4689643115127053184	&	4499	&	1.86	&	1.46	&	6.78	&	--0.77	&	5.66	&	0.26	&	2.17	&	0.32	&	F-31510	&	073.D-0211(A)	\\
		4689642423637641984	&	4786	&	2.44	&	1.21	&	6.76	&	--0.79	&	5.69	&	0.31	&	2.20	&	0.37	&	F-31519	&	073.D-0211(A)	\\
		4689643145217009792	&	4799	&	2.47	&	1.22	&	6.84	&	--0.71	&	5.86	&	0.40	&	2.27	&	0.36	&	F-31758	&	073.D-0211(A)	\\
		4689628134286385792	&	4772	&	2.41	&	1.26	&	6.80	&	--0.75	&	6.12	&	0.70	&	2.14	&	0.27	&	F-33686	&	073.D-0211(A)	\\
		4689639640518059136	&	4757	&	2.43	&	1.23	&	6.72	&	--0.83	&	5.65	&	0.31	&	2.11	&	0.32	&	F-35206	&	073.D-0211(A)	\\
		4689625041910730112	&	4755	&	2.31	&	1.20	&	6.84	&	--0.71	&	5.94	&	0.48	&	2.16	&	0.25	&	F-3616	&	073.D-0211(A)	\\
		4689639709218245632	&	4586	&	2.00	&	1.32	&	6.81	&	--0.74	&	6.01	&	0.58	&	2.26	&	0.38	&	F-37781	&	073.D-0211(A)	\\
		4689640671311342720	&	4792	&	2.46	&	1.24	&	6.80	&	--0.75	&	5.98	&	0.56	&	2.25	&	0.38	&	F-40001	&	073.D-0211(A)	\\
		4689640538154028672	&	4783	&	2.45	&	1.15	&	6.86	&	--0.69	&	6.03	&	0.55	&	2.18	&	0.25	&	F-40035	&	073.D-0211(A)	\\
		4689641122269520640	&	4759	&	2.43	&	1.12	&	6.84	&	--0.71	&	6.00	&	0.54	&	2.13	&	0.22	&	F-42112	&	073.D-0211(A)	\\
		4689642389271520512	&	4755	&	2.35	&	1.24	&	6.82	&	--0.73	&	5.90	&	0.46	&	2.29	&	0.40	&	F-42188	&	073.D-0211(A)	\\
		4689641190988972288	&	4766	&	2.44	&	1.30	&	6.81	&	--0.74	&	6.18	&	0.75	&	2.49	&	0.61	&	F-43415	&	073.D-0211(A)	\\
		4689625076270567168	&	4595	&	2.03	&	1.29	&	6.76	&	--0.79	&	5.60	&	0.22	&	2.08	&	0.25	&	F-4754	&	073.D-0211(A)	\\
		4689624732673041664	&	4740	&	2.35	&	1.27	&	6.80	&	--0.75	&	5.72	&	0.30	&	2.21	&	0.34	&	F-4974	&	073.D-0211(A)	\\
		4689622533650157056	&	4759	&	2.50	&	1.08	&	6.83	&	--0.72	&	5.98	&	0.53	&	2.18	&	0.28	&	F-5404	&	073.D-0211(A)	\\
		4689623255203461632	&	4702	&	2.31	&	1.21	&	6.83	&	--0.72	&	5.90	&	0.45	&	2.21	&	0.31	&	F-5576	&	073.D-0211(A)	\\
		4689628546604996608	&	4772	&	2.50	&	1.10	&	6.76	&	--0.79	&	6.13	&	0.75	&	2.15	&	0.32	&	F-6363	&	073.D-0211(A)	\\
		4689625484312800640	&	4729	&	2.32	&	1.14	&	6.79	&	--0.76	&	5.71	&	0.30	&	2.31	&	0.45	&	F-6420	&	073.D-0211(A)	\\
		4689623152124232576	&	4764	&	2.46	&	1.11	&	6.81	&	--0.74	&	5.90	&	0.47	&	2.11	&	0.23	&	F-7200	&	073.D-0211(A)	\\
		4689625793517924480	&	4733	&	2.27	&	1.23	&	6.86	&	--0.69	&	5.85	&	0.37	&	2.16	&	0.23	&	F-7711	&	073.D-0211(A)	\\
		&		&		&	1.27	&	6.88	&	--0.67	&	5.89	&	0.39	&	2.41	&	0.46	&	7711	&	073.D-0211(A)	\\
		4689625763465150336	&	4751	&	2.29	&	1.31	&	6.81	&	--0.74	&	6.15	&	0.72	&	2.28	&	0.40	&	F-8025	&	073.D-0211(A)	\\
		4689628684043911040	&	4710	&	2.42	&	1.02	&	6.86	&	--0.69	&	5.75	&	0.27	&	2.08	&	0.15	&	F-8173	&	073.D-0211(A)	\\
		4689626725536875904	&	4764	&	2.44	&	1.12	&	6.82	&	--0.73	&	5.93	&	0.49	&	2.19	&	0.30	&	F-9325	&	073.D-0211(A)	\\
		4689625832184604160	&	4755	&	2.44	&	1.16	&	6.78	&	--0.77	&	5.64	&	0.24	&	2.14	&	0.29	&	F-9733	&	073.D-0211(A)	\\
		4689626313220598400	&	4605	&	2.01	&	1.26	&	6.89	&	--0.66	&	6.04	&	0.53	&	2.31	&	0.35	&	F-9734	&	073.D-0211(A)	\\
		4689618891515852160	&	4704	&	2.19	&	1.44	&	6.81	&	--0.74	&	6.10	&	0.67	&	2.29	&	0.41	&	R198	&	088.D-0026(A)	\\
		4689619166393741824	&	4512	&	1.96	&	1.46	&	6.82	&	--0.73	&	5.73	&	0.29	&	2.22	&	0.33	&	R199	&	088.D-0026(A)	\\
		4689830508836705664	&	4217	&	1.32	&	1.61	&	6.73	&	--0.82	&	5.63	&	0.28	&	2.18	&	0.38	&	R213	&	088.D-0026(A)	\\
		4689641465868395648	&	4599	&	2.06	&	1.40	&	6.79	&	--0.76	&	5.60	&	0.19	&	2.09	&	0.23	&	R214	&	088.D-0026(A)	\\
		4689638132973884160	&	4545	&	1.92	&	1.43	&	6.75	&	--0.80	&	5.68	&	0.31	&	2.08	&	0.26	&	R221	&	088.D-0026(A)	\\
		4689638407851742464	&	4665	&	2.12	&	1.36	&	6.77	&	--0.78	&	5.92	&	0.53	&	2.13	&	0.29	&	R222	&	088.D-0026(A)	\\
		4689830684933831552	&	4620	&	2.17	&	1.37	&	6.78	&	--0.77	&	5.72	&	0.32	&	2.23	&	0.38	&	R228	&	088.D-0026(A)	\\
		4689642737171879936	&	4563	&	2.00	&	1.45	&	6.75	&	--0.80	&	5.55	&	0.18	&	2.15	&	0.33	&	R231	&	088.D-0026(A)	\\
		&		&		&	1.37	&	6.69	&	--0.86	&	5.57	&	0.26	&	2.09	&	0.33	&	29490	&	073.D-0211(A)	\\
		4689638300460480640	&	4690	&	2.11	&	1.35	&	6.79	&	--0.76	&	6.04	&	0.63	&	2.29	&	0.43	&	R234	&	088.D-0026(A)	\\
		4689625316788603648	&	4569	&	1.99	&	1.39	&	6.78	&	--0.77	&	5.75	&	0.35	&	2.21	&	0.36	&	R238	&	088.D-0026(A)	\\
		4689637514498506240	&	4727	&	2.28	&	1.31	&	6.85	&	--0.70	&	5.90	&	0.43	&	2.48	&	0.56	&	R239	&	088.D-0026(A)	\\
		4689642286190119680	&	4395	&	1.63	&	1.62	&	6.75	&	--0.80	&	5.84	&	0.47	&	2.23	&	0.41	&	R247	&	088.D-0026(A)	\\
		4689638609700267264	&	4647	&	2.03	&	1.47	&	6.84	&	--0.71	&	6.20	&	0.74	&	2.37	&	0.46	&	R249	&	088.D-0026(A)	\\
		&		&		&	1.46	&	6.85	&	--0.70	&	6.21	&	0.74	&	2.39	&	0.47	&	21369	&	073.D-0211(A)	\\
		4689642290501499904	&	4241	&	1.36	&	1.55	&	6.82	&	--0.73	&	5.91	&	0.47	&	2.28	&	0.39	&	R256	&	088.D-0026(A)	\\
		&		&		&	1.55	&	6.81	&	--0.74	&	5.95	&	0.52	&	2.28	&	0.40	&	B-29146	&	073.D-0211(A)	\\
		4689637548858307584	&	4657	&	2.22	&	1.30	&	6.81	&	--0.74	&	5.69	&	0.26	&	2.26	&	0.38	&	R258	&	088.D-0026(A)	\\
		4689831165970072448	&	4486	&	1.77	&	1.65	&	6.79	&	--0.76	&	5.97	&	0.56	&	2.23	&	0.37	&	R264	&	088.D-0026(A)	\\
		4689643351342005888	&	4417	&	1.71	&	1.51	&	6.73	&	--0.82	&	5.58	&	0.23	&	2.11	&	0.31	&	R266	&	088.D-0026(A)	\\
		4689625454227561984	&	4667	&	2.26	&	1.34	&	6.78	&	--0.77	&	5.71	&	0.31	&	2.14	&	0.29	&	R269	&	088.D-0026(A)	\\
		4689638682729411712	&	4612	&	1.95	&	1.44	&	6.78	&	--0.77	&	6.00	&	0.60	&	2.26	&	0.41	&	R273	&	088.D-0026(A)	\\
		4689637922511631872	&	4667	&	2.14	&	1.32	&	6.76	&	--0.79	&	5.61	&	0.23	&	2.11	&	0.28	&	R274	&	088.D-0026(A)	\\
		4689642251832561280	&	4573	&	1.98	&	1.49	&	6.78	&	--0.77	&	5.73	&	0.33	&	2.26	&	0.41	&	R279	&	088.D-0026(A)	\\
		4689643145185459072	&	4635	&	2.10	&	1.45	&	6.87	&	--0.68	&	5.99	&	0.50	&	2.21	&	0.27	&	R281	&	088.D-0026(A)	\\
		4689831264750937344	&	4237	&	1.32	&	1.87	&	6.82	&	--0.73	&	5.74	&	0.30	&	2.09	&	0.20	&	R283	&	088.D-0026(A)	\\
		4689637926815132032	&	4639	&	2.08	&	1.36	&	6.87	&	--0.68	&	5.96	&	0.47	&	2.29	&	0.35	&	R288	&	088.D-0026(A)	\\
		4689643218214134016	&	4649	&	2.12	&	1.38	&	6.86	&	--0.69	&	5.96	&	0.48	&	2.17	&	0.24	&	R310	&	088.D-0026(A)	\\
		4689624556566726912	&	4725	&	2.29	&	1.37	&	6.82	&	--0.73	&	5.88	&	0.44	&	2.08	&	0.19	&	R317	&	088.D-0026(A)	\\
		&		&		&	1.29	&	6.82	&	--0.73	&	5.92	&	0.48	&	2.33	&	0.44	&	F-3476	&	073.D-0211(A)	\\
		4689639159462267392	&	4426	&	1.65	&	1.56	&	6.79	&	--0.76	&	5.87	&	0.46	&	2.24	&	0.38	&	R330	&	088.D-0026(A)	\\
		4689637991231168768	&	4466	&	1.79	&	1.51	&	6.75	&	--0.80	&	5.68	&	0.31	&	2.23	&	0.41	&	R342	&	088.D-0026(A)	\\
		4689638820168905088	&	4702	&	2.10	&	1.37	&	6.78	&	--0.77	&	5.96	&	0.56	&	2.21	&	0.36	&	R346	&	088.D-0026(A)	\\
		4689638815864893312	&	4624	&	2.11	&	1.32	&	6.83	&	--0.72	&	5.77	&	0.32	&	2.36	&	0.46	&	R348	&	088.D-0026(A)	\\
		4689624835742003840	&	4575	&	2.06	&	1.43	&	6.74	&	--0.81	&	5.75	&	0.39	&	2.23	&	0.42	&	R352	&	088.D-0026(A)	\\
		4689625969623521152	&	4663	&	2.06	&	1.41	&	6.79	&	--0.76	&	5.90	&	0.49	&	2.27	&	0.41	&	R356	&	088.D-0026(A)	\\
		4689637785072541952	&	4663	&	2.15	&	1.38	&	6.81	&	--0.74	&	5.87	&	0.44	&	2.33	&	0.45	&	R360	&	088.D-0026(A)	\\
		4689637995526918400	&	4535	&	1.80	&	1.38	&	6.82	&	--0.73	&	5.98	&	0.54	&	2.32	&	0.43	&	R377	&	088.D-0026(A)	\\
		4689832712158218880	&	4498	&	1.80	&	1.60	&	6.77	&	--0.78	&	6.02	&	0.63	&	2.17	&	0.33	&	R390	&	088.D-0026(A)	\\
		4689624797099838080	&	4267	&	1.35	&	1.72	&	6.73	&	--0.82	&	6.05	&	0.70	&	2.26	&	0.46	&	R408	&	088.D-0026(A)	\\
		4689643905408840320	&	4426	&	1.68	&	1.52	&	6.82	&	--0.73	&	5.98	&	0.54	&	2.21	&	0.32	&	R450	&	088.D-0026(A)	\\
		&		&		&	1.54	&	6.84	&	--0.71	&	6.04	&	0.58	&	2.24	&	0.33	&	B-43889	&	073.D-0211(A)	\\
		&		&		&	1.51	&	6.81	&	--0.74	&	6.01	&	0.58	&	2.29	&	0.41	&	43889	&	073.D-0211(A)	\\
		4689643905408839424	&	4545	&	1.90	&	1.37	&	6.80	&	--0.75	&	5.92	&	0.50	&	2.23	&	0.36	&	R512	&	088.D-0026(A)	\\
		&		&		&	1.54	&	6.78	&	--0.77	&	5.94	&	0.54	&	2.20	&	0.35	&	F-43632	&	073.D-0211(A)	\\
		&		&		&	1.38	&	6.79	&	--0.76	&	5.95	&	0.54	&	2.18	&	0.32	&	43632	&	073.D-0211(A)	\\
		4689645107999550720	&	4288	&	1.50	&	1.52	&	6.80	&	--0.75	&	5.70	&	0.28	&	2.19	&	0.32	&	R563	&	088.D-0026(A)	\\
		&		&		&	1.56	&	6.77	&	--0.78	&	5.69	&	0.30	&	2.22	&	0.38	&	B-3449	&	073.D-0211(A)	\\
		4689643729302626688	&	4489	&	1.84	&	1.28	&	6.81	&	--0.74	&	5.57	&	0.14	&	2.23	&	0.35	&	R613	&	088.D-0026(A)	\\
		4689623049045585280	&	4717	&	2.26	&	1.28	&	6.88	&	--0.67	&	5.93	&	0.43	&	2.27	&	0.32	&	R615	&	088.D-0026(A)	\\
		4689626072702505856	&	4558	&	1.94	&	1.36	&	6.82	&	--0.73	&	5.95	&	0.51	&	2.34	&	0.45	&	R617	&	088.D-0026(A)	\\
		4689621880814574592	&	4498	&	1.88	&	1.38	&	6.74	&	--0.81	&	5.54	&	0.18	&	2.14	&	0.33	&	R640	&	088.D-0026(A)	\\
		4689627481451470208	&	4748	&	2.22	&	1.51	&	6.75	&	--0.80	&	6.10	&	0.73	&	2.34	&	0.52	&	R643	&	088.D-0026(A)	\\
		4689639674858443136	&	4270	&	1.38	&	1.75	&	6.76	&	--0.79	&	5.85	&	0.47	&	2.30	&	0.47	&	R650	&	088.D-0026(A)	\\
		4689622739807987328	&	4641	&	2.16	&	1.31	&	6.76	&	--0.79	&	5.60	&	0.22	&	2.03	&	0.20	&	R664	&	088.D-0026(A)	\\
		4689621709015302400	&	4560	&	2.04	&	1.29	&	6.81	&	--0.74	&	5.88	&	0.45	&	1.99	&	0.11	&	R677	&	088.D-0026(A)	\\
		4689627515811202176	&	4669	&	2.12	&	1.39	&	6.76	&	--0.79	&	5.88	&	0.50	&	2.22	&	0.39	&	R680	&	088.D-0026(A)	\\
		4689639331261085824	&	4526	&	1.94	&	1.41	&	6.75	&	--0.80	&	5.72	&	0.35	&	2.19	&	0.37	&	R683	&	088.D-0026(A)	\\
		4689640744312947072	&	4527	&	1.90	&	1.30	&	6.81	&	--0.74	&	5.64	&	0.21	&	2.09	&	0.21	&	R689	&	088.D-0026(A)	\\
		4689641190988963456	&	4632	&	2.13	&	1.41	&	6.76	&	--0.79	&	5.89	&	0.51	&	2.06	&	0.23	&	R717	&	088.D-0026(A)	\\
		4689626240194530560	&	4710	&	2.24	&	1.33	&	6.81	&	--0.74	&	6.21	&	0.78	&	2.28	&	0.40	&	R730	&	088.D-0026(A)	\\
		4689640843083316480	&	4677	&	2.19	&	1.45	&	6.83	&	--0.72	&	5.94	&	0.49	&	2.29	&	0.39	&	R735	&	088.D-0026(A)	\\
		4689640057117702272	&	4665	&	2.21	&	1.30	&	6.83	&	--0.72	&	5.75	&	0.30	&	2.18	&	0.28	&	R738	&	088.D-0026(A)	\\
		4689621846454233216	&	4589	&	2.11	&	1.26	&	6.80	&	--0.75	&	5.60	&	0.18	&	2.04	&	0.17	&	R740	&	088.D-0026(A)	\\
		4689639850959315072	&	4634	&	2.04	&	1.29	&	6.81	&	--0.74	&	5.97	&	0.54	&	2.28	&	0.40	&	R744	&	088.D-0026(A)	\\
		4689640946182241792	&	4538	&	1.89	&	1.44	&	6.79	&	--0.76	&	6.01	&	0.60	&	2.28	&	0.42	&	R745	&	088.D-0026(A)	\\
		4689623560134016256	&	4595	&	2.01	&	1.50	&	6.79	&	--0.76	&	5.76	&	0.35	&	2.24	&	0.38	&	R756	&	088.D-0026(A)	\\
		&		&		&	1.42	&	6.76	&	--0.79	&	5.79	&	0.41	&	2.23	&	0.40	&	F-10198	&	073.D-0211(A)	\\
		4689575185928019840	&	4275	&	1.45	&	1.47	&	6.77	&	--0.78	&	5.58	&	0.19	&	2.12	&	0.28	&	R761	&	088.D-0026(A)	\\
		4689641019190234496	&	4569	&	1.97	&	1.43	&	6.82	&	--0.73	&	6.00	&	0.56	&	2.20	&	0.31	&	R767	&	088.D-0026(A)	\\
		4689626755588392192	&	4645	&	2.17	&	1.34	&	6.86	&	--0.69	&	6.17	&	0.69	&	2.35	&	0.42	&	R769	&	088.D-0026(A)	\\
		4689622288822455168	&	4342	&	1.66	&	1.52	&	6.78	&	--0.77	&	5.74	&	0.34	&	2.18	&	0.33	&	R770	&	088.D-0026(A)	\\
		4689575289007212416	&	4231	&	1.34	&	1.56	&	6.77	&	--0.78	&	5.60	&	0.21	&	2.17	&	0.33	&	R772	&	088.D-0026(A)	\\
		4689622361850255488	&	4446	&	1.72	&	1.45	&	6.78	&	--0.77	&	5.92	&	0.52	&	2.24	&	0.39	&	R774	&	088.D-0026(A)	\\
		4689645795196398208	&	4586	&	2.03	&	1.36	&	6.70	&	--0.85	&	5.63	&	0.31	&	2.02	&	0.25	&	R777	&	088.D-0026(A)	\\
		4689632669773531520	&	4671	&	2.24	&	1.27	&	6.82	&	--0.73	&	5.85	&	0.41	&	2.10	&	0.21	&	R782	&	088.D-0026(A)	\\
		&		&		&	1.26	&	6.82	&	--0.73	&	5.96	&	0.52	&	2.10	&	0.21	&	F-10527	&	073.D-0211(A)	\\
		4689581263316313856	&	4553	&	2.07	&	1.53	&	6.77	&	--0.78	&	5.63	&	0.24	&	2.05	&	0.21	&	R786	&	088.D-0026(A)	\\
		4689646001354810368	&	4554	&	2.00	&	1.50	&	6.81	&	--0.74	&	6.07	&	0.64	&	2.17	&	0.29	&	R788	&	088.D-0026(A)	\\
		4689634181601863296	&	4512	&	1.88	&	1.45	&	6.78	&	--0.77	&	5.90	&	0.50	&	2.12	&	0.27	&	R789	&	088.D-0026(A)	\\
		4689633941083703424	&	4501	&	1.96	&	1.38	&	6.77	&	--0.78	&	5.66	&	0.27	&	2.10	&	0.26	&	R793	&	088.D-0026(A)	\\
	\end{longtable}
\end{center}
Note; In cases where abundances were determined using data from several observing programmes we provide microturbulence velocities and abundances obtained using spectra from each individual programme. The IDs of individual stars and IDs of the corresponding observing programmes are given in the last two columns, respectively (072.D-0777(A), PI: P.~Fran\c{c}ois; 073.D-0211(A), PI: E.~Carretta; 088.D-0026(A), PI: I.~McDonald).
\twocolumn

\end{appendix}

\end{document}